\documentclass[twocolumn,iop]{openjournal}
\usepackage{graphicx}
\usepackage{amsmath}
\usepackage{amssymb}
\usepackage{xspace}
\usepackage{color}
\usepackage[colorlinks=true,allcolors=blue]{hyperref}
\usepackage{orcidlink}

\begin{document}

\title{A Unified Halo Mass Function Across Dark Matter Models from High-Resolution Multi-Scale Simulations}
\shorttitle{Unified Halo Mass Function Across Dark Matter Models}

\shortauthors{Benson, Nadler, Du \& Gluscevic}
\author{Andrew J. Benson$^{*}$\,\orcidlink{0000-0001-5501-6008}}
\affiliation{Carnegie Institution for Science, 813 Santa Barbara Street, Pasadena, CA 91101, USA}
\email{${}^{*}$\text{abenson@carnegiescience.edu}}

\author{Ethan O. Nadler\,\orcidlink{0000-0002-1182-3825}}
\affiliation{Department of Astronomy \& Astrophysics, University of California, San Diego, La Jolla, CA 92093, USA}

\author{Xiaolong Du\,\orcidlink{0000-0003-0728-2533}}
\affiliation{Department of Physics and Astronomy, University of California, Los Angeles, CA, 90095, USA}

\author{Vera Gluscevic\,\orcidlink{0000-0002-3589-8637}}
\affiliation{Department of Physics \& Astronomy, University of Southern California, Los Angeles, CA 90007, USA}
\affiliation{Institute for Advanced Study, 1 Einstein Drive, Princeton, NJ 08540, USA}
\affiliation{Center for Computational Astrophysics, Flatiron Institute, 162 5th Avenue, New York, NY, 10010, USA}

\begin{abstract}
We measure the dark matter halo mass function, with backsplash halos removed, from a wide range of cosmological-box and zoom-in simulations. These include the MultiDark Planck boxes, along with a suite of zoom-in simulations of Group, Milky Way, and LMC-mass halos. The Milky Way simulations include both CDM and non-CDM initial conditions. Using these measurements, we calibrate the parameters of flexible fitting functions for the halo mass function and the window function, along with parameterized models for various systematics, including finite box size effects, halo isolation criteria, halo detection efficiency, and contamination by artificial halos (objects forming from particle noise in the initial conditions). We show that this model shows remarkable consistency with N-body simulations over a broad range of redshifts, and ten orders of magnitude in halo mass ($10^6\mathrm{M}_\odot$ to $10^{16}\mathrm{M}_\odot$). Our model typically maintains a high precision of 12\% and captures complex behaviors, including small-scale cut-offs, oscillations, and enhancements. In specific mass intervals for certain power spectra, we see larger deviations of 40--50\%. Furthermore, when integrated with a simple model for environmental dependence, this fitting function provides a robust description of how environmental density influences the halo mass function. This precision model captures a wide variety of dark matter paradigms (including thermal relics, axions, and models with dark-sector interactions), is accurate for halo masses down to $10^7\mathrm{M}_\odot$, and is a critical ingredient for model-independent dark-matter inference from forthcoming data.
\end{abstract}

\section{Introduction}

The gravity-induced growth of large-scale structure leads to the formation of non-linear, self-gravitating, approximately virialized structures known as ``dark matter halos,'' within which galaxies form \citep{2018ARA&A..56..435W}. The number of these halos formed, on average, per unit halo mass and per unit volume is characterized by the halo mass function, $n(M|z,\delta_\mathrm{e})$, where $M$ is the halo mass, $z$ the redshift, and $\delta_\mathrm{e}$ is the environmental overdensity, and which is often averaged over all environments $n(M|z) = \int_{-\infty}^{+\infty} p(\delta_\mathrm{e}) n(M|z,\delta_\mathrm{e}) \mathrm{d}\delta_\mathrm{e}$. The mass function is a fundamental statistic characterizing the population of halos, and is a key ingredient in a wide variety of modeling approaches, including the halo model \citep{2002PhR...372....1C,2023OJAp....6E..39A}, and small-scale structure analyses \citep[e.g.,][]{2017ARA&A..55..343B}.

The halo mass function is typically modeled using methods based on the approach of \cite{1974ApJ...187..425P}, who derived an approximation to the halo mass function based on the statistics of Gaussian random fields, and simple, spherical collapse models of halo formation. While the model of \cite{1974ApJ...187..425P} is accurate only to order unity, improvements to this model have received substantial attention, resulting in accurate fitting functions to $n(M,z)$ measured from high-resolution cosmological simulations \citep{2001MNRAS.323....1S,2003MNRAS.346..565R,2007MNRAS.374....2R,2008ApJ...688..709T,2011ApJ...732..122B,2016MNRAS.456.2486D,2016MNRAS.462..893R,ford2026functionalformgalaxyhalo}. Such fitting functions generally consist of two pieces. The first is the fitting function for the halo mass function itself, often a modified version of the original \cite{1974ApJ...187..425P} result:
\begin{equation}
    n(M|z) = \sqrt{\frac{2}{\pi}} \frac{\Omega_\mathrm{M} \rho_\mathrm{crit}}{M^2} \left| \frac{\mathrm{d}\log \sigma}{\mathrm{d}\log M} \right| \nu \exp\left( - \frac{\nu^2}{2} \right),
\end{equation}
where $\Omega_\mathrm{M}$ is the matter density, $\rho_\mathrm{crit}$ is the critical density, $\sigma(M)$ is a measure of the fractional variance in the linear theory density field on scales corresponding to halo mass $M$, and $\nu = \delta_\mathrm{c}/\sigma(M)$ is the ``peak height,'' where $\delta_\mathrm{c}$ is the linear theory critical overdensity for collapse of spherical perturbations. The second piece, often less explicitly stated, is the window function, $W(k|M)$, which appears in the definition of $\sigma(M)$:
\begin{equation}
\sigma^2(M) = \int_0^\infty 4 \pi k^2 P(k) W^2(k|M) \mathrm{d}k,
\end{equation}
where $P(k)$ is the linear theory power spectrum. The window function has important consequences for the halo mass function, as we will discuss throughout.

Recent analyses of quadruply-lensed quasars have been used to constrain the particle nature of dark matter \citep{2020MNRAS.491.6077G,2024MNRAS.530.2960N,2023MNRAS.524.6159K,2025arXiv251107513G}, and the maximum mass at which the halo mass function could be sharply truncated\footnote{That is, if the halo mass function is as predicted by CDM above some mass, $m_\mathrm{low}$, and truncated to zero below that mass, what is the largest value of $m_\mathrm{low}$ that remains consistent with observations of quad lenses? \cite{2026arXiv260405237N} find a constraint of $m_\mathrm{low}=10^{8.3}\mathrm{M}_\odot$ at a 10:1 odds ratio using current data.}. This involves forward-modeling the effects of the halo (and subhalo) population on a lensed system, and so the halo mass function is a crucial ingredient. These methods are sensitive to halos with masses down to $\sim 10^7\mathrm{M}_\odot$. With the expanded sample sizes of quad lenses from LSST \citep{2025RSPTA.38340117S,2025arXiv251107513G}, the statistical power of such analyses will increase, driving constraints on $m_\mathrm{low}$ to lower values~\citep{2026arXiv260405237N}. As such, accurate calibrations of halo mass functions to lower masses are desirable. Similarly, models for the mass of halos infalling into the Milky Way underlie a wide range of analyses that use nearby dwarf satellite galaxies to derive dark matter constraints \citep[e.g.,][]{2019ApJ...878L..32N,2021ApJ...907L..46M,2021PhRvL.126i1101N,2022PhRvD.106l3026D,2023JCAP...11..037A,2025PhRvD.111b3530C}.

As these experiments aim to probe the nature of dark matter and the primordial power spectrum, the halo mass function model that they employ must be applicable across a wide range of dark matter microphysics and power spectrum shapes---for example, prior work has explored the constraints that can be derived on thermal warm \citep{2020MNRAS.491.6077G}, fuzzy dark matter \citep{2022MNRAS.517.1867L}, and running spectral indices \citep{2022MNRAS.512.3163G}, and future analyses will further explore these and similar models. For many such analyses, the model used must be able to \emph{rapidly} predict halo mass functions, for example, to allow rapid generation of tens of millions of halo populations needed in forward modeling approaches \citep[e.g.,][]{2020MNRAS.491.6077G}. This requirement necessitates a model that is applicable, calibrated, and accurate across all dark matter microphysics and power spectrum shapes of interest. If, instead, a new calibration is needed for each dark matter/power spectrum model considered---which typically involves running expensive N-body simulations---this speed advantage is lost, and analyses become less tractable.

Here, we will focus on finding a halo mass function model that is applicable across a wide range of power spectra, representative of those found with these and other dark matter microphysics. We will not, for now, explore the effects of late-time physics (e.g., self-interactions) on the halo mass function. Prior related work has focused either on much larger mass scales \citep[e.g.,][]{2016MNRAS.456.2486D}, or on specific classes of small-scale power spectrum behavior \citep[e.g.,][who focused on models with strong dark acoustic oscillations]{2021MNRAS.506..128B}. As we will show below, these previous models work well in the regimes for which they were calibrated, but often fail when applied outside of those regimes.

In this work, we provide a new calibration of the halo mass function across the mass range $10^6$ to $10^{16}\mathrm{M}_\odot$, and the redshift range $z=0$ to $z=4$. To achieve this, we combine measurements of the halo mass function from a series of cosmological box simulations with measurements from zoom-in re-simulations of the regions around target halos (taking care to account for the fact that these zoom-in simulations represent a biased subset of environments) to span a broad mass range. We further explore numerous non-CDM simulations to assess the extent to which a single fitting function can predict the halo mass function for a wide variety of input power spectra.

The remainder of this paper is organized as follows. In \S\ref{sec:methods:sims} we describe the set of simulations used in this work, and in \S\ref{sec:methods:models} and \S\ref{sec:likelihood} we describe how they are analyzed to measure the halo mass function, and how we constrain a fitting function to match these data. We present the results of this analysis in \S\ref{sec:results}, and examine the implications of these results in \S\ref{sec:discussion}. Finally, in \S\ref{sec:conclusion} we present our conclusions.

\section{N-body Simulations}\label{sec:methods:sims}

In this section, we describe the set of N-body simulations that we use to constrain the mass function. In order to span a wide range of halo masses, we make use of both cosmological volume simulations and high-resolution zoom-ins around individual halos, as described in \S\ref{sec:mdpl} and \S\ref{sec:symphony} respectively.

\subsection{MultiDark Planck suite}\label{sec:mdpl}

\begin{table}
\caption{Box sizes, particle masses and minimum halo masses (used in our likelihood analysis) of the simulations in the MultiDark Planck simulation suite.}
\label{tb:multiDarkResolutions}
\centering
\begin{tabular}{cr@{}c@{}lr@{.}lr@{.}l}
\hline
{\bf Simulation} & \multicolumn{3}{c}{\boldmath{$L_\mathrm{box}/h^{-1}\mathrm{Mpc}$}} & \multicolumn{2}{c}{\boldmath{$m_\mathrm{p}/h^{-1}\mathrm{M}_\odot$}} & \multicolumn{2}{c}{\boldmath{$M_\mathrm{min}/h^{-1}\mathrm{M}_\odot$}} \\
\hline
VSMDPL   &              & &160 & 6&$20\times10^{6}$  & 1&$86\times10^{9}$ \\
SMDPL    &              & &400 & 9&$63\times10^{7}$  & 2&$89\times10^{10}$ \\
MDPL2    &             1&,&000 & 1&$51\times10^{9}$  & 4&$53\times10^{11}$ \\
BigMDPL  &             2&,&500 & 2&$36\times10^{10}$ & 7&$08\times10^{12}$ \\
HugeMDPL & \hspace{2em}4&,&000 & 7&$90\times10^{10}$ & 2&$37\times10^{13}$ \\
\hline
\end{tabular}
\end{table}

We make use of the MultiDark Planck (MDPL) suite of cosmological simulations \citep{2013AN....334..691R,2016MNRAS.457.4340K}. These consist of five cosmological boxes, with different sizes and particle masses (summarized in Table~\ref{tb:multiDarkResolutions}), but all run with the same cosmological and power spectrum parameters---specifically $(h,\Omega_\Lambda,\Omega_\mathrm{m},\Omega_\mathrm{b},n_\mathrm{s},\sigma_8)=(0.6777,0.693,0.307,0.0482,0.96,0.8228)$ where $h=H_0/(100 \hbox{km/s/Mpc})$ is the Hubble constant, $\Omega_\Lambda$, $\Omega_\mathrm{m}$, and $\Omega_\mathrm{b}$ are the fractions of the critical density in dark energy, matter, and baryons respectively, $n_\mathrm{s}$ is the logarithmic slope of the primordial power spectrum, and $\sigma_8$ is the linear theory amplitude of mass fluctuations in 8~Mpc$/h$ radius spheres at $z=0$.

As we will discuss in more detail in \S\ref{sec:nbodyAnalysis}, in this work we analyze only isolated halos---that is, we exclude subhalos and backsplash halos (i.e., halos which were, at some time in the past, a subhalo in a larger halo). Subhalos and backsplash halos have typically experienced strong tidal interactions with their host, which significantly alter their masses. Additionally, in semi-analytic models of (sub)halo evolution, the backsplash population is a distinct population, accounted for in the merger trees, and therefore should be excluded from our present analysis, focusing only on truly isolated halos.

We extract halos consisting of 30 particles or more, \emph{but} consider only those halos consisting of 300 or more particles in our likelihood analysis, resulting in a minimum halo mass, $M_\mathrm{min}$, as shown in Table~\ref{tb:multiDarkResolutions}. Halos resolved with fewer particles are more prone to numerical effects, including random fluctuations in their masses \citep{2010ApJ...711.1198T,2017MNRAS.467.3454B}, and contamination by ``artificial halos'' \citep{2007MNRAS.380...93W}. In this work, we choose 300 particles as a threshold above which these effects are expected to be minimized. \cite{2016MNRAS.456.2486D} also adopted a cut of 300 particles, motivated by the earlier works of \cite{2007MNRAS.378...55M}, \cite{2008MNRAS.391.1940M}, and \cite{2015MNRAS.454.3328V}, to avoid numerical biases in their halo catalogs. Similarly, \citet{2025ApJ...986..127N}, using the same Symphony and COZMIC simulations as we employ in this work, found converged results using a threshold of 300 particles (although this was for subhalos).

While setting a conservative threshold on the number of particles per halo, such as described above, minimizes numerical effects, it does not entirely remove them. For example, \citet[][see also \citealt{2010ApJ...711.1198T} and \citealt{2017MNRAS.467.3454B}]{2017MNRAS.471.2871B} show that a halo consisting of 1,000 particles has an approximately 15\% uncertainty in mass due to particle discreteness noise. Similarly, while the mass function of artificial halos is a strongly decreasing function of mass \citep{2007MNRAS.380...93W}, even above our threshold of 300 particles there will be \emph{some} contamination by artificial halos. Therefore, in this work, our approach is to supplement our conservative threshold of 300 particles with forward modeling of these numerical effects using parametric models. Once those models have been calibrated to match the available data, they can be removed, leaving a prediction for the true, ``intrinsic'' halo mass function.

\begin{table}
\caption{Snapshots of the MDPL simulations used in this work. For each simulation (first column), we identify the redshifts of the snapshots used (second column), the number of snapshots available at this redshift and higher (third column), and the spacing between snapshots at this redshift (fourth column).}
\label{tb:snapshots}
\centering
\begin{tabular}{lrrr}
\hline
Simulation & $z$ & $N(\ge z)$ & $\Delta z/(1+z)$ \\
\hline
VSMDPL & 0.000 & 138 & 0.0200 \\
VSMDPL & 0.490 & 120 & 0.0201 \\
VSMDPL & 0.990 & 107 & 0.0201 \\
VSMDPL & 2.030 & 88 & 0.0229 \\
VSMDPL & 3.040 & 75 & 0.0223 \\
\hline
SMDPL & 0.000 & 117 & 0.0302 \\
SMDPL & 0.505 & 71 & 0.0091 \\
SMDPL & 1.000 & 49 & 0.0417 \\
SMDPL & 2.021 & 39 & 0.0409 \\
SMDPL & 3.032 & 32 & 0.0420 \\
\hline
MDPL2 & 0.000 & 119 & 0.0224 \\
MDPL2 & 0.490 & 101 & 0.0224 \\
MDPL2 & 0.987 & 88 & 0.0223 \\
MDPL2 & 2.028 & 69 & 0.0223 \\
MDPL2 & 3.127 & 55 & 0.0228 \\
\hline
BigMDPL & 0.000 & 79 & 0.0302 \\
BigMDPL & 0.492 & 34 & 0.0092 \\
BigMDPL & 1.000 & 11 & 0.2225 \\
BigMDPL & 2.145 & 9 & 0.1080 \\
BigMDPL & 2.891 & 7 & 0.1322 \\
\hline
HugeMDPL & 0.000 & 100 & 0.0224 \\
HugeMDPL & 0.490 & 82 & 0.0224 \\
HugeMDPL & 0.987 & 69 & 0.0223 \\
HugeMDPL & 2.028 & 50 & 0.0223 \\
HugeMDPL & 3.037 & 37 & 0.0223 \\
\hline
\end{tabular}
\end{table}

In Table~\ref{tb:snapshots} we identify the snapshots of the MDPL simulations that we utilize in this work. The five MDPL simulations do not have precisely the same set of snapshot times---we therefore choose the snapshots closest to $z=0.0$, 0.5, 1.0, 2.0, and 3.0 for each simulation. Table~\ref{tb:snapshots} also lists the number of snapshots available at each redshift and higher, and the spacing of snapshots at each redshift. These details are important when identifying backsplash halos (see \S\ref{sec:nbodyAnalysis}).

\subsection{Symphony/Milky Way-est/COZMIC suites}\label{sec:symphony}

To study the halo mass function at very low masses, we make use of a suite of zoom-in simulations\footnote{We note that zoom-in simulations are more typically employed to study the subhalo population of the target halo---which we explicitly exclude here. However, the zoom-in region outside of the target halo contains valuable, high-resolution information about the broader halo population. As we will describe, if regions contaminated by low resolution particles are carefully excluded, the non-random environment resulting from target halo selection is accounted for, and possible biases arising from target halo selection are considered, zoom-in simulations can be used to study more than just the target halo and its population of subhalos.} of LMC, Milky Way, and group-mass halos. These simulations are taken from the \emph{Symphony} \citep{2022arXiv220902675N}, \emph{Milky Way-est} \citep{2024arXiv240408043B} and COZMIC \citep{2025ApJ...986..127N,2025ApJ...986..128A,2025ApJ...986..129N,2025ApJ...993...17N} suites. The LMC-mass halos have virial masses in the range $10^{11.02\pm0.05}\mathrm{M}_\odot$, the Milky Way-mass halos have virial masses in the range $10^{12.09\pm0.02}\mathrm{M}_\odot$, and the group-mass halos have virial masses in the range $10^{13.15\pm0.1}\mathrm{M}_\odot$ \citep{2022arXiv220902675N}. In addition to the standard resolution zoom-in simulations from those works (which we will refer to as the ``X1'' resolution), some of the simulations have been repeated at higher resolutions (8 times lower particle mass, ``X8'', and, in one case, 64 times lower particle mass, ``X64'', from \citealt{2025ApJ...983L..23N}). For each zoom-in system, we use only the highest resolution resimulation available in this work.

The non-CDM simulations in the COZMIC suite include thermal warm dark matter (WDM; \citealt{2025ApJ...986..127N}), models with dark matter--baryon interactions (IDM; \citealt{2025ApJ...986..127N}), fuzzy dark matter models (FDM; \citealt{2025ApJ...986..127N}), models with a mixture of thermal warm and cold dark matter \citep{2025ApJ...986..128A}, models with coupling to a dark fermion (characterized by their temperature at kinetic decoupling, $T_\mathrm{kd}$; \citealt{2025ApJ...986..129N}), and models with enhancements in the power spectrum~\citep{2025ApJ...993...17N}.\footnote{For the dark fermion models from \cite{2025ApJ...986..129N}, we use the simulations with power spectrum modification but without dark matter self-interactions.} In each case, simulations are conducted for a range of dark matter particle masses or interaction strengths. In these simulations, only the initial linear power spectrum is modified in accordance with the beyond-CDM physics (i.e., late-time interactions in models with interactions, and quantum pressure effects in FDM models are ignored).

Thermal warm dark matter (WDM) models are characterized by the particle mass ($m_\mathrm{WDM}$; ranging from 3 to 10~keV in the COZMIC simulations), which results in a half-mode mass (defined as the halo mass corresponding to the scale at which the transfer function is a factor 2 below that for CDM) of \cite{2025ApJ...986..127N}:
\begin{equation}
M_\mathrm{hm}(m_\mathrm{WDM}) = 4.3\times 10^8\mathrm{M}_\odot \left(\frac{m_\mathrm{WDM}}{3~\mathrm{keV}}\right)^{-3.564}.
\end{equation}
In the case of mixed WDM/CDM models, in addition to the particle mass, a second parameter, $f_\mathrm{WDM}$, characterizes the mass fraction of the dark matter that is WDM (with a fraction $1-f_\mathrm{WDM}$ being CDM). For the WDM models with enhancements, a Gaussian bump is added to the power spectrum:
\begin{equation}
\mathcal{T}_\mathrm{bump}(k) = \sqrt{\frac{P_\mathrm{bump}(k)}{P_\mathrm{CDM}(k)}} = 1 + A \exp \left[ - \frac{[\log(k/k_0)]^2}{\sigma_k^2} \right],
\end{equation}
where $A=2$, $\sigma_k=0.5$, and $k_0$ is chosen to match the half-mode wavenumber of a corresponding WDM model; hence, we will often refer to these models by their equivalent WDM mass plus a ``bump,'' ``cutoff,'' or ``bump + cutoff'' label \citep[see][for details]{2025ApJ...993...17N}. Three cases are considered, combining different choices of power enhancement (``bumps'') and suppression (``cutoffs''):
\begin{equation}
\left\{
\begin{aligned}
&\mathcal{T}_{\mathrm{Bump+Cutoff}}(k) = \mathcal{T}_{\mathrm{Bump}}(k)\times \mathcal{T}_{\mathrm{WDM}}(k), && \text{Bump} \\
& && \text{\,\,+ Cutoff}; \\
&\mathcal{T}_{\mathrm{Cutoff}}(k) = \min\left[\mathcal{T}_{\mathrm{Bump+Cutoff}}(k), 1\right], && \text{Cutoff}; \\
&\mathcal{T}_{\mathrm{Bump}}(k)=\max\left[\mathcal{T}_{\mathrm{Bump+Cutoff}}(k), 1\right], && \text{Bump}.
\end{aligned}
\right.
\end{equation}

For dark fermion models, the power spectrum is characterized by the temperature of the visible sector at the time at which the dark fermion kinematically decouples with the dark matter, which depends on the ratio of dark-to-visible sector temperatures, $\xi$ \citep{nadler_2025_14666735}:
\begin{equation}
T_\mathrm{kd} = 0.26~\mathrm{keV} \left(\frac{\xi}{0.5}\right)^{-1/2}.
\end{equation}
Values of $\xi$ were chosen to match the half-mode wavenumbers of thermal-relic WDM for $m_\mathrm{WDM}=3.5$, 6.5, and 10~keV, resulting in $T_\mathrm{kd}=0.72$, 1.46, and 2.32~keV, respectively \citep{nadler_2025_14666735}.

Fuzzy dark matter models are characterized by the particle mass, $m_\mathrm{FDM}$, which determines the half-mode mass \citep{2025ApJ...986..127N}:
\begin{equation}
M_\mathrm{hm}(m_\mathrm{FDM}) = 4.5 \times 10^{10} \mathrm{M}_\odot \left(\frac{m_\mathrm{FDM}}{10^{-22}~\hbox{eV}}\right)^{-1.41}.
\end{equation}

Finally, interacting dark matter models are considered in which a dark matter particle of mass $m_\mathrm{IDM}$ is able to scatter with protons. These models are characterized by $m_\mathrm{IDM}$ and their velocity-dependent momentum transfer cross-section, $\sigma = \sigma_0 v^n$, where $n$ controls the velocity dependence, and $\sigma_0$ the scattering amplitude. The scattering amplitude, $\sigma_0$, is chosen such that the resulting transfer function either matches the half-mode scale of a thermal relic WDM model with $m_\mathrm{WDM}=6.5$~keV, or is strictly always more suppressed than that model---these are referred to as ``half-mode'' and ``envelope'' models respectively \citep[see][for details]{2025ApJ...986..127N}.

\begin{figure*}
\begin{center}
\includegraphics[width=6.5in]{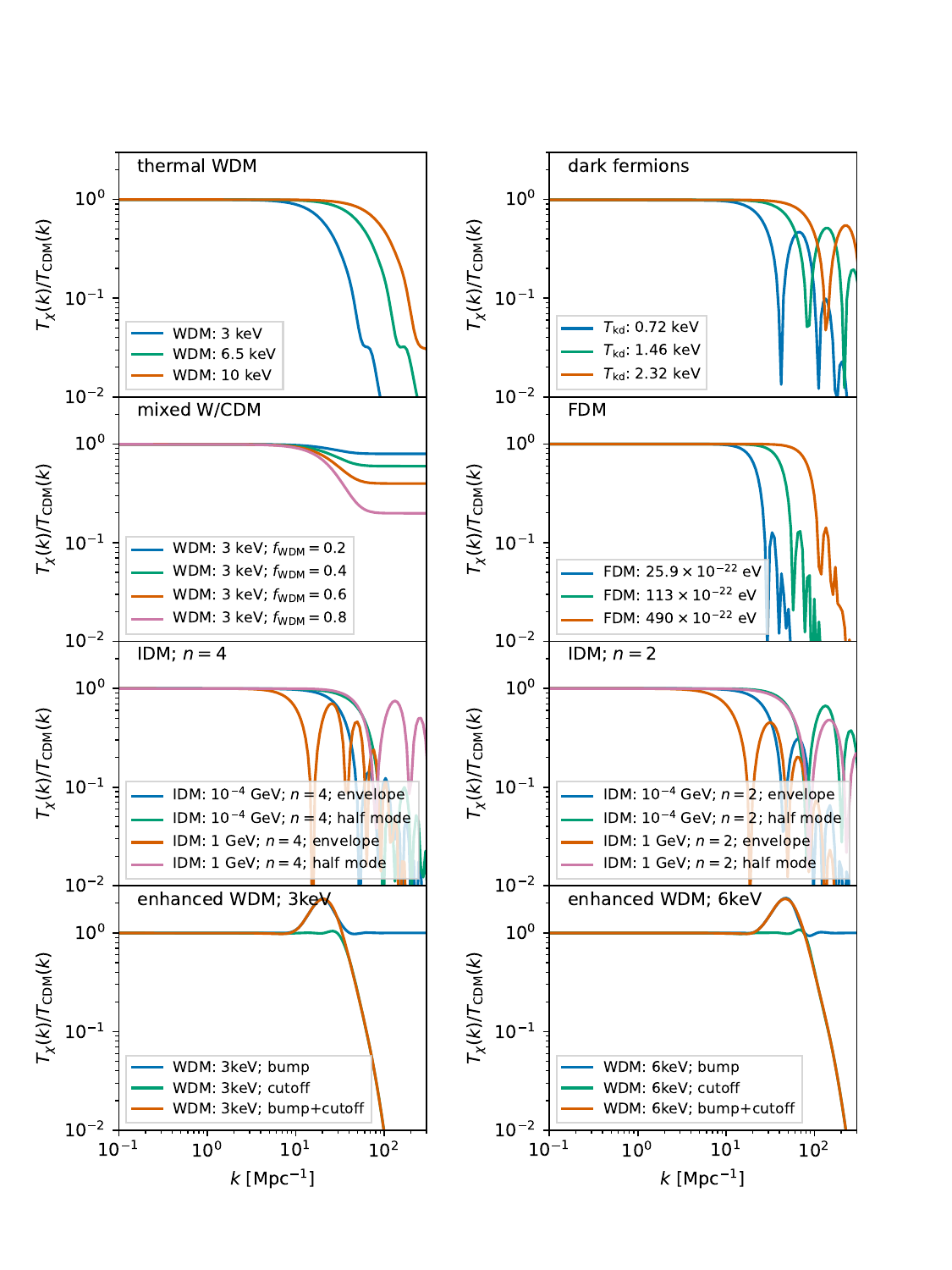}
\end{center}
\vspace{-15mm}\caption{Transfer functions for dark matter (relative to CDM) for the beyond-CDM simulations considered in this work: WDM (first row, left), dark fermion (first row, right), mixed W/CDM (second row, left), FDM (second row, right), IDM models (third row), and WDM with bumps/cutoffs (fourth row). In each panel, colors indicate different dark matter particle masses/WDM fractions/decoupling temperatures as indicated in the panel (in many cases, only a representative subsample is shown).}
\label{fig:transferFunctions}
\end{figure*}

Figure~\ref{fig:transferFunctions} shows examples of transfer functions for dark matter (relative to CDM) for these beyond-CDM simulations for WDM (first row, left), dark fermion (first row, right), mixed W/CDM (second row, left), FDM (second row, right), IDM models (third row), and WDM with bumps/cutoffs (fourth row). In each panel, the transfer function is shown for a representative sample of dark matter properties considered in this work, as indicated in each panel.

In the case of IDM models, we show results for both the ``envelope'' and ``half-mode'' models considered by \cite{2025ApJ...986..127N}. In the ``half-mode'' models, the IDM transfer function is chosen such that the squared transfer function has a half-mode scale matched to that of a $m_\chi = 6.5~\hbox{keV}$ WDM model (close to current observational limits), while the ``envelope'' models are chosen to be strictly more suppressed than the same $m_\chi = 6.5~\hbox{keV}$ WDM model. In these IDM models, the interaction cross-section has a velocity dependence $\sigma \propto v^n$, and we include both $n=4$ and $n=2$ cases considered by \cite{2025ApJ...986..127N}.

Together, these beyond-CDM simulations incorporate a range of qualitatively different behaviors in the transfer functions, including smooth cut-offs, partial cut-offs, cut-offs with oscillations, and localized bumps.

Table~\ref{tb:zoomIns} gives details of this suite of simulations, including the N-body particle mass, minimum halo mass ($M_\mathrm{min}$; again corresponding to 300 particles), transfer function type, dark matter particle mass, and the number of zoom-in realizations available. We refer the reader to \cite{2025ApJ...986..127N}, \cite{2025ApJ...986..128A}, \cite{2025ApJ...986..129N}, and \cite{2025ApJ...993...17N} for full details of these simulations.

\begin{table*}
\caption{Properties of the zoom-in simulations used in this work. Column 1 specifies the type of host system targeted by the zoom-in (an LMC-, Milky Way-, or group-mass halo, with virial masses at $z=0$ of approximately $10^{11}\mathrm{M}_\odot$, $10^{12}\mathrm{M}_\odot$, and $10^{13}\mathrm{M}_\odot$ respectively). Column 2 gives the N-body particle mass. Column 3 gives the minimum halo mass used in likelihood analyses (corresponding to 300 particles). Column 4 describes the transfer function used. Column 5 gives the set of dark matter particle masses/WDM fractions/kinetic decoupling temperatures used. Column 6 lists the number of each simulation available (written as the number of dark matter particle masses considered, multiplied by the number of target host halos simulated).}
\label{tb:zoomIns}
\centering
\begin{tabular}{cccccr@{$\times$}r@{$=$}r}
\hline
{\bf Host type} & \boldmath{$m_\mathrm{p}$} & \boldmath{$M_\mathrm{min}$} & {\bf Transfer function} & \boldmath{$m_\chi$} & \multicolumn{3}{c}{\bf Number} \\
\hline
LMC       & $4.03\times 10^5\mathrm{M}_\odot$ & $1.21\times 10^8\mathrm{M}_\odot$ & CDM & N/A                                                & 1 & 38 & 38 \\
Milky Way & $4.03\times 10^5\mathrm{M}_\odot$ & $1.21\times 10^8\mathrm{M}_\odot$ & CDM & N/A                                                & 1 & 46 & 46 \\
Milky Way & $6.29\times 10^3\mathrm{M}_\odot$ & $1.89\times 10^6\mathrm{M}_\odot$ & CDM & N/A                                                & 1 &  1 &  1 \\
Milky Way & $4.03\times 10^5\mathrm{M}_\odot$ & $1.21\times 10^8\mathrm{M}_\odot$ & WDM & 3, 4, 5, 6, 6.5, 10~keV                            & 6 &  2 & 12 \\
Milky Way & $5.04\times 10^4\mathrm{M}_\odot$ & $1.51\times 10^7\mathrm{M}_\odot$ & WDM & 3, 4, 5, 6, 6.5, 10~keV                            & 6 &  1 &  6 \\
Milky Way & $5.04\times 10^4\mathrm{M}_\odot$ & $1.51\times 10^7\mathrm{M}_\odot$ & WDM & 3~keV; $f_\mathrm{WDM}=0.2$, 0.4, 0.6, 0.8         & 4 &  1 &  4 \\
Milky Way & $5.04\times 10^4\mathrm{M}_\odot$ & $1.51\times 10^7\mathrm{M}_\odot$ & WDM & 3~keV; bump, cutoff, bump+cutoff                   & 3 &  1 &  3 \\
Milky Way & $5.04\times 10^4\mathrm{M}_\odot$ & $1.51\times 10^7\mathrm{M}_\odot$ & $T_\mathrm{kd}$ & 0.72, 1.46, 2.32~keV                   & 3 &  1 &  3 \\
Milky Way & $4.03\times 10^5\mathrm{M}_\odot$ & $1.21\times 10^8\mathrm{M}_\odot$ & IDM (half-mode) & $10^{-4}$, $10^{-2}$, $1$~GeV          & 3 &  2 &  6 \\
Milky Way & $5.04\times 10^4\mathrm{M}_\odot$ & $1.51\times 10^7\mathrm{M}_\odot$ & IDM (half-mode) & $10^{-4}$, $10^{-2}$, $1$~GeV          & 3 &  1 &  3 \\
Milky Way & $4.03\times 10^5\mathrm{M}_\odot$ & $1.21\times 10^8\mathrm{M}_\odot$ & IDM (envelope) & $10^{-4}$, $10^{-2}$, $1$~GeV           & 3 &  2 &  6 \\
Milky Way & $5.04\times 10^4\mathrm{M}_\odot$ & $1.51\times 10^7\mathrm{M}_\odot$ & IDM (envelope) & $10^{-4}$, $10^{-2}$, $1$~GeV           & 3 &  1 &  3 \\
Milky Way & $4.03\times 10^5\mathrm{M}_\odot$ & $1.21\times 10^8\mathrm{M}_\odot$ & IDM ($n=2$; half-mode) & $10^{-4}$, $10^{-2}$, $1$~GeV   & 3 &  2 &  6 \\
Milky Way & $5.04\times 10^4\mathrm{M}_\odot$ & $1.51\times 10^7\mathrm{M}_\odot$ & IDM ($n=2$; half-mode) & $10^{-4}$, $10^{-2}$, $1$~GeV   & 3 &  1 &  3 \\
Milky Way & $4.03\times 10^5\mathrm{M}_\odot$ & $1.21\times 10^8\mathrm{M}_\odot$ & IDM ($n=2$; envelope) & $10^{-4}$, $10^{-2}$, $1$~GeV    & 3 &  2 &  6 \\
Milky Way & $5.04\times 10^4\mathrm{M}_\odot$ & $1.51\times 10^7\mathrm{M}_\odot$ & IDM ($n=2$; envelope) & $10^{-4}$, $10^{-2}$, $1$~GeV    & 3 &  1 &  3 \\
Milky Way & $4.03\times 10^5\mathrm{M}_\odot$ & $1.21\times 10^8\mathrm{M}_\odot$ & FDM & 25.9, 69.4, 113, 151, 185, $490\times 10^{-22}$~eV & 6 &  2 & 12 \\
Milky Way & $5.04\times 10^4\mathrm{M}_\odot$ & $1.51\times 10^7\mathrm{M}_\odot$ & FDM & 25.9, 69.4, 113, 151, 185, $490\times 10^{-22}$~eV & 6 &  1 &  6 \\
Group     & $4.03\times 10^5\mathrm{M}_\odot$ & $1.21\times 10^8\mathrm{M}_\odot$ & CDM & N/A                                                & 1 & 49 & 49 \\\hline
\end{tabular}
\end{table*}

\section{Halo Mass Function Models and Analysis}\label{sec:methods:models}

In this section we describe how the N-body simulations described in \S\ref{sec:methods:sims} are analyzed (\S\ref{sec:nbodyAnalysis}), how we construct a mass function model (\S\ref{sec:hmfModel}) and a likelihood function (\S\ref{sec:likelihood}), and how this model is fit to the N-body data (\S\ref{sec:mcmc}).

\subsection{N-body Analysis}\label{sec:nbodyAnalysis}

Each simulation was analyzed using the Rockstar \citep{2013ApJ...762..109B} halo finder and consistent-trees merger tree algorithm \citep{2013ApJ...763...18B} to identify dark matter halos and measure their properties. For the MDPL simulations, this process is described by \cite{2016MNRAS.462..893R}, while for the Symphony/Milky Way-est/COZMIC suites, this is described by \cite{2023ApJ...945..159N,2024arXiv240408043B,2025ApJ...986..127N}. 

The resulting halo catalogs contain both subhalos and backsplash halos \citep{2000ApJ...540..113B}. As discussed by \cite{2017MNRAS.467.3454B}, backsplash halos have distinct properties from non-backsplash halos, as they have been processed through the potential of a host halo (resulting in tidal evolution), and so should be treated as a separate population. Additionally, in semi-analytic models of (sub)halo evolution, the backsplash population is a distinct population, accounted for in the merger trees. Including backsplash halos would therefore lead to double-counting of mass if our mass functions were utilized by such models. Following \cite{2017MNRAS.467.3454B}, we remove backsplash halos from the halo catalogs before computing halo mass functions. Subhalos (identified as any halo within the virial radius of a larger halo at the redshift of interest) are also removed. Consequently, in this work (and notably unlike most applications of zoom-in simulations), we focus on only those halos which are not, and never have been, a subhalo in any larger halo.

As described by Sarnaaik \& Benson (in prep.), we identify and exclude backsplash halos using the algorithm of \cite{2020MNRAS.493.4763M}. That is, briefly, we begin from the snapshot of interest, at redshift $z=z_\mathrm{f}$, considering each halo in turn. We follow the main-line branch of the halo back through the merger tree. If any of these main-line halos are within the virial radius of another halo, at any redshift, we flag the $z=z_\mathrm{f}$ halo as a backsplash (or ``fly-by'') halo if three conditions are met:
\begin{enumerate}
    \item the host halo (within whose virial radius a main-line halo intrudes) must have a descendant at $z=z_\mathrm{f}$;
    \item that $z=z_\mathrm{f}$ descendant must not be within the virial radius of the $z=z_\mathrm{f}$ halo of interest;
    \item the $z=z_\mathrm{f}$ descendant of the host halo must have a larger mass than our halo of interest.
\end{enumerate}
This is different from the approach of \cite{2017MNRAS.467.3454B}---\cite{2020MNRAS.493.4763M} have demonstrated their approach is quite robust, so we use it here. We find that the two approaches, \cite{2017MNRAS.467.3454B} and \cite{2020MNRAS.493.4763M}, give very similar results. For example, in the MDPL2 simulation at $z=0$ we find a total of 32,528 halos in the mass range $10^{11}$--$10^{12}\mathrm{M}_\odot$ before any backsplash removal. The algorithm of \cite{2017MNRAS.467.3454B} identifies 13.2\% of these as backsplash halos, leaving 28,240 halos, while that of \cite{2020MNRAS.493.4763M} identifies 12.6\% as backsplash halos, leaving 28,428. This suggests that backsplash identification is quite robust. At $z=3.13$ in MDPL2 we find that the two methods, \cite{2017MNRAS.467.3454B} and \cite{2020MNRAS.493.4763M}, identify 6.1\% and 5.9\% of halos as backsplash, respectively. For BigMDPL, in the mass range $10^{12}\mathrm{M}_\odot$ to $10^{13} \mathrm{M}_\odot$, at $z=0$ the two methods identify 11.6\% and 11.6\% of halos as backsplash halos respectively, while at $z=2.89$ in BigMDPL the two methods identify only 2.1\% and 2.2\% of halos as backsplash halos respectively. Therefore, at $z=0$, backsplash fractions are very similar in MDPL2 and BigMDPL. However, at $z\approx 3$, these differ by a factor of almost 3, with BigMDPL having a much lower backsplash fraction. From Table~\ref{tb:multiDarkResolutions} it can be seen that BigMDPL has many fewer snapshots prior to its $z\approx 3$ snapshot than does MDPL2 (7 compared to 55 respectively), which are then spaced more widely in redshift ($\Delta z/(1+z) = 0.1322$ compared to 0.0228 respectively). This much lower fidelity information on the history of $z\approx 3$ halos in BigMDPL likely makes it more difficult to identify backsplash halos. However, at $z\approx 3$ the backsplash fraction is small even in MDPL2, so it is not a large systematic effect in our determination of halo mass functions. Nevertheless, this requirement for backsplash halo identification should be kept in mind when planning snapshot distributions in future simulations. With the \cite{2020MNRAS.493.4763M} approach, backsplash halos make up around 30\% of the halo population in the high-resolution Symphony CDM Milky Way simulation, for example.

The efficiency of this algorithm will be sensitive to the number and spacing of simulation snapshots at redshifts $z > z_\mathrm{f}$. Fewer and/or more widely spaced snapshots may mean that fly-by encounters are missed, resulting in backsplash halos being missed. This is particularly problematic for the BigMDPL simulation---in Table~\ref{tb:snapshots} we show the number and spacing of snapshots in the MDPL simulations. The BigMDPL simulation has relatively few snapshots---particularly at the higher redshifts we consider, there are only around 10 or fewer snapshots at earlier times. We will return to this point when discussing the halo mass function from this simulation.

Having removed backsplash halos, for the MDPL simulations, it is straightforward to compute a halo mass function by counting halos into bins in halo mass and dividing by the volume of the simulation cube. Throughout this work, we utilize the spherical collapse density contrast \citep{1980lssu.book.....P,1996MNRAS.282..263E}, $\Delta_\mathrm{vir}$,  to define halo mass. This choice has been shown to result in the most ``universal'' halo mass function \citep{2016MNRAS.456.2486D} when parameterized in terms of the peak-height parameter, $\nu$.

\subsubsection{Zoom-in Analysis}\label{sec:zoominAnalysis}

\begin{figure*}
\includegraphics[width=\textwidth]{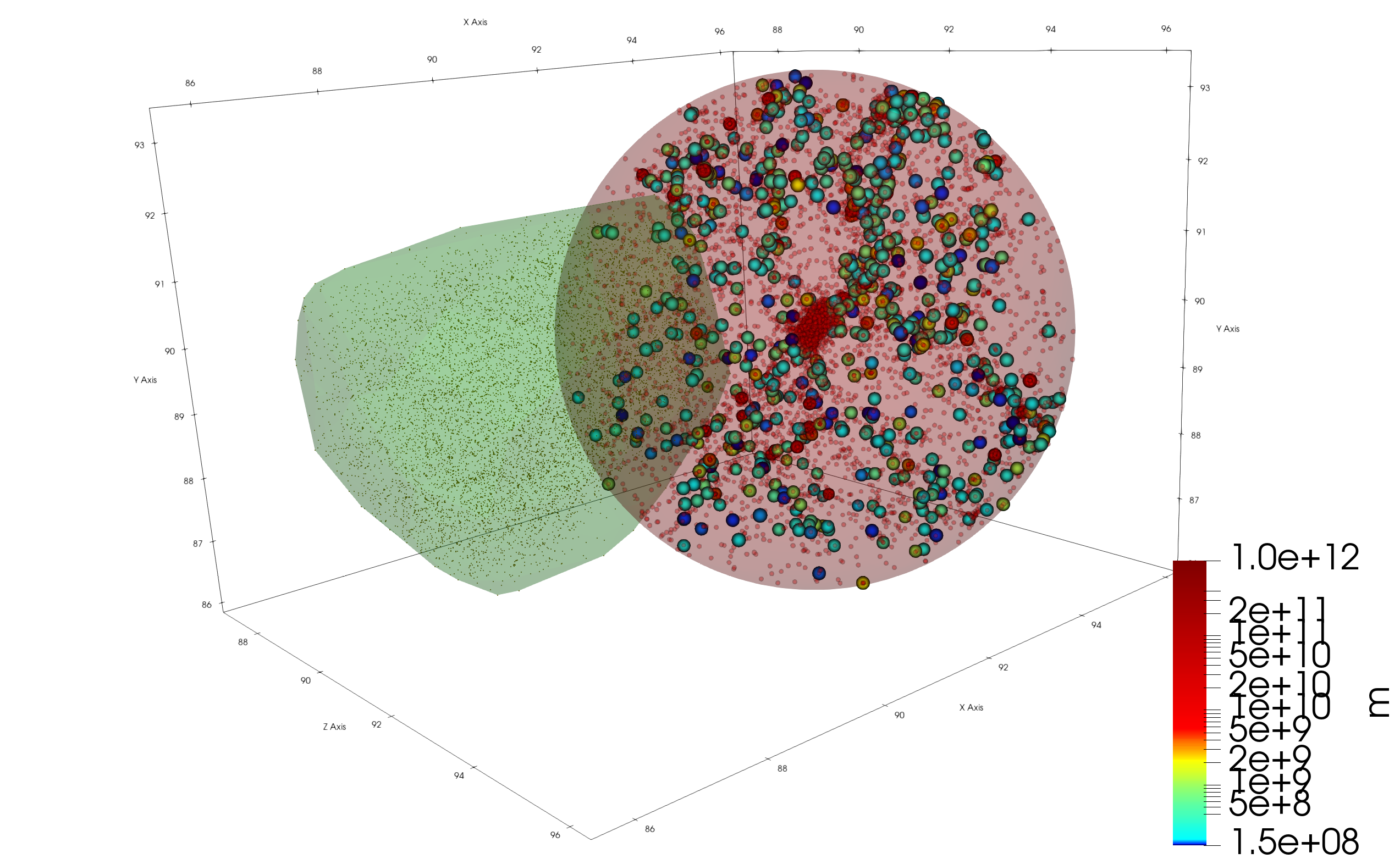}
\caption{Example of our analysis of zoom-in simulations. The pink-shaded sphere represents the region around an example $z=0$ Milky Way target halo that is uncontaminated by low-resolution particles. Small red balls within this sphere show the locations of the centers of all non-backsplash halos within this region, while larger, colored balls show the locations of non-backsplash halos in this region that contain more than 3,000 particles, and are colored by mass as indicated in the figure. The green points show the corresponding locations of particles corresponding to halo centers (i.e., the small red balls) in the initial conditions, with the green shaded region showing the convex hull bounding these points.}
\label{fig:convexHull}
\end{figure*}

Extracting the halo mass functions from the zoom-in simulations is more complicated and requires additional steps. Specifically, we must determine a region from which to measure the halo mass function---this must exclude any parts of the simulation containing too many low-resolution particles\footnote{In zoom-in simulations, the surrounding environment of the region of interest is typically modeled at lower resolution, with higher mass particles, for computational efficiency.}. Additionally, we must determine the Lagrangian volume of this region (since our mass function will be defined per Lagrangian volume), and the properties (mass and overdensity) of the ``background'' needed for the peak-background split method (see \S\ref{sec:peakBackgroundSplit}).

To meet these requirements, the simulations are processed, as usual, by the Rockstar halo finder and ConsistentTrees tree builder \citep{2013ApJ...762..109B,2013ApJ...763...18B}. Using these halo catalogs, merger trees, \emph{and} the full particle data for each snapshot of interest and the initial conditions, we then proceed as follows:
\begin{enumerate}
    \item identify fly-by halos using the ``fly-by'' algorithm of \cite{2020MNRAS.493.4763M}, as described in \S\ref{sec:nbodyAnalysis};
    \item identify the primary halo (i.e., that which was the target of the zoom-in) at $z=0$ and determine its properties (position, mass, virial radius) at each snapshot;
    \item at the snapshots of interest, determine the radius of the largest sphere, centered on the primary halo, that contains no more than 10\% of its mass in low-resolution particles---halos in this region are then made almost entirely of the highest resolution particles;
    \item extract all non-backsplash halos in this uncontaminated sphere;
    \item extract all particles within this uncontaminated sphere;
    \item using the IDs of the extracted particles within the uncontaminated sphere, find the positions of these particles in the initial conditions;
    \item construct the convex hull of these particles in the initial conditions;
    \item select all particles within the convex hull in the initial conditions, determine the total mass, $M_\mathrm{e}$, within the convex hull and its volume, $V_\mathrm{e}$, from which the overdensity, $\delta_\mathrm{e}$, of the convex hull (linearly extrapolated to $z=0$) can be determined;
    \item using the non-backsplash halos in the uncontaminated sphere, construct the halo mass function, using $V_\mathrm{L} = V_\mathrm{e}/(1+\hat{\delta}_\mathrm{e})$, where $\hat{\delta}_\mathrm{e}$ is the overdensity of the convex hull \emph{not} extrapolated to $z=0$, as the Lagrangian volume of this sphere.
\end{enumerate}
This procedure, illustrated in Figure~\ref{fig:convexHull}, provides us both with the halo mass function (of non-backsplash halos, normalized using the Lagrangian volume of the convex hull) and information about the environment of the zoom-in region ($M_\mathrm{e},\delta_\mathrm{e}$).

This convex hull-based approach necessarily results in some ``background'' particles (i.e., particles that are \emph{not} within the uncontaminated sphere, but which \emph{are} within the convex hull of those particles in the initial conditions) being included in the estimate of the initial overdensity. Figure~A1 of \cite{2024MNRAS.528.7300Z} shows that for background particle contamination fractions of more than around 20\% the linear overdensity inferred from the region may be too biased, or show to much scatter, to accurately represent the environment of the final region from which the halo mass function is being measured. We have computed the excess fraction of particles contained within the convex hull in the initial conditions relative to those particles used to define the hull. For the Symphony resolution X1 simulations the median for the Milky Way sample is 14\% (with a minimum/maximum over the set of 4\%/74\%), for the LMC sample the median is 22\% (minimum/maximum of 5\%/78\%), and for the Group sample the median is 18\% (minimum/maximum of 5\%/108\%). From this we conclude that the median contamination fraction is comparable to (or slightly less than) the $\sim$20\% threshold found by \cite{2024MNRAS.528.7300Z}---which of course means that roughly half of the realizations we consider exceed this. While alternatives to a convex hull are possible, they are necessarily more complicated (typically requiring some regularization metric to limit the curvature of the surface). As such, we retain the convex hull approach and explore its consequences in \S\ref{sec:environmentalTrends}.

\subsection{Halo Mass Function}\label{sec:hmfModel}

Our goal is to measure the mass function of dark matter halos, that is, the true distribution of halo masses after accounting for any numerical effects. What we measure (as described above) from N-body simulations is the true (or ``intrinsic'') mass function modified by various numerical effects. From here on, we will refer to the distribution of halo masses directly measured from an N-body simulation as the ``measured'' halo mass function, and the true halo mass function (that which we wish to constrain, after removal of all numerical effects) as the ``intrinsic'' halo mass function. We therefore construct forward models of all numerical effects, apply these to a model of the intrinsic halo mass function, and fit the result to the measured halo mass functions obtained from the N-body simulations, allowing us to recover constraints on the intrinsic halo mass function. We detail the numerical effects and how we forward model them in the following subsections, and will comment on the relative importance of these effects in \S\ref{sec:resultsCDM}.

\subsubsection{Intrinsic halo mass function}

\begin{figure}
\centering
\includegraphics[width=85mm]{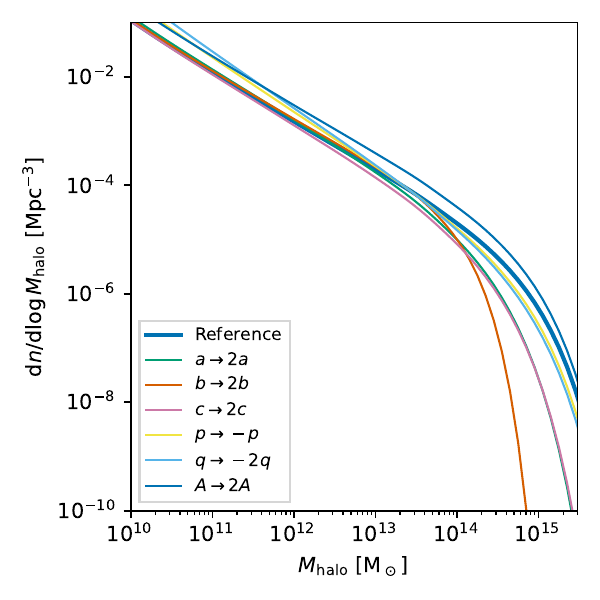}
\caption{The effects on the halo mass function when varying individual parameters in our extended version of the \cite{2011ApJ...732..122B} fitting function. The heavy blue line shows a reference model (using the best-fit parameters determined later in this work), while each colored line shows the effects of transforming a single parameter of the fitting function as indicated in the legend. Note that the change in each parameter is chosen to be quite extreme to clearly illustrate its effects---no real halo mass function considered in this work displays such large variations in the best-fit parameters.}
\label{fig:functionalForm}
\end{figure}

As our intrinsic halo mass function, we adopt an extension of the parametric form proposed by \cite{2011ApJ...732..122B} for the halo mass function:
\begin{eqnarray}
 n(M,z) &=& \sqrt{\frac{2}{\pi}} A \frac{\bar{\rho}}{M^2} \alpha(M) \left( 1 + [a \nu^2]^{-p} \right) \left(a \nu^2 \right)^{q/2} \nonumber \\
 & & \exp\left( - c \frac{[a \nu^2]^b}{2} \right),
 \label{eq:hmfIntrinsic}
\end{eqnarray}
where $\nu = \delta_\mathrm{c}(z)/\sigma(M)$ is the usual peak-height parameter\footnote{The peak-height parameter, $\nu$, measures the significance of a fluctuation in the density field. Specifically, it measures the number of standard deviations above the mean density. For example, a region with $\nu=3$ has a density of $\bar{\rho}[1+3\sigma(M)]$. For any chosen mass scale, regions with a large, positive peak-height parameter will undergo gravitational collapse earlier due to their greater overdensities.}, $\alpha(M) = |\mathrm{d}\log\sigma/\mathrm{d}\log M|$ and $A$, $a$, $b$, $c$, $p$, and $q$ are parameters to be constrained by fitting to the N-body simulation data. Dependence on the environment enters via the $\delta_\mathrm{c}$ and $\sigma(M)$ terms under the peak-background split as will be described in \S\ref{sec:peakBackgroundSplit}. The original model of \cite{2011ApJ...732..122B} had $b=c=1$ in the exponential term. We introduce these two new parameters to allow for additional flexibility in fitting the high-mass cut-off in the mass function. Figure~\ref{fig:functionalForm} illustrates this functional form and how it responds to changes in the various parameters. The heavy blue line indicates a reference model (with parameters set equal to the best-fit results from later in this work), with each colored line showing the effects of transforming a single parameter of the fitting function as indicated in the legend. The parameter $a$ scales the peak height---increasing it therefore reduces the mass at which the exponential cut-off in the halo mass function begins to become important. The parameter $b$ affects the scaling of the exponential cut-off term with peak height---increasing it makes this cut-off substantially sharper. The parameter $c$ also appears in the exponential cut-off term, where it has an identical effect to $a$, but since it does not affect the power-law part of the fitting function its effects are restricted to the high-mass region (unlike those of $a$). Parameters $p$ and $q$ largely control the low mass slope, although they also have an effect at higher masses. Lastly, the normalization parameter $A$ simply scales the overall amplitude of the halo mass function.

\subsubsection{Peak-Background Split}\label{sec:peakBackgroundSplit}

The zoom-in simulations, by their nature, are drawn from a biased set of regions in their parent simulation volumes. That is, the overdensities of their Lagrangian volumes in the initial conditions will not be normally distributed (as expected for an unbiased sample). Since the halo mass function is expected to vary with environmental overdensity \citep[e.g.][]{1996MNRAS.282..347M} we therefore cannot simply fit the halo mass function to these collectively. Instead, we fit the halo mass function of each zoom-in individually, using the peak-background split approach \citep{1991ApJ...379..440B,1996ApJS..103....1B} to construct a halo mass function that captures the dependence on the environmental overdensity. While each zoom-in individually has relatively little constraining power (due to the relatively small number of halos present), we treat them as independent data sets and combine their likelihoods. Collectively, they provide strong constraints on the halo mass function.

As noted by \cite{2019MNRAS.485.5010B}, when the halo mass function does not have the Press-Schechter form, averaging the mass function predicted by the peak-background split approach over the expected distribution of environmental densities (which differs from a normal distribution due to the imposition of the condition that the region of interest has not collapsed to become a halo on any larger scale;  \citealt{1991ApJ...379..440B}) does not exactly reproduce the mass function predicted in the absence of the peak background split, as it should do. In this work, we take the view that this simply indicates that the actual distribution of environmental overdensities over which to average must differ from that derived by \cite{1991ApJ...379..440B}, presumably due to the usual limitations of Press-Schechter-type analyses.

In the peak-background split approach, we make the mappings:
\begin{equation}
    \delta_\mathrm{c} \rightarrow \delta_\mathrm{c} - \delta_\mathrm{e},
\end{equation}
and
\begin{equation}
    \sigma^2(M) \rightarrow \sigma^2(M) - \sigma^2(M_\mathrm{e}),
    \label{eq:peackBackgroundVariance}
\end{equation}
where $\delta_\mathrm{e}$ is the linear overdensity (extrapolated to $z=0$ as usual) and $M_\mathrm{e}$ is the mass of the ``background''. The first of these implies that, in regions of enhanced environmental overdensity, it becomes easier for halos to collapse (i.e., the critical overdensity for collapse is lowered). The second implies that the variance in the density field on any scale is reduced because we have fixed the large-scale (background) overdensity. From hereon, this transformation will be denoted by the operator $\mathcal{P}_\mathrm{peak-background}$ such that the effects of the peak-background split are described by $\sigma^\prime(M),\delta^\prime_\mathrm{c}(z) = \mathcal{P}_\mathrm{peak-background}[\sigma(M),\delta_\mathrm{c}(z)]$.

As described in \S\ref{sec:zoominAnalysis}, we set $M_\mathrm{e}$ equal to the total mass contained in the Lagrangian volume of the zoom-in simulation from which we extract the halo mass function, and set $\delta_\mathrm{e}$ equal to the overdensity of that Lagrangian volume in the initial conditions (extrapolated to $z=0$).

We note that equation~(\ref{eq:peackBackgroundVariance}) is formally valid only if the window function is sharp in $k$-space (since then the variances at each scale are simply additive). In practice, we use a different window function, so this equation will be only approximate. We also note that we apply this to the entire region of the convex hull in which we measure the mass function. However, the formation of halos in this region (particularly near the outer edges) will be sensitive to structure nearby. We will later assess the ability of the peak-background split to describe the environmental dependence of the halo mass function with these caveats in mind.

\subsubsection{Missing power from simulations}\label{sec:missingPower}

For cosmological boxes, such as the MDPL simulations used here, there is power missing on large scales due to the finite volume of the box, i.e., long wavelength modes are excluded. Furthermore, for modes of a given wavenumber, only a finite number of modes are present in the volume, meaning that there will be statistical deviations from the expected mean power at a given wavenumber.

Prior works have sometimes accounted for this by using the specific realization of the initial conditions to compute a power spectrum specific to the simulation \citep[see, for example,][]{2016MNRAS.456.2486D}. We do not have access to the initial conditions for the MDPL suite, and so we adopt a different approach here.

To account for the missing large scale power we assume that wave-vectors with all components ($k_x,k_y,k_z$) smaller than $\Delta k = 2 \pi f / L$, where $L$ is the simulation cube length, and $f=1/2$ specifies the minimum wavenumber, in units of $2 \pi / L$, to which the power spectrum is integrated, are missing from the power spectrum. We then compute, as a function of wavenumber magnitude $k$, the fraction of power missed due to these missing wavenumbers. The total power at that wavenumber is then reduced by that amount when computing halo mass functions.

The fractional suppression in power, $f(x)$, as a function of $x=k/\Delta k$ is given by:
\begin{widetext}
\begin{equation}
 f(x) = \left\{ \begin{array}{ll} 0 & \hbox{ for } x \le 1, \\ 3(1-x^{-1}) & \hbox{ for } 1 < x \le \sqrt{2}, \\ 3 x^{-1} [ 1 + 2 \pi^{-1} x \sin^{-1}(\{x^2-1\}^{-1}) - 4 \pi^{-1} x \sin^{-1}(\{1-x^2\}^{-1/2})  ] & \hbox{ for } \sqrt{2} < x \le \sqrt{3}, \\ 1 & \hbox{ for } \sqrt{3} < x. \end{array} \right. 
\end{equation}
\end{widetext}
For $1 < x \le \sqrt{2}$, the solution is found by considering the six spherical caps of the sphere that protrude from the faces of the cube. The solution for $\sqrt{2} < x \le \sqrt{3}$ is more complicated and follows the solution given by \cite{achille2013}. We thus define an operator $\mathcal{P}_\mathrm{cube}$ such that this suppression of the power spectrum is described by $P(k) \rightarrow \mathcal{P}_\mathrm{cube}[P(k)]$.

For the second issue concerning the finite number of modes, we calculate the variance within the cube and assume that the actual realization of the power spectrum in the simulation is normally-distributed around the mean expectation with this variance. We further assume that the number of halos of given mass is modified by this deviation as predicted by linear bias theory \citep{1996MNRAS.282..347M}. We therefore write:
\begin{equation}
    n(M,z) \rightarrow n(M,z) \left[ 1 + \epsilon b(M,z) \sigma_\mathrm{cube} \right],
    \label{eq:sigmaCube}
\end{equation}
where $b(M,z)$ is the bias of halos of mass $M$ and redshift $z$ (for which we adopt the model of \citealt{2010ApJ...724..878T}), $\sigma^2_\mathrm{cube}$ is the variance within the cube
\begin{equation}
   \sigma^2_\mathrm{cube} =  \int \mathrm{d^3}\mathbf{k} P(k) W^2(\mathbf{k}),
\end{equation}
where
\begin{equation}
    W(\mathbf{k}) = \prod_{i=1}^3 \frac{2 \sin(k_i/2)}{\sqrt{2 \pi} k_i},
\end{equation}
is the window function of a cube, where subscript $i$ runs over the three Cartesian axes, $(x,y,z)$. The parameter $\epsilon$ then quantifies the additional power in this specific realization as a fraction of the root-variance on the scale of the cube, and will be treated as a parameter to fit for in our analysis (a separate $\epsilon$ is fit for each MDPL simulation, labeled $\epsilon_\mathrm{cube,VSMDPL}$, $\epsilon_\mathrm{cube,SMDPL}$, etc.). While the smallest cube (VSMDPL) will have the largest variance\footnote{We find $\sigma_\mathrm{cube}=4.3\times 10^{-2}$, $1.3\times 10^{-2}$, $3.1\times10^{-3}$, $6.3\times 10^{-4}$ and $2.7\times 10^{-4}$ for VSMDPL, SMDPL, MDPL2, BigMDPL, and HugeMDPL respectively.}, $\sigma_\mathrm{cube}$, because $\epsilon$ is defined in this way, multiplying $\sigma_\mathrm{cube}$, we expect it to be of order unity for all simulation cubes. More precisely, we expect $\epsilon$ to be a random variable following a standard normal distribution. In this and the following subsections, we define operators, $\mathcal{F}$, as a shorthand notation for these transformations of the halo mass function. In this case, we define $n_\mathrm{cube}(M,z) = \mathcal{F}_\mathrm{cube}[n(M,z)]$.

\subsubsection{Isolation Bias}

The Symphony and COZMIC suites select halos for resimulation based on both a mass criterion and an isolation criterion. For the LMC-mass halos, the isolation criterion requires there to be no more massive halo within $3 h^{-1}$~Mpc in the parent simulation box, while for the Symphony Milky Way halos the requirement is that there is no more massive halo within $4R_\mathrm{vir}$ in the parent simulation box, and for the group-mass halos the requirement is that there is no halo more massive than $10^{13} \mathrm{M}_\odot$ within $3 h^{-1}$~Mpc \citep{2023ApJ...945..159N}.\footnote{Milky Way-est halos do not have an explicit isolation cut applied~~\citep{2024arXiv240408043B}, although we will still apply an isolation bias term to model the specific environments of these systems.} To allow for the possibility that this isolation selection criterion may result in some bias in the halo mass function in the resimulated region, we adopt a simple model in which this isolation bias simply scales the halo mass function by a redshift-dependent factor:
\begin{equation}
    n(M,z) \rightarrow \mathcal{I} (1+z)^\alpha n(M,z),
\end{equation}
denoted hereafter by the operator $\mathcal{F}_\mathrm{iso}$. We will allow the parameters $\mathcal{I}$ and $\alpha$ to differ for the LMC, Milky Way, and group simulations as they have different isolation criteria, referred to as $\mathcal{I}_\mathrm{LMC}$, $\alpha_\mathrm{iso,LMC}$, $\mathcal{I}_\mathrm{MW}$, $\alpha_\mathrm{iso,MW}$, $\mathcal{I}_\mathrm{group}$, and $\alpha_\mathrm{iso,group}$.

\subsubsection{Convolution with halo mass errors}

As shown by \cite{2010ApJ...711.1198T} and \cite{2017MNRAS.467.3454B}, halo masses determined from N-body simulations are subject to measurement uncertainties. Therefore, we convolve the intrinsic mass function with a model of the distribution of measured masses. Specifically,
\begin{eqnarray}
    n(M,z) &\rightarrow& \int_0^\infty \frac{\mathrm{d}M^\prime}{\sqrt{2\pi}s(M^\prime)} \exp\left( - \frac{1}{2} \left[ \frac{M-M^\prime}{s(M^\prime)}\right]^2\right) n(M,z) \nonumber \\
    && \left/ \int_0^\infty \frac{\mathrm{d}M^\prime}{\sqrt{2\pi}s(M^\prime)} \exp\left( - \frac{1}{2} \left[ \frac{M-M^\prime}{s(M^\prime)}\right]^2\right) \right. ,
\end{eqnarray}
denoted hereafter by the operator $\mathcal{F}_\mathrm{err}$, where $s(M)$ is the uncertainty in the measured mass of a halo of mass $M$.

For $s(M)$ we follow \cite{2017MNRAS.471.2871B} and use a modified form of the result found by \cite{2010ApJ...711.1198T}, specifically:
\begin{equation}
    s(M) = 0.135 M \left(\frac{M}{1000 m_\mathrm{p}}\right)^{-1/3},
\end{equation}
where $m_\mathrm{p}$ is the mass of a particle in the simulation.

\subsubsection{Contribution of artificial halos}

It is well established that N-body simulations with power spectra exhibiting a small-scale cutoff exhibit ``artificial halos'' below this scale, which result from discreteness noise in the initial conditions \citep{2007MNRAS.380...93W}. Rather than attempt to identify and remove these artificial halos\footnote{While approaches to removing artificial halos, such as that of \cite{2014MNRAS.439..300L}, are quite successful, they require access to the full particle distribution from the simulation snapshots. Since this is frequently not available in publicly-released simulation products, we instead adopt a forward modeling approach here.} \citep{2014MNRAS.439..300L}, we instead construct a general model for their number density and constrain the parameters of this model as part of our fitting procedure. We note that this is necessary because---unlike in the original COZMIC analyses---we do not make a conservative mass cut that removes artificial halos from the outset.

Specifically, we assume that artificial halos arise from an extra contribution to the variance, $\sigma^2(M)$, arising from particle noise in the initial conditions, and described by a simple form:
\begin{equation}
     \sigma^2(M,t) \rightarrow \sigma^2(M,t) + \left(\frac{M}{M_0}\right)^{\alpha_\mathrm{art}} D^{2{\beta_\mathrm{art}}}(t),
\end{equation}
denoted hereafter by the operator $\mathcal{P}_\mathrm{art}$. Here, the parameter $M_0$ controls the mass scale at which the contribution to the variance is 1 at $z=0$, while $\alpha_\mathrm{art}$ controls how this depends on mass, and $\beta_\mathrm{art}$ how this depends on redshift. We do not assume that the variance of this noise grows exactly according to linear theory, which would correspond to $\beta_\mathrm{art}=1$. Artificial halos arise from fluctuations in the number of particles in a given small volume of the simulation's initial conditions, resulting in fluctuations in overdensity, but the velocities of these particles are not initialized to follow the linear theory expectations given those overdensities. At late times, we may expect linear theory to be approximately followed, but at early times, this is unlikely to be the case. As such, we allow for some modification of the growth rate via the $\beta_\mathrm{art}$ parameter. We assume that $M_0$ scales with the particle mass in each simulation such that $M_0(m_\mathrm{p}) = N_\mathrm{art} m_\mathrm{p}$. The parameters $\alpha_\mathrm{art}$, $\beta_\mathrm{art}$, and $N_\mathrm{art}$ will be constrained as part of our fitting procedure. Note that this effect is included for all simulations used, not just those with small-scale cutoffs in their power spectra. For CDM simulations, its effects will, however, be negligible at low halo masses, where the CDM halo mass function rises fast enough that the artificial halo contribution will be small (at least down to the particle number thresholds considered in this work. At extremely high halo masses it is possible that our artificial halo model will eventually exceed the intrinsic halo mass function in CDM (and all other models)---the intrinsic $\sigma(M)$ may decline fast enough that the artificial halo term dominates. We will examine this further in \S\ref{sec:resultsCDM}.

\subsubsection{Numerical halo-detection efficiency}\label{sec:detectionEfficiency}

We allow for the possibility that, for halos resolved by relatively small numbers of particles, halo finder algorithms may be imperfect, resulting in an incomplete census of halos. We find that such behavior is apparent in the MDPL simulations, for which the mass functions of lower resolution realizations show deviations from their higher resolution counterparts for halos containing relatively few particles, as we will show in \S\ref{sec:resultsCDM}.

To model this effect, we assume that the fraction of halos detected is given by:
\begin{equation}
    \epsilon(M) = 1 - (1-\epsilon_\mathrm{min}) \left(\frac{M}{M_\mathrm{min}}\right)^\alpha (1+z)^\beta,
\end{equation}
where $\epsilon_\mathrm{min}$ gives the detection efficiency at the minimum mass $M_\mathrm{min}$ at $z=0$ and parameters $\alpha$ and $\beta$ control the mass and redshift dependence of the efficiency. We set $M_\mathrm{min} = N_\mathrm{min} m_\mathrm{p}$ and treat $N_\mathrm{min}$ as a parameter. Then,
\begin{equation}
    n(M,z) \rightarrow \epsilon(M) n(M,z),
\end{equation}
denoted hereafter by the operator $\mathcal{F}_\mathrm{det}$.

The various MDPL simulations were analyzed by different versions of the Rockstar halo finder (as evidenced by differences in the file formats of the resulting data files), and the Symphony/Milky Way-est/COZMIC simulations by yet another version. We therefore conservatively fit for separate values of the parameters, $(\epsilon_\mathrm{min}, N_\mathrm{min}, \alpha, \beta)$, of this detection efficiency for each VSMDPL (``MDPLRockstar4''), SMDPL/MDPL2 (these two have matching format Rockstar files and so are treated with the same detection efficiency model;  ``MDPLRockstar3''), BigMDPL (``MDPLRockstar2''), HugeMDPL (``MDPLRockstar1''), and Symphony/Milky Way-est/COZMIC (``MDPLSymphony''), where the names in quotes refer to the subscripts used to identify the parameters of the detection efficiency algorithm in each case.

\subsubsection{Choice of Window Function}\label{sec:windowFunction}

\cite{2013MNRAS.428.1774B} showed that, for power spectra with a small-scale cut-off, using a sharp-$k$ window function has the advantage that it avoids spurious halos in the HMF below the cut-off scale which arise because of the dependence of the HMF on $\mathrm{d} W(k|M)/\mathrm{d}k$ (through the $\alpha = |\mathrm{d}\log \sigma/\mathrm{d}\log M|$ term).

However, \cite{2018JCAP...04..010L} show that some halos \emph{do} form below the cut-off scale. A sharp-$k$ window function is insufficient as these may not be captured. The reasons for this are not fully understood. One possibility is non-linear coupling between modes (although see section 3.f of \citealt{1994ApJ...431..495J}). Alternatively, the lack of such halos when a sharp-$k$ filter is used may simply reflect the oversimplifications inherent in the Press-Schechter approach. 

\cite{2021MNRAS.506..128B}, using simulations run within the ``ETHOS'' structure formation framework, show that a window function intermediate between sharp-$k$ and top-hat window functions of the form:
\begin{equation}
 W(kR) = \left[ 1+\left(\frac{kR}{c_\mathrm{W}}\right)^\beta \right]^{-1}
\end{equation}
with $c_\mathrm{W} = 3.78$, $\beta = 3.46$ performs well in matching the halo mass function resulting from power spectra containing dark acoustic oscillations.

In this work, to allow for more flexibility, we modify the \cite{2021MNRAS.506..128B} window function by allowing the parameters $c_\mathrm{W}$ and $\beta$ to be functions of the local slope of the smoothed power spectrum, and introduce a minimum scaled wavenumber below which the window function is constant. Specifically, we modify the window function to
\begin{equation}
W(kR) = \left\{ \begin{array}{ll} 1 & \hbox{if } \frac{kR}{c_\mathrm{W}} < x_\mathrm{min}, \\ \left[ 1+ \left(\frac{kR}{c_\mathrm{W}} - x_\mathrm{min}\right)^\beta \right]^{-1} & \hbox{otherwise,} \end{array} \right.
\end{equation}
and we choose
\begin{equation}
    y = y_0 y_1^{n-n_0},
    \label{eq:windowFunctionScaling}
\end{equation}
where $y$ is either $c_\mathrm{W}$ or $\beta$, $n = \mathrm{d}\log \tilde{P} / \mathrm{d} \log k$ is the logarithmic derivative of the smoothed linear theory power spectrum
\begin{eqnarray}
\tilde{P}(k) &=& \int_{-\infty}^\infty \frac{1}{\sqrt{2 \pi} \log k_\mathrm{width}} \nonumber \\
 & & \exp\left[-\frac{1}{2} \left(\frac{\log k - \log k^\prime}{\log k_\mathrm{width}}\right)^2\right] P(k^\prime) \mathrm{d} \log k^\prime,
\end{eqnarray}
and $n_0 = -2.6$ is a convenient zero-point. In the above, $x_\mathrm{min}$ and $\log k_\mathrm{width}$ are free parameters to be constrained. Given the origin of this window function, we refer to our model as the ``extended ETHOS'' window function.

In models with small-scale cut-offs in the power spectrum, $\sigma(M)$ approaches a constant for small $M$. The halo mass function is then largely determined by the $\alpha(M) = |\mathrm{d}\log\sigma/\mathrm{d}\log M|$ term in equation~(\ref{eq:hmfIntrinsic}), which is found by integrating $k^3 P(k) W(k) \mathrm{d}W(k)/\mathrm{d}k$ over wavenumber.

\begin{figure}
\centering
\includegraphics[width=85mm]{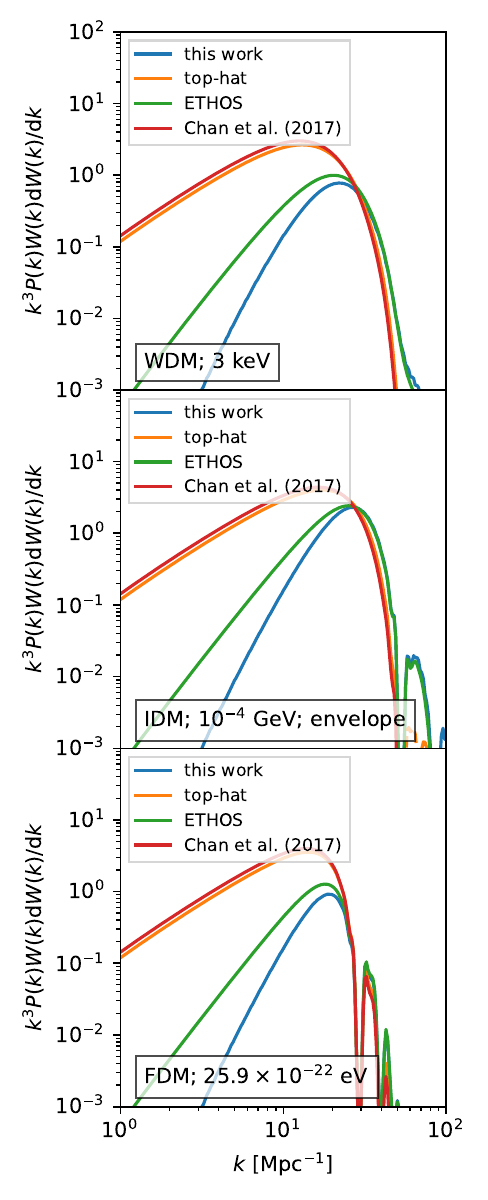}
\caption{The combination $k^3 P(k) W(k) \mathrm{d}W(k)/\mathrm{d}k$---the integral of which the halo mass function is proportional to for low halo masses---is shown for four different choices of window function evaluated for a mass of $10^8\mathrm{M}_\odot$: the extended ETHOS window function used in this work (blue line; with parameters taken from our analysis in \S\ref{sec:results}), top-hat (orange line), the smooth-$k$ window function of \protect\citeauthor{2021MNRAS.506..128B}~(\protect\citeyear{2021MNRAS.506..128B}; with their preferred parameters $c_\mathrm{W} = 3.78$, $\beta = 3.46$; green line), and the smooth $k$-space window function of \protect\citeauthor{2017PhRvD..96j3543C}~(\protect\citeyear{2017PhRvD..96j3543C}; with their preferred parameters $(\hat{c},\hat{\beta})=(3.3,4.8)$; red line). Solid lines indicate where this function is negative, with dashed lines indicating where this function is positive. Results are shown for 3~keV WDM (upper panel), $10^{-4}$~GeV IDM (middle panel), and $25.9\times 10^{-22}$~eV FDM (lower panel).}
\label{fig:windowFunctions}
\end{figure}

Figure~\ref{fig:windowFunctions} shows this integrand for three of our models, which exhibit small-scale cut-offs: 3keV WDM (upper panel), $10^{-4}$~GeV IDM (middle panel), and $25.9\times 10^{-22}$~eV FDM (lower panel). In each panel, the blue line shows the preferred window function found in this work (see \S\ref{sec:results}), while other lines\footnote{Note that the sharp-$k$ window function is not shown here as the integrand is zero everywhere except at the point corresponding to the cut-off wavenumber for this window function.} show the spherical top-hat window function (orange line), that of \citeauthor{2021MNRAS.506..128B}~(\citeyear{2021MNRAS.506..128B}; green line), and that proposed by \citeauthor{2017PhRvD..96j3543C}~(\citeyear{2017PhRvD..96j3543C}; red line). Notably, all four cases shown have similar cut-offs at large wavenumber (albeit with different amplitudes of the secondary peaks at even higher wavenumbers). However, at low wavenumbers, there are distinct differences between the models, with the top-hat and \cite{2017PhRvD..96j3543C} models agreeing quite closely, and the \cite{2021MNRAS.506..128B} model and that used in this work having much lower amplitude (our model lower than that of \citealt{2021MNRAS.506..128B}). Both our model and that of \cite{2021MNRAS.506..128B} were constructed to allow good matches to be obtained for halo mass functions arising from power spectra with features below their cut-off scale. Figure~\ref{fig:windowFunctions} clearly shows that both models find it necessary to suppress the contributions of large-scale modes to the halo mass functions to achieve a good match in these cases. Without this strong suppression, leakage of power from large-scale modes will dominate over the small-scale power in such models, resulting in a significant overproduction of low-mass halos below the cut-off scale.

\subsubsection{Final halo mass function}

The various effects described in the preceding subsections are combined to give us a final ``measured'' halo mass function. For the MDPL simulations, this is:
\begin{eqnarray}
    P^\prime(k) &=& \mathcal{P}_\mathrm{cube}[P(k)], \nonumber \\
    \sigma^\prime(M) &=& \mathcal{P}_\mathrm{art}[\sigma(M)], \nonumber \\
    n^\prime(M,z) &=& \mathcal{F}_\mathrm{det}[\mathcal{F}_\mathrm{err}[\mathcal{F}_\mathrm{cube}[n(M,z)]]],
\end{eqnarray}
while for the Symphony/Milky Way-est/COZMIC zoom-in simulations we have:
\begin{eqnarray}
    \sigma^\prime(M),\delta^\prime_\mathrm{c}(z) &=& \mathcal{P}_\mathrm{peak-background}[\mathcal{P}_\mathrm{art}[\sigma(M)],\delta_\mathrm{c}(z)], \nonumber \\
    n^\prime(M,z) &=& \mathcal{F}_\mathrm{det}[\mathcal{F}_\mathrm{err}[\mathcal{F}_\mathrm{iso}[n(M,z)]]],
\end{eqnarray}
It is this ``measured'' halo mass function model that we will fit to those measured from N-body simulations.

\section{Likelihood analysis}\label{sec:likelihood}

Many prior studies of the halo mass function \citep{2007MNRAS.374....2R,2008ApJ...688..709T,2016MNRAS.456.2486D} have performed their fits in the $\nu$--$f(\nu)$ plane. While the functional form of our intrinsic halo mass function, and the environmental dependence adopted in this work are both formulated in the $\nu$--$f(\nu)$ plane, once we add the various numerical and nuisance modifiers to the halo mass function (many of which are formulated in the space of halo masses, and therefore $n(M)$) the form is no longer universal in $f(\nu)$. While we \emph{could} still perform the fit in the $f(\nu)$ space, the motivation to do so is greatly reduced by this fact. As such, we elect to fit in the $n(M)$ plane directly, but will show the resulting intrinsic halo mass function in the $\nu$--$f(\nu)$ plane below (see Figure~\ref{fig:globalMassFunction}).

We assume that the number of N-body halos in each bin of the measured halo mass function follows a Poisson distribution, noting that possible correlations between the number of halos in each bin due to correlated large-scale structure are accounted for by our approach described in \S\ref{sec:missingPower}, specifically equation~(\ref{eq:sigmaCube})\footnote{As $\epsilon$ in equation~(\ref{eq:sigmaCube}) is changed, the mass function responds coherently across all bins, resulting in correlated changes in the mass function. Physically, this accounts for the effect of having a finite number of large-scale modes in the simulation box---if one such mode happens to have been sampled with a higher amplitude than average, for example, then it would increase/decrease the halo mass function coherently on scales where the bias is positive/negative.}

Prior work has been able to fit halo mass functions to an accuracy of 5--10\% (see, for example, Figure~2 of \citealt{2016MNRAS.456.2486D}), indicating that the functional forms used are imperfect. To allow for this fact, and for the possibility that the non-CDM models considered in this work will be even more challenging to fit to high accuracy, we introduce a model discrepancy factor, $\mathcal{C}_\mathrm{disc}$, which represents the additional variance between model and data due to these imperfections. To incorporate this into our likelihood analysis, we replace the Poisson distribution with a negative binomial distribution with stopping parameter $r=1/\mathcal{C}_\mathrm{disc}$. The negative binomial distribution here plays the role of an overdispersed Poisson-like distribution---in the limit $\mathcal{C}_\mathrm{sc} \rightarrow 0$ the negative binomial distribution approaches the Poisson distribution. Such a model discrepancy term acts to account for the fact that our model is not a perfect description of the simulation data, and provides a quantitative estimate of the difference between model and data. Increasing $\mathrm{C}_\mathrm{disc}$ from zero will initially cause the likelihood of a model to increase, as the negative binomial distribution becomes broader. Above some value, however, further increase in $\mathrm{C}_\mathrm{disc}$ (and the continued broadening of the negative binomial distribution) will result in a lowering of the likelihood due to the decreased amplitude of the negative binomial distribution (which, of course, is normalized to unity). As such, there will be a preferred scale for $\mathrm{C}_\mathrm{disc}$ which maximizes the likelihood (for fixed values of the other model parameters), and provides an estimate of the discrepancy between model and simulation data. Note that our choice of a single value of $\mathrm{C}_\mathrm{disc}$, and not, for example, one which varies as a function of halo mass or redshift, is merely the simplest possible choice, providing an estimate of the global discrepancy between model and data.

The likelihood function is then given by:
\begin{eqnarray}
    \log \mathcal{L} &=& \sum_{i=1}^{N_\mathrm{sims}} \sum_{j=1}^{N_{\mathrm{bins},i}} N_{i,j} \log \hat{N}_{i,j} - \log N_{i,j}!  \nonumber \\
    & & + \log \Gamma(r+N_{i,j}) - \log \Gamma(r) \nonumber \\
    & & - N_{i,j} \log(r+\hat{N}_{i,j}) - r \log (\hat{N}_{i,j}/r+1)
    \label{eq:likelihood}
\end{eqnarray}
where subscript $i$ corresponds to different simulations, and subscript $j$ to bins in the halo mass function of a given simulation. Here, $N_{ij}$ is the number of halos measured in the $j^\mathrm{th}$ bin of the halo mass function in the $i^\mathrm{th}$ simulation, and $\hat{N}_{ij}$ is the corresponding number of halos predicted for that bin by our model, which we find by integrating the model ``measured'' mass function across the width of the bin and multiplying by the relevant simulation volume. In equation~(\ref{eq:likelihood}) we sum over only those mass bins corresponding to at least 300 particles. Previous work \citep{2025ApJ...986..127N} using the Symphony and COZMIC simulations has found converged results using a threshold of 300 particles (although this was for subhalos).

\subsection{Correlations in the mass function}\label{sec:correlations}

In the likelihood function described above, we have assumed that all bins in each mass function are independent and that each simulation is independent of every other simulation. This is not true---correlations arise for at least two reasons, which we will describe below.

First, in the Symphony/Milky Way-est/COZMIC simulations, we have instances where the same halo was resimulated for multiple different dark matter power spectra. Specifically, when generating the initial conditions for the COZMIC simulations, the same phases were used as for the Symphony simulations---only the amplitudes of the modes were adjusted to reflect the different power spectra assumed in COZMIC. Since the COZMIC models are all designed to have power spectra close to that of CDM at large masses, this means that the larger scale structure (and, therefore, the more massive halos) in COZMIC will be strongly correlated with that in the corresponding Symphony (or Milky Way-est) simulation. To mitigate the effects of these correlations, we exclude from each COZMIC halo mass function those bins above a mass threshold, $M_\mathrm{phase}$ chosen such that, at all lower masses, the number of halos in each bin differs from that in the corresponding Symphony simulation by at least $1\sigma$ (assuming Poisson statistics). This ensures that only bins in which the COZMIC power spectrum results in a significant difference in the number of halos present are included.

We note that, for both Symphony and COZMIC, some halos were resimulated multiple times, at different resolutions. In these cases, we use only the highest resolution resimulation, so there is no concern about correlations between resimulations at different resolutions.

The second source of correlations arises from our use of halo mass functions measured at multiple snapshots from the same simulation. Clearly, if we were to measure the halo mass function from a simulation at some redshift, $z$, and at another redshift $z+\Delta z$, then, as $\Delta z \rightarrow 0$, the two mass functions must become perfectly correlated. For larger $\Delta z$, the strength of the correlation presumably decreases, but will be non-zero even for large $\Delta z$. 

Halos at $z=0$ have progenitors at $z>0$. If there is a fluctuation in the number of $z=0$ halos in some bins, we expect a corresponding fluctuation in the number of lower mass progenitor halos at $z>0$---so there will be correlations between halo mass functions at different redshifts. We can therefore estimate the correlation structure using bootstrap resampling of the $z=0$ halos, and including only the bootstrapped resampled halos and their progenitors in our halo mass function measures. Specifically, we take the $z=0$ halo population in the MDPL2 simulation. In each bootstrap realization we assign a weight to each halo drawn from a Poisson distribution with mean $\mu=1$, such that the total number of halos is unchanged on average. We apply this same weight to all $z>0$ progenitors of that halo. From this, we constructed weighted estimates of the MDPL2 halo mass functions at all redshifts of interest. This process is repeated 1,000 times (using just 1\% of the MDPL2 volume for computational speed---this still provides sufficient signal to accurately quantify the covariances we seek). Using the resulting set of 1,000 halo mass functions across redshifts we estimate the covariance between bins (both within a redshift, and between redshifts).

\begin{figure*}
\begin{center}
\includegraphics[]{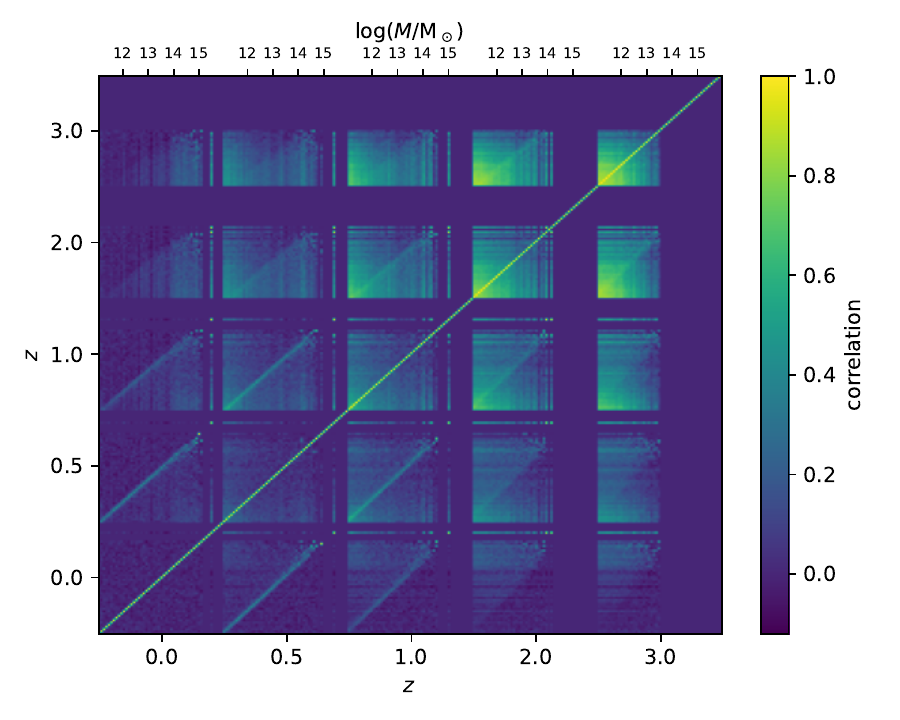}
\end{center}
\vspace{-8mm}\caption{The correlation matrix for MDPL2 halo mass functions. Correlations were measured by taking 1\% (randomly selected) of halos at $z=0$ in MDPL2 and bootstrap resampling them. At higher redshifts, only those progenitor halos of the $z=0$ halos which were included in each bootstrap resample were considered, and used to construct the halo mass function at that redshift. This process was repeated a large number of times, and the resulting correlations between bins across redshifts were measured. The matrix shown in this figure indicates correlations between all bins of the halo mass function at all redshifts considered (approximately $z=0.0$, $0.5$, $1.0$, $2.0$, and $3.0$. The $5\times5$ block structure of the matrix corresponds to these five redshifts (as indicated on the bottom and left axes), with elements within each block corresponding to halo mass (as shown along the top axis).}
\label{fig:correlation}
\end{figure*}

Figure~\ref{fig:correlation} shows the correlation matrix of halo mass functions at different redshifts obtained in this way from the MDPL2 simulation, using only 1\% of the total MDPL2 simulation for speed. At $z=0$ there are no off-diagonal correlations by construction---correlations that should be present due to large-scale structure are not captured by this calculation (but are accounted for elsewhere as described in \S\ref{sec:missingPower}). At higher redshifts, we find correlations between bins of the halo mass function within the same snapshot. These arise because, for example, if the bootstrapped sample happens to include an excess of massive halos at $z=0$, those halos will have progenitors across many mass bins at higher $z$, resulting in an excess in multiple bins in the higher $z$ halo mass function, and, therefore, a correlation. Looking at correlations between different redshifts, we see that there are clear correlations between, for example, bins at $z=0$ and lower mass bins at $z>0$, as expected (i.e., progenitor halos are almost always lower mass than their $z=0$ descendant).

Ideally, such correlations should be accounted for in our likelihood function. However, we have not found a (computationally-)viable model for this---Appendix~\ref{app:correlations} summarizes our attempts and their reasons for failure. Therefore, in the present work, we simply ignore these correlations. This will mean that our model parameters are somewhat over-constrained---a point that we will revisit in \S\ref{sec:discussion}. In future work, such correlations could potentially be accounted for using likelihood-free inference methods and forward modeling of the number of halos in each bin/snapshot, provided that a sufficiently simple and accurate form describing the correlation matrix could be found.

\subsection{MCMC methodology}\label{sec:mcmc}

To determine constraints on the parameters of our model, we perform an MCMC sampling. Specifically, we use a differential evolution approach, running a total of 160 chains. Chains are initialized tightly clustered in a Gaussian sphere, thereby allowing the chains to diffuse out from this initial state and explore the likelihood surface. To reduce the computational burden we make an initial MCMC run with chains starting close to the median of the prior on each parameter, using a simplified version of our halo mass function model in which we ignore all of the effects outlined in \S\ref{sec:missingPower}--\ref{sec:detectionEfficiency} (since some of those effects involve computationally slow steps such as integration). After this initial MCMC has converged, we start a new MCMC, including the full model, with chains starting from the maximum posterior probability point from the first run. We expect that this new starting point will be reasonably close to the true maximum posterior in the full model, thereby aiding convergence.

Proposals are constructed by selecting two chains, $j$ and $k$, at random and using a fraction of their difference as the proposal step:
\begin{eqnarray}
    \theta_i \rightarrow \theta_i + \gamma(\theta_j-\theta_k) +\epsilon,
\end{eqnarray}
where $\gamma$ is a parameter controlling the size of the proposals, and $\epsilon$ is a random vector drawn from a Cauchy distribution with median zero and width parameter equal to $10^{-4}$ of the current range of parameter values spanned by the ensemble of chains. This random vector ensures that chains are positively recurrent. For a multivariate normal likelihood in $N$ dimensions, the optimal choice for $\gamma$ is $\gamma_0 = 2.38/\sqrt{N}$ \citep{terr_braak_markov_2006}. Our model has $N=47$, which gives $\gamma_0=0.35$. Using this as approximate guidance, we initially choose $\gamma=0.5$, and then adaptively adjust $\gamma$ to keep the acceptance rate of proposals in the range $0.1$ to $0.9$.

Since our chains are initialized in an under-dispersed state\footnote{For differential evolution, MCMC starting in an overdispersed state is problematic, as the chains would have to converge into a smaller region of parameter space. Instead, starting them from an under-dispersed state allows them to naturally diffuse out until they reach equilibrium in the likelihood surface. (We thank Andrew Robertson for making this point clear to us.)}, convergence metrics such as the Gelman-Rubin $\hat{R}$ statistic \citep{gelman_a._inference_1992}, can not be applied. Therefore, we use simple visual inspection of the chains to judge when convergence is reached. This is explored in Appendix~\ref{app:convergence}, where we show that a burn-in of 7,600 steps is sufficient to reach convergence, and where we also identify one outlier chain that we remove from our analysis.

Once convergence is reached, the simulation is allowed to run for a further 10,000 steps. We measure correlation lengths in these post-convergence steps, finding a maximum of 36 steps across all parameters. This results in a total of around $10,000 \times 160 / 36 \approx 44,000$ independent samples from the posterior distribution on our model parameter space.

\subsubsection{Priors}

Priors for all model parameters are listed in Table~\ref{tb:priors}. We adopt broad, uninformative priors for most parameters. 
\begin{itemize}
    \item For parameters of the intrinsic mass function, priors are broad and encompass the values found by \cite{2011ApJ...732..122B}, which were $(A,a,p,q)=(0.333,0.788,0.807,1.795)$. For the new parameters, $b$ and $c$, we adopt broad, uniform priors encompassing the plausible range based on an initial inspection of the N-body halo mass functions.
    
    \item For our artificial halo model, broad priors on all parameters are adopted to encompass the plausible range based on an initial inspection of the artificial halo mass function seen in simulations with suppressed power at low halo mass.
    
    \item For the window function we adopt broad priors for $c_\mathrm{W,0}$ and $\beta_0$ encompassing the prior best-fit values reported by \cite{2021MNRAS.506..128B}. For  $c_\mathrm{W,1}$ and $\beta_1$ we adopt broad, positive priors. These parameters must be positive to ensure that the equation~(\ref{eq:windowFunctionScaling}) results in real values for $c_\mathrm{w}$ and $\beta$. For $x_\mathrm{min}$ and $k_\mathrm{width}$ we similarly adopt broad, uninformative priors.
    
    \item For the simulation perturbation parameters, a standard normal prior is expected by construction. We limit the range to $\pm10$ for computational convenience.
    
    \item For the isolation bias parameters, we expect $\mathcal{I}$ to be approximately 1 if isolation bias effects are weak, and $\mathcal{I} > 0$ is required to ensure a positive mass function. We therefore adopt a lognormal prior centered at $1$ and root-variance of $0.2$~dex. For the exponents, $\alpha$, we adopt a broad uniform prior between $-3$ and $0$---we expect this parameter to be negative as the effects of the isolation selection should diminish with increasing redshift.
 
    \item For the detection efficiency model, we choose a uniform prior on $\log_{10} N_\mathrm{min}$ between 1 and 100 particles, a uniform prior on $\epsilon_\mathrm{min}$ from 0 to 1, and broad priors on the exponents $\alpha_\mathrm{min}$ and $\beta_\mathrm{min}$.
\end{itemize}

\begin{table}[]
    \caption{Priors on model parameters. The first column gives the parameter name, while the second defines the prior distribution used for the parameter. Note that in some cases, we work with the logarithm of the parameter in our MCMC simulation. In those cases, the prior applies to the base-10 logarithm of the parameter, as indicated in the table. In the table ``Uniform($a,b$)'' indicates a uniform distribution between $a$ and $b$, and ``Normal($\mu,\sigma,a,b$)'' indicates a normal distribution with mean $\mu$, variance $\sigma^2$, truncated below $a$ and above $b$. Parameters are grouped according to the component of our model to which they apply.}
    \label{tb:priors}
    \centering
    \begin{tabular}{ll}
\hline
\textbf{Parameter} & \textbf{Prior} \\
\hline
\multicolumn{2}{c}{\emph{Intrinsic mass function}} \\
$a$ & \hbox{Uniform}(+0.03,+10.00) \\
$b$ & \hbox{Uniform}(-3.00,+3.00) \\
$c$ & \hbox{Uniform}(+0.10,+5.00) \\
$p$ & \hbox{Uniform}(-3.00,+3.00) \\
$q$ & \hbox{Uniform}(-3.00,+3.00) \\
$\log_{10} A$ & \hbox{Uniform}(-3.00,+3.00) \\
\hline
\multicolumn{2}{c}{\emph{Artificial halos}} \\
$\alpha_\mathrm{art}$ & \hbox{Uniform}(-3.00,+0.00) \\
$\beta_\mathrm{art}$ & \hbox{Uniform}(-3.00,+3.00) \\
$\log_{10} N_\mathrm{art}$ & \hbox{Uniform}(+0.00,+3.00) \\
\hline
\multicolumn{2}{c}{\emph{Window function}} \\
$c_\mathrm{W,0}$ & \hbox{Uniform}(+0.50,+6.00) \\
$\beta_0$ & \hbox{Uniform}(+0.50,+10.00) \\
$\log_{10} c_\mathrm{W,1}$ & \hbox{Uniform}(-1.00,+0.70) \\
$\log_{10} \beta_1$ & \hbox{Uniform}(-1.00,+0.70) \\
$x_\mathrm{min}$ & \hbox{Uniform}(+0.00,+5.00) \\
$\log_{10} k_\mathrm{width}$ & \hbox{Uniform}(-2.00,+1.00) \\
\hline
\multicolumn{2}{c}{\emph{Model discrepancy}} \\
$\log_{10} \mathcal{C}_\mathrm{disc}$ & \hbox{Uniform}(-6.00,+0.00) \\
\hline
\multicolumn{2}{c}{\emph{Simulation perturbation}} \\
$\epsilon_\mathrm{cube,BigMDPL}$ & \hbox{Normal}(+0.00,+1.00,-10.00,+10.00) \\
$\epsilon_\mathrm{cube,HugeMDPL}$ & \hbox{Normal}(+0.00,+1.00,-10.00,+10.00) \\
$\epsilon_\mathrm{cube,MDPL2}$ & \hbox{Normal}(+0.00,+1.00,-10.00,+10.00) \\
$\epsilon_\mathrm{cube,SMDPL}$ & \hbox{Normal}(+0.00,+1.00,-10.00,+10.00) \\
$\epsilon_\mathrm{cube,VSMDPL}$ & \hbox{Normal}(+0.00,+1.00,-10.00,+10.00) \\
\hline
\multicolumn{2}{c}{\emph{Isolation bias}} \\
$\log_{10} \mathcal{I}_\mathrm{LMC}$ & \hbox{Normal}(+0.00,+0.10,-3.00,+3.00) \\
$\alpha_\mathrm{iso,LMC}$ & \hbox{Uniform}(-3.00,+0.00) \\
$\log_{10} \mathcal{I}_\mathrm{MW}$ & \hbox{Normal}(+0.00,+0.10,-3.00,+3.00) \\
$\alpha_\mathrm{iso,MW}$ & \hbox{Uniform}(-3.00,+0.00) \\
$\log_{10} \mathcal{I}_\mathrm{Group}$ & \hbox{Normal}(+0.00,+0.10,-3.00,+3.00) \\
$\alpha_\mathrm{iso,Group}$ & \hbox{Uniform}(-3.00,+0.00) \\
\hline
\multicolumn{2}{c}{\emph{In-simulation detection efficiency}} \\
$\log_{10} N_\mathrm{min,MDPL1}$ & \hbox{Uniform}(+0.00,+2.00) \\
$\epsilon_\mathrm{min,MDPL1}$ & \hbox{Uniform}(+0.00,+1.00) \\
$\alpha_\mathrm{det,MDPL1}$ & \hbox{Uniform}(-3.00,+0.00) \\
$\beta_\mathrm{det,MDPL1}$ & \hbox{Uniform}(-3.00,+3.00) \\
$\log_{10} N_\mathrm{min,MDPL2}$ & \hbox{Uniform}(+0.00,+2.00) \\
$\epsilon_\mathrm{min,MDPL2}$ & \hbox{Uniform}(+0.00,+1.00) \\
$\alpha_\mathrm{det,MDPLr2}$ & \hbox{Uniform}(-3.00,+0.00) \\
$\beta_\mathrm{det,MDPL2}$ & \hbox{Uniform}(-3.00,+3.00) \\
$\log_{10} N_\mathrm{min,MDPL3}$ & \hbox{Uniform}(+0.00,+2.00) \\
$\epsilon_\mathrm{min,MDPL3}$ & \hbox{Uniform}(+0.00,+1.00) \\
$\alpha_\mathrm{det,MDPL3}$ & \hbox{Uniform}(-3.00,+0.00) \\
$\beta_\mathrm{det,MDPL3}$ & \hbox{Uniform}(-3.00,+3.00) \\
$\log_{10} N_\mathrm{min,MDPL4}$ & \hbox{Uniform}(+0.00,+2.00) \\
$\epsilon_\mathrm{min,MDPL4}$ & \hbox{Uniform}(+0.00,+1.00) \\
$\alpha_\mathrm{det,MDPL4}$ & \hbox{Uniform}(-3.00,+0.00) \\
$\beta_\mathrm{det,MDPL4}$ & \hbox{Uniform}(-3.00,+3.00) \\
$\log_{10} N_\mathrm{min,Symphony}$ & \hbox{Uniform}(+0.00,+2.00) \\
$\epsilon_\mathrm{min,Symphony}$ & \hbox{Uniform}(+0.00,+1.00) \\
$\alpha_\mathrm{det,Symphony}$ & \hbox{Uniform}(-3.00,+0.00) \\
$\beta_\mathrm{det,Symphony}$ & \hbox{Uniform}(-3.00,+3.00) \\
\hline
\end{tabular}

\end{table}

\section{Results}\label{sec:results}

We will begin by focusing on how well our model is able to match the halo mass function in CDM and non-CDM models, and trends with redshift and environment. We will then explore our results for artificial halos, isolation bias, and detection efficiency. Note that while only halos of mass corresponding to at least 300 particles were used in our likelihood analysis, when showing halo mass functions, we plot N-body results down to masses corresponding to 30 particles.

\subsection{Parameters of the intrinsic halo mass function}\label{sec:resultsIntrinsic}

\begin{figure*}
\includegraphics[width=7.0in]{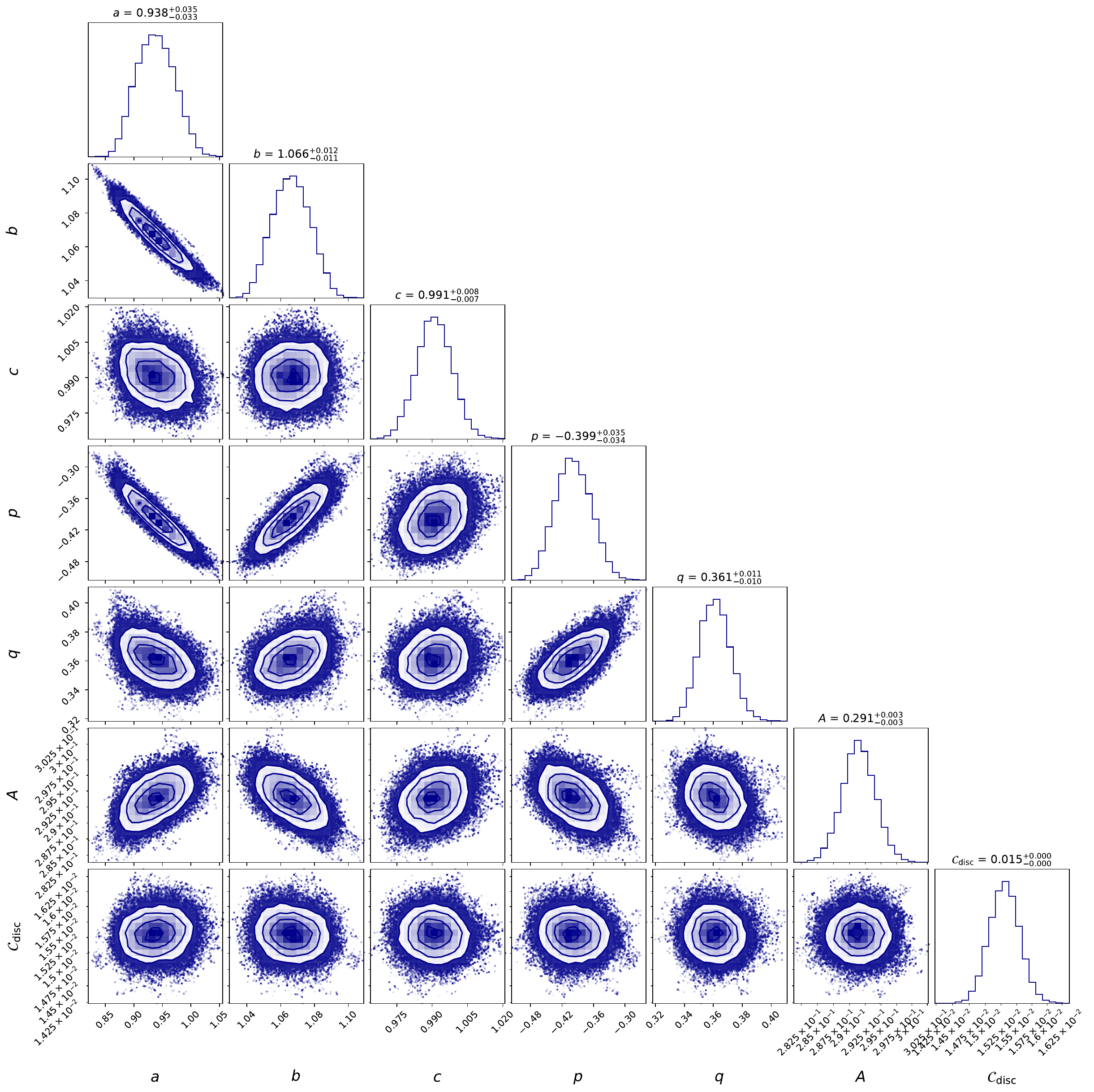} 
\caption{Posterior distributions for the parameters of the intrinsic halo mass function resulting from our MCMC simulation. We also include the model discrepancy term, $\mathcal{C}_\mathrm{disc}$. Panels on the diagonal show the 1D marginalized distributions for each parameter (with the median and $16^\mathrm{th}$ and $84^\mathrm{th}$ percentiles shown above each panel), while off-diagonal panels show joint distributions for each pair of parameters. All posterior distributions are compact and unimodal. Clear correlations are seen between several pairs of parameters.}
\label{fig:hmfCornerPlot}
\end{figure*}

Figure~\ref{fig:hmfCornerPlot} shows the posterior distributions for the parameters of the intrinsic halo mass function, plus the model discrepancy term, $\mathcal{C}_\mathrm{disc}$, resulting from our MCMC sampling. Panels on the diagonal show the 1D marginalized distributions for each parameter (with the median and $16^\mathrm{th}$ and $84^\mathrm{th}$ percentiles shown above each panel), while off-diagonal panels show joint distributions for each pair of parameters. All posterior distributions are compact and unimodal, with several pairs of parameters showing clear correlations. Our posteriors exclude the best-fit values found by \cite{2011ApJ...732..122B}, but this is not surprising as we are using a modified form of their mass function and include several additional numerical/systematic effects in our analysis. The model discrepancy term has a median of $\mathcal{C}_\mathrm{disc}=0.015$. As this describes the fractional variance added to the likelihood function, this suggests that our model is, overall, good to within around 12\
---although we will see cases below where the model performs less well than this. This accuracy is comparable to that quoted by \cite{2016MNRAS.456.2486D} in their analysis. However, we emphasize that here we fit to a much wider range of halo masses and dark matter models than was considered by \cite{2016MNRAS.456.2486D}---the ability of our model to maintain this level of agreement with the N-body data over this much broader span of mass and dark matter physics represents a significant advance.

Considering the new parameters added to the \cite{2011ApJ...732..122B} mass function ($b$ and $c$---see \S\ref{sec:hmfModel}), Figure~\ref{fig:hmfCornerPlot} shows that $b$ has a mild preference to be very slightly larger than $1$, while $c$ is consistent with $1$. Recalling that $b=c=1$ results in the original \cite{2011ApJ...732..122B}, this suggests that the addition of these parameters is only marginally advantageous in fitting the N-body halo mass functions. We retain these parameters in our model, but it is clear that the data do not strongly require them, and they could be dropped from future analyses for simplicity. We will explore the constraints on the window function parameters when discussing non-CDM models in \S\ref{sec:hmfNonCDM} below.

\subsection{Cold Dark Matter}\label{sec:resultsCDM}

\begin{figure}
\includegraphics[width=3.5in]{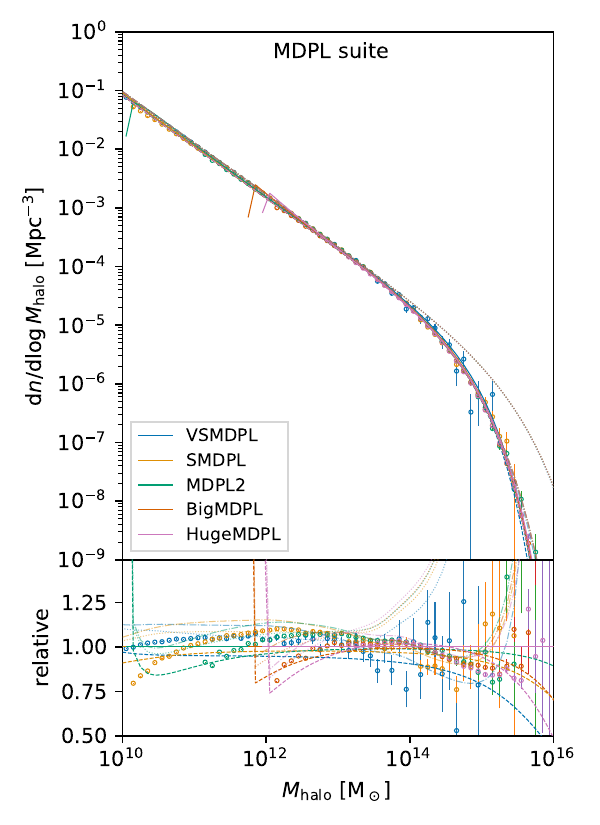}
\caption{Halo mass functions at $z=0$ in the five MDPL cosmological volumes (upper panel). Circles indicate the halo mass function measured from each simulation, with error bars indicating Poisson uncertainties for reference. Solid lines indicate our best-fit model (including all numerical effects), with colors corresponding to each MDPL volume as indicated in the figure. Dashed lines show the intrinsic halo mass function after numerical effects are removed, while dotted and dot-dashed lines show the halo mass function fits of \cite{2021MNRAS.506..128B} and \cite{2016MNRAS.456.2486D}, respectively, following the same color coding. The lower panel shows the same halo mass functions but normalized to the model expectation for each MDPL volume, to emphasize the relative offsets between N-body data and our model.}
\label{fig:hmfMDPLBoxes}
\end{figure}

The upper panel of Figure~\ref{fig:hmfMDPLBoxes} shows halo mass functions at $z=0$ from the MDPL simulation suite. Circles indicate the halo mass function measured from each simulation, with error bars indicating Poisson uncertainties for reference. Solid lines indicate our best-fit model (including all numerical effects), with colors corresponding to each MDPL volume as indicated in the figure. Dashed lines show the intrinsic halo mass function after numerical effects are removed, while dotted and dot-dashed lines show the halo mass function fits of \cite{2021MNRAS.506..128B} and \cite{2016MNRAS.456.2486D}, respectively, following the same color coding. The lower panel of this figure shows the same results, but normalized to the model expectation for each MDPL volume, to more clearly show the relative offsets.

First, we note that while the halo mass functions measured from the five MDPL boxes are largely consistent, they do exhibit clear deviations as we approach the resolution limit of each simulation. For example, at the lowest masses shown, the SMDPL halo mass function (orange circles) lies systematically below that of the higher resolution VSMDPL halo mass function (blue circles). This motivates the need for the detection efficiency model of \S\ref{sec:detectionEfficiency}, which can absorb such systematic offsets.

Our model results in a very good fit to the data across the full $\sim 5$ orders of magnitude in halo mass probed by the combined MDPL simulations. The \cite{2021MNRAS.506..128B} halo mass function model significantly overpredicts the measured halo mass function at both low and high masses, making it inappropriate in this high mass regime (recall that \citealt{2021MNRAS.506..128B} calibrated their model to much lower mass halos in a model with dark acoustic oscillations). The \cite{2016MNRAS.456.2486D} halo mass function model is an excellent match to our measurements from the MDPL suite, although it slightly deviates from our fit at the highest masses (but remains largely within the uncertainties of the data in this regime).

\begin{figure*}
\begin{tabular}{cc}
\includegraphics[width=3.5in]{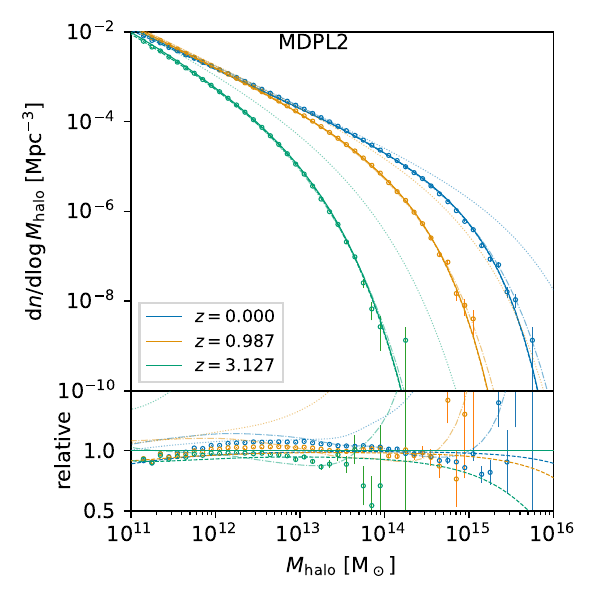} & \includegraphics[width=3.5in]{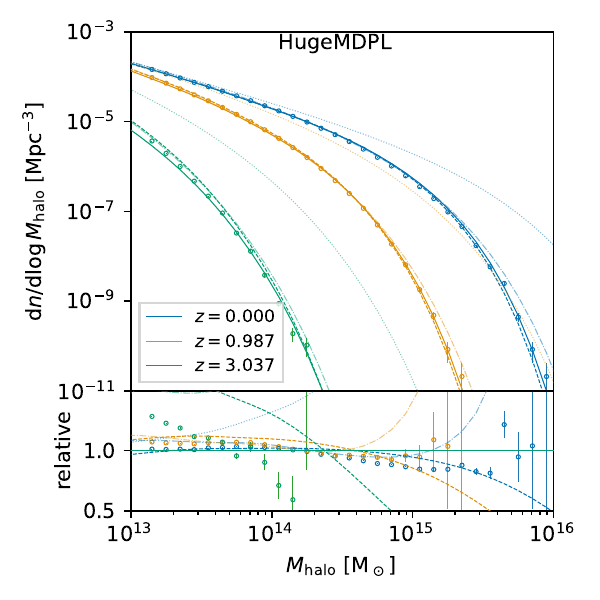}
\end{tabular}
\caption{Halo mass functions from the MDPL2 (left panel) and HugeMDPL (right panel) simulations from $z=0$ to $z\approx 3$. Symbols and line types follow those in Figure~\ref{fig:hmfMDPLBoxes}. The lower panels show the same halo mass functions but normalized to the model expectation for each MDPL volume, to emphasize the relative offsets between N-body data and our model.}
\label{fig:hmfMDPLRedshifts}
\end{figure*}

Figure~\ref{fig:hmfMDPLRedshifts} shows the evolution of the halo mass function from MDPL2 (left panel) and HugeMDPL (right panel) from $z=0$ to $z\approx 3$. Our model accurately matches the evolution across this range of redshifts, except for the lowest mass halos in HugeMDPL at $z\approx 3$, where it slightly underpredicts. The model of \cite{2016MNRAS.456.2486D} agrees well with the MDPL2 data, but noticeably overpredicts the HugeMDPL data. \cite{2016MNRAS.456.2486D} fit to peak heights\footnote{Note that \cite{2016MNRAS.456.2486D} define $\nu = \delta^2_\mathrm{c}(z)/\sigma^2(M)$, while in this work we use $\nu = \delta_\mathrm{c}(z)/\sigma(M)$. We have converted to our definition for the purposes of this comparison.} of around $\nu = 4$--5, which corresponds to a halo mass of around $10^{16}\mathrm{M}_\odot$ in the MDPL cosmology at $z=0$. The range of masses fit here is therefore within the range for which the \cite{2016MNRAS.456.2486D} mass function was fit. Viewed as a difference in halo mass (i.e., looking at the offset between our model and that of \cite{2016MNRAS.456.2486D} along the $x$-axis), the offset is around 0.05~dex (around 12\%). We note that \cite{2016MNRAS.456.2486D} used a different algorithm (a spherical overdensity approach, starting from peaks in the density field identified using an $N^\mathrm{th}$-nearest neighbor approach) to identify halos and characterize their masses, while we use masses estimated by Rockstar, which uses a friends-of-friends approach to find halos ``seeds'' to which particles are assigned, and then uses spherical overdensity definitions to assign halo masses. These differences in algorithm to locate and characterize halos may explain the differences in halo masses between our results and those of \cite{2016MNRAS.456.2486D} at high masses.

The right panel of Figure~\ref{fig:hmfMDPLRedshifts} also illustrates the differences between our fitted halo mass function (solid lines) and the intrinsic halo mass function (dashed lines). The intrinsic halo mass function is lower by up to around 50\%. We find that this shift is due almost entirely to the presence of artificial halos in the fitted halo mass function\footnote{The other numerical effects (detection efficiency, convolution with the uncertainty distribution of halo mass, and the variance on the scale of the simulation volume) cause shifts of only 5--10\% in the halo mass function in this regime}. While our model predicts that artificial halos are extremely rare at these masses, so are real halos. Whether the presence of artificial halos at these extremely high masses/extremely low abundances is correct remains and open question---our model for artificial halos is empirical in nature, but we note that on large scales the power spectrum is approximately $P(k) \propto k$, such that physical perturbations must become sub-Poissonian. Testing our empirical model for artificial halos would require access to the particle distribution in the initial conditions of a very large volume cosmological simulation, and the application of methods such as those developed by \cite{2014MNRAS.439..300L} to identify artificial halos directly. Such an investigation is beyond the scope of this work, but would be a valuable direction for future study. We note that the higher resolution MDPL2 simulation (left panel of Figure~\ref{fig:hmfMDPLRedshifts}) shows a much smaller difference between the fitted and true halo mass function. The particle mass is approximately 50 times smaller in MDPL2 than in HugeMDPL which suppresses the number of artificial halos.

\begin{figure}
\includegraphics[width=3.5in]{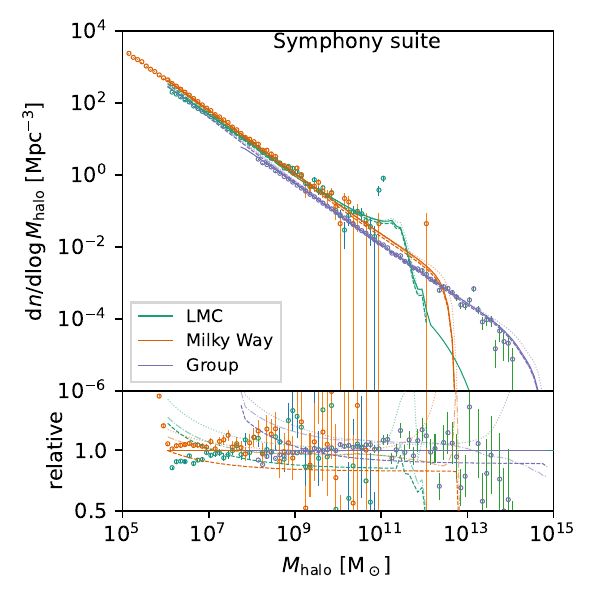}
\caption{Halo mass functions at $z=0$ for the Symphony LMC, Milky Way, and group simulations. Symbols and line types follow those in Figure~\ref{fig:hmfMDPLBoxes}. For the Milky Way halo mass function, we show results from the highest resolution (``X64'') simulation. The lower panel shows the same halo mass functions but normalized to the model expectation for each Symphony sample, to emphasize the relative offsets between N-body data and our model.}
\label{fig:hmfSymphonyScales}
\end{figure}

Figure~\ref{fig:hmfSymphonyScales} shows the halo mass functions obtained from the Symphony LMC, Milky Way, and group simulations, averaged over all available realizations\footnote{The non-smooth features in the LMC halo mass function arise from the fact that we average over all Symphony LMC realizations. In the peak-background split model the halo mass function has a sharp peak just below the scale of the environment. Given the finite number of Symphony LMC realizations over which we average, this results in the non-smooth features (there are 4 LMC realizations which happen to have an environment scale very similar to each other, resulting in the peak just above $10^{11}\mathrm{M}_\odot$, and 1 realization which happens to have a significantly larger environment mass than any other, resulting in the peak closer to $10^{12}\mathrm{M}_\odot$). This same effect is present in the Milky Way and Group halo mass functions but is less prominent because the halo mass function is already more suppressed on the relevant scales, and because their realizations happen to not have the same statistical features as the LMC realizations (and for the Milky Way case we show only the single X64 resolution realization).}. For the Milky Way halo mass function, we show results from the highest resolution (``X64'') simulation. Our model produces a very good match to the N-body halo mass functions, as do those of \cite{2021MNRAS.506..128B} and \cite{2016MNRAS.456.2486D}, except at the lowest masses shown.

\begin{figure}
\includegraphics[width=3.5in]{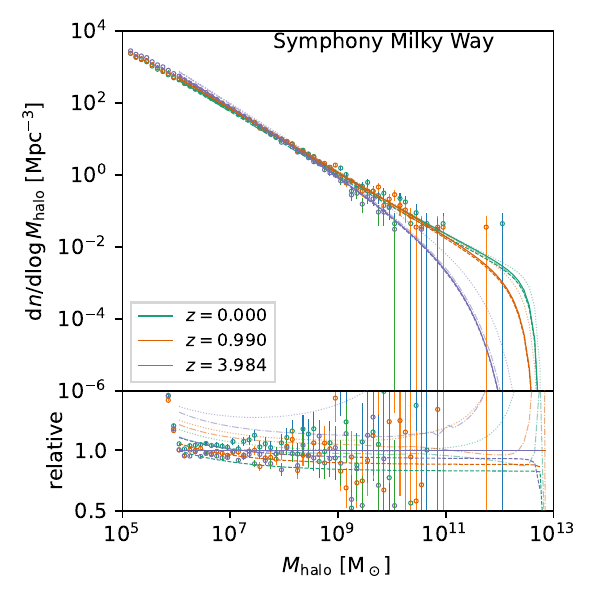}
\caption{Halo mass functions from $z=0$ to $z\approx 4$ for the Symphony Milky Way X64 simulation. Symbols and line types follow those in Figure~\ref{fig:hmfMDPLBoxes}. The lower panel shows the same halo mass functions but normalized to the model expectation for each Symphony sample, to emphasize the relative offsets between N-body data and our model.}
\label{fig:hmfSymphonyRedshifts}
\end{figure}

Figure~\ref{fig:hmfSymphonyRedshifts} shows the halo mass functions obtained from the Symphony Milky Way X64 simulation from $z=0$ to $z\approx 4$. Our model produces an excellent match to the evolution of the mass function at masses greater than around $10^9\mathrm{M}_\odot$, and to the lack of evolution at lower masses.

\subsection{Non-CDM}\label{sec:hmfNonCDM}

\begin{figure*}
\includegraphics[width=7in]{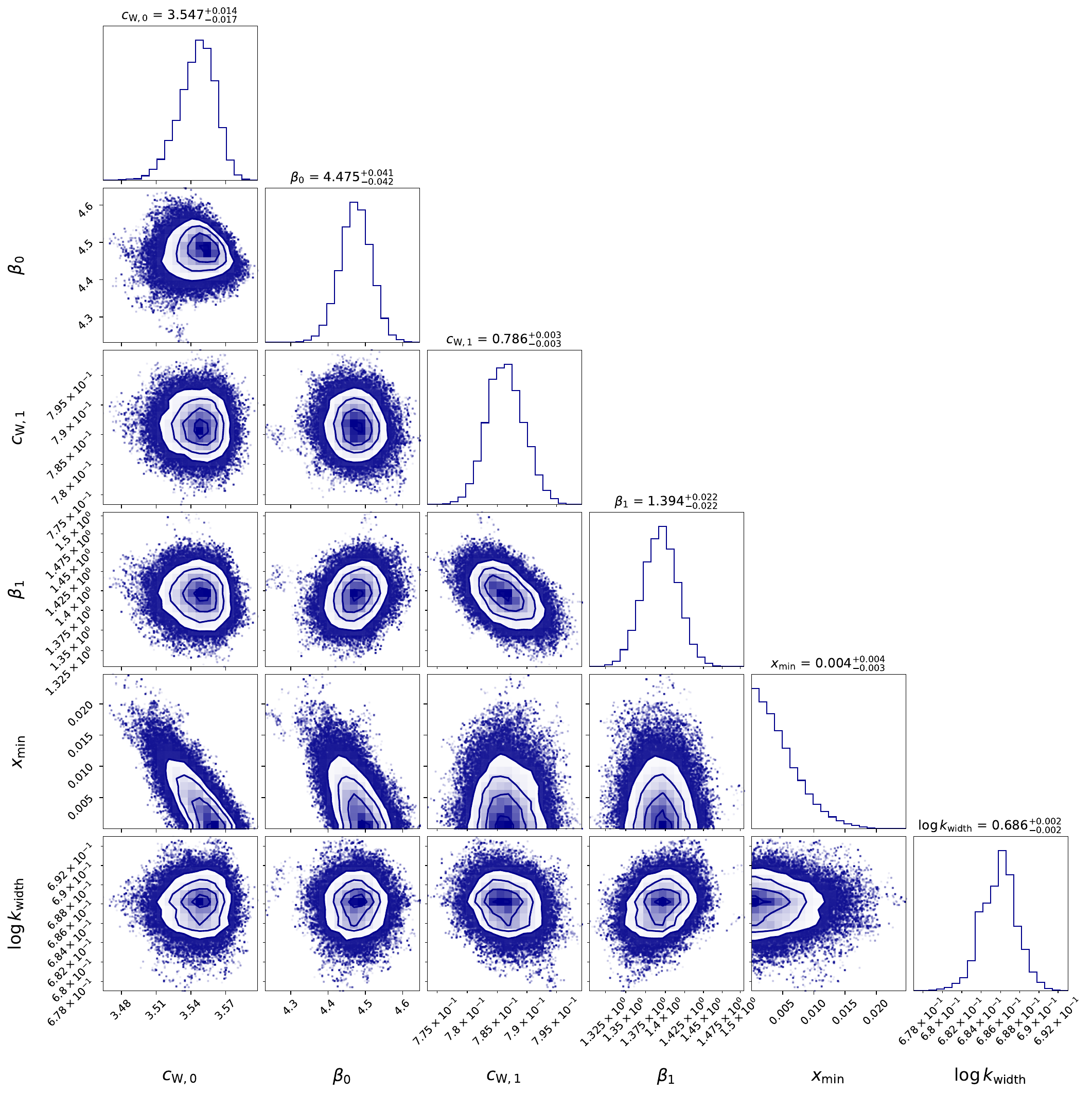} 
\caption{Posterior distributions for the parameters of the window function resulting from our MCMC simulation, following the format of Figure~\ref{fig:hmfCornerPlot}.}
\label{fig:wkCornerPlot}
\end{figure*}

Before exploring halo mass functions of non-CDM models, we examine the constraints on the parameters of our window function model. Recall that our window function builds upon that proposed by \cite{2021MNRAS.506..128B} by allowing the parameters, $(c_\mathrm{w},\beta)$ to depend on the slope of the power spectrum (averaged over some interval of wavenumber). \cite{2021MNRAS.506..128B} found best-fit values of $(c_\mathrm{w},\beta)=(3.78, 3.46)$ to match the halo mass functions of N-body simulations in models with dark acoustic oscillations. At $n = -2.6$ (which, for the CDM models considered in this work is reached at a halo mass of around $10^7\mathrm{M}_\odot$), the reference power spectrum slope in our window function, we find maximum posterior likelihood values of $(c_\mathrm{w},\beta)=(c_\mathrm{w,0},\beta_0)=(3.552^{+0.015}_{-0.018},4.474^{+0.040}_{-0.043})$. At higher masses, $n$ is less negative and our model prefers lower values of $c_\mathrm{W}$ and higher $\beta$. For example, at $n=-2$ (corresponding to halo masses of around $10^{13}\mathrm{M}_\odot$) we obtain $(c_\mathrm{w},\beta)=(3.077,5.444)$. Therefore, our analysis finds that, for less negative power spectrum slopes, the window function must cut off at a slightly smaller wavenumber, and more sharply.

The $x_\mathrm{min}$ parameter, which controls the wavenumber below which our window function becomes constant, is consistent with zero (i.e., the data does not require any range of wavenumber over which the window function is truly constant), but small values up to around $0.01$ are not ruled out. This parameter is also strongly correlated with $c_\mathrm{W,0}$ and $\beta_0$, with higher values of both of those parameters preferring a smaller $x_\mathrm{min}$. The effect of $x_\mathrm{min}$ is to make the derivative of the window function at small wavenumbers precisely zero. Both larger $c_\mathrm{w}$ and $\beta$ similarly drive the derivative of the window function at small wavenumbers toward zero---as such, the presence of this correlation is likely due to the requirement that the derivative of the window function (to which the halo mass function at low masses is proportional) at small wavenumbers be sufficiently small. The smoothing parameter $\log_\mathrm{e} k_\mathrm{width}$ has a median of $0.685$, and is tightly constrained, indicating that smoothing of the power spectrum over around a factor of 2 is preferred for our window function model.

\begin{figure*}
\includegraphics[width=7in]{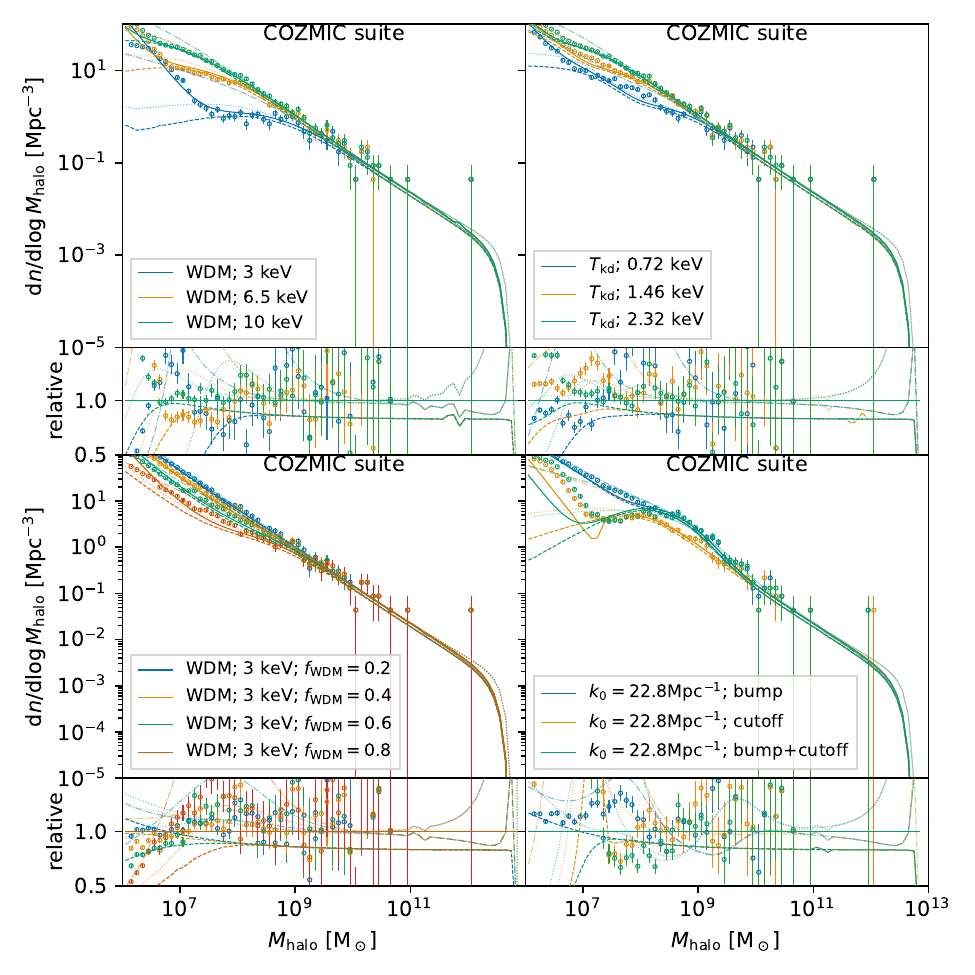} 
\caption{Halo mass functions at $z=0$ for COZMIC WDM (upper left) and $T_\mathrm{kd}$ (upper right) models, plus mixed W/CDM (lower left) and enhanced power spectrum models with bumps and/or cutoffs (lower right) models. Symbols and line types follow those in Figure~\ref{fig:hmfMDPLBoxes}. The lower panels show the same halo mass functions but normalized to the model expectation for each COZMIC sample, to emphasize the relative offsets between N-body data and our model.}
\label{fig:hmfCOZMICWDM}
\end{figure*}

Figure~\ref{fig:hmfCOZMICWDM} shows halo mass functions at $z=0$ for COZMIC WDM (upper left), $T_\mathrm{kd}$ (upper right), mixed W/CDM (lower left), and enhanced power spectrum models with bumps and/or cutoffs (lower right). In each case, a representative subset of all available models is shown. Considering first the WDM models (top left panel), we see clear suppression in the halo mass function at lower masses, as expected due to the suppressed power spectrum in WDM. The suppression is more extreme in the lower mass models, as expected. In all cases, there is a clear upturn in the halo mass function at lower masses, which we associated with artificial halos, arising from particle noise in the simulations. Our models (solid lines) perform very well in matching the halo mass function at masses above where artificial halos dominate. At lower masses, our model, which includes contributions from artificial halos, approximately matches the upturn, but does not capture it in detail. This suggests that our model for the contribution from artificial halos is insufficient. Nevertheless, we can also examine the intrinsic halo mass function (i.e., that with artificial halos and other numerical effects removed), shown by the dashed lines in the figure. This shows a clear turnover and decline of the halo mass function toward low masses. The \cite{2021MNRAS.506..128B} halo mass functions (dotted lines) somewhat overpredict the COZMIC halo mass functions at and below the turnover, while the \cite{2016MNRAS.456.2486D} halo mass function (dot-dashed lines) largely fails to produce a turnover at all. We find that this is a direct consequence of the spherical top-hat window function used for the \cite{2016MNRAS.456.2486D} halo mass function---if we instead retain the ``extended ETHOS'' window function of this work, the \cite{2016MNRAS.456.2486D} halo mass function does show a turnover and is in much better agreement with the N-body data for warm dark matter.

Considering next the $T_\mathrm{kd}$ models (upper right panel) we also see suppression, albeit less extreme, in the COZMIC N-body simulation halo mass functions. Our model (solid lines) is able to capture this suppression moderately well, matching well at masses just above where artificial halos begin to dominate, but somewhat underpredicting at masses slightly larger still, likely corresponding to the ``bump'' in the transfer function for these models (see Figure~\ref{fig:transferFunctions}). The \cite{2021MNRAS.506..128B} halo mass function similarly underpredicts in this mass range for the $T_\mathrm{kd}=0.72$~keV model, but does somewhat better than our model for $T_\mathrm{kd}=1.46$~keV.

For mixed W/CDM models (lower left panels), a similar story emerges. For the CDM-dominated cases (e.g., $f_\mathrm{WDM}=0.2$), our model, and that of \cite{2021MNRAS.506..128B}, perform very well in matching the N-body halo mass function. This is not surprising, of course, and in these cases the transfer function and halo mass function are very close to that for pure CDM, which we have already seen is well-matched by our model. For WDM-dominated cases (e.g., $f_\mathrm{WDM}=0.8$), our model tends to underpredict the halo mass function at low masses, while that of \cite{2021MNRAS.506..128B} performs better. Our model predicts that at low masses, below the scale where artificial halos dominate, the intrinsic halo mass function should continue to rise, but with an amplitude reduced relative to a pure CDM model.

Turning next to enhanced power models, with a bump and/or cutoff (lower right panel), we see that our model is able to match the overall differences between the three cases very well, matching the location of the peak in the halo mass function due to the presence of a bump in the transfer function (both with and without a cutoff). Here, the \cite{2021MNRAS.506..128B} model performs less well in matching the decline in the halo mass function at masses below this peak, producing too many halos in this regime, and also underpredicts the amplitude of the peak in the bump+cutoff case. Both our model and that of \cite{2021MNRAS.506..128B} slightly underpredict the halo mass function in the bump model at scales around the bump. Interestingly, the \cite{2016MNRAS.456.2486D} performs better in this region.

\begin{figure}
\includegraphics[height=8.5in]{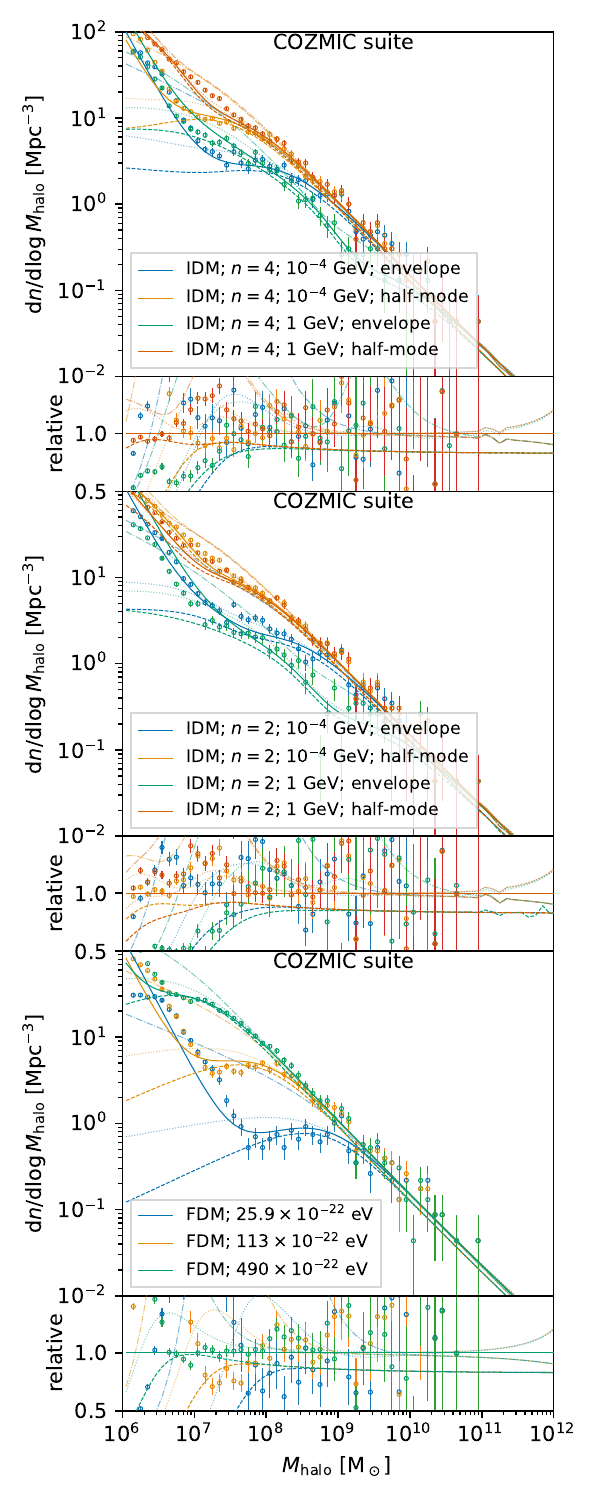} 
\caption{Halo mass functions at $z=0$ for COZMIC IDM $n=4$ (upper), IDM $n=2$ (middle), and FDM (lower) models. Symbols and line types follow those in Figure~\ref{fig:hmfMDPLBoxes}. The lower panel shows the same halo mass functions but normalized to the model expectation for each Symphony sample, to emphasize the relative offsets between N-body data and our model.}
\label{fig:hmfCOZMICXDM}
\end{figure}

Figure~\ref{fig:hmfCOZMICXDM} shows $z=0$ halo mass functions from COZMIC IDM and FDM models. In each case, a representative subset of available simulations is shown. For IDM models (upper and middle panels), we find good overall agreement for our model, but once again see that the \cite{2021MNRAS.506..128B} model performs slightly better in some cases (e.g., in the $n=4$, 1~Gev, half-mode model in the upper panel). For FDM models (lower panel), our model performs well in matching the COZMIC halo mass function for all particle masses. The \cite{2021MNRAS.506..128B} significantly overpredicts the halo mass function at masses below the peak (and before artificial halos begin to dominate), while the \cite{2016MNRAS.456.2486D} largely fails to capture the cutoff in the halo mass function for these models.

Overall, we find that our model performs reasonably well across all of the non-CDM models considered---largely reproducing the shapes of cutoffs, locations of peaks, and other details of the halo mass function. The \cite{2021MNRAS.506..128B} model performs somewhat better in models with strong oscillations (notably the $T_\mathrm{kd}$ and IDM models)---this is not surprising as it was constructed for and calibrated to models with strong dark acoustic oscillations. However, it performs noticeably worse than our model for cases with a strong cut-off in the halo mass function (WDM, FDM, and the cutoff models). The \cite{2016MNRAS.456.2486D} model performs quite poorly in all cases considered in this section, typically predicting too many halos at low masses and failing to capture the detailed shape of the halo mass function. This failure arises because of the use of the spherical top-hat window function, which produces too much coupling between large-scale modes and the low mass halo regime, as can be seen in Figure~\ref{fig:windowFunctions}.

\subsection{Environmental trends}\label{sec:environmentalTrends}
\begin{figure}
\includegraphics[height=7.5in]{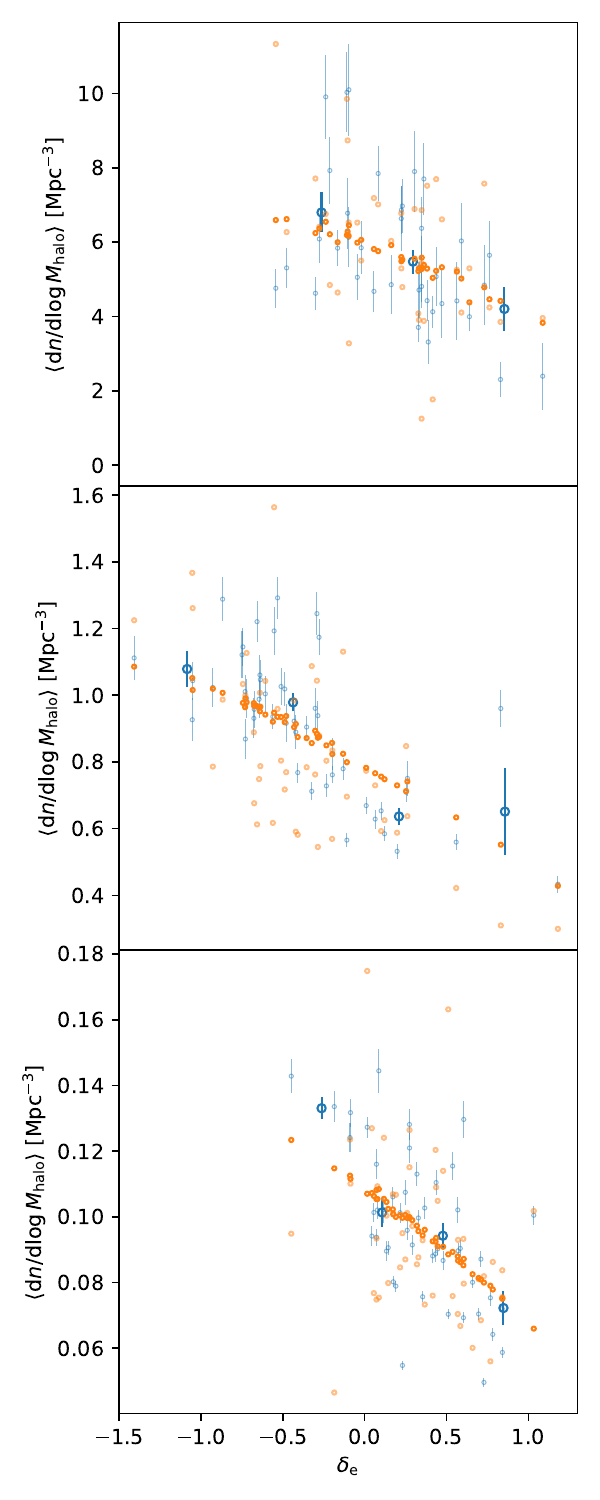} 
\caption{Environmental dependence of the $z=0$ halo mass function. Panels show results for Symphony LMC, Milky Way, and Group simulations from top to bottom. In each panel, points indicate the halo mass function averaged over the mass range $5 \times 10^7$--$5 \times 10^8 \mathrm{M}_\odot$, $4 \times 10^8$--$4 \times 10^9 \mathrm{M}_\odot$, and $3 \times 10^9$--$3 \times 10^{10} \mathrm{M}_\odot$ for LMC, Milky Way, and Group simulations respectively, corresponding to approximately 1,000--10,000 particles in each case, as a function of environmental overdensity, $\delta_\mathrm{e}$ measured within the convex hull from which our halo mass functions are determined (see \S\ref{sec:zoominAnalysis}) and linearly extrapolated to $z=0$. Faint blue points show measurements from the N-body simulations, with error bars indicating Poisson uncertainties, while the bold blue points indicate binned means of these results with error bars showing the uncertainties on the mean. Orange points show results from our model, with faint orange points showing model results with added scatter to mimic the expected scatter in the N-body data.}
\label{fig:environmentalDependence}
\end{figure}

Figure~\ref{fig:environmentalDependence} shows the environmental dependence of the halo mass function. Panels show results for Symphony LMC, Milky Way, and Group simulations from top to bottom. In each panel, points indicate the halo mass function averaged over the mass range $5 \times 10^7$--$5 \times 10^8 \mathrm{M}_\odot$, $4 \times 10^8$--$4 \times 10^9 \mathrm{M}_\odot$, and $3 \times 10^9$--$3 \times 10^{10} \mathrm{M}_\odot$ for LMC, Milky Way, and Group simulations respectively, corresponding to approximately 1,000--10,000 particles in each case, as a function of environmental overdensity, $\delta_\mathrm{e}$ measured within the convex hull from which our halo mass functions are determined (see \S\ref{sec:zoominAnalysis}) and linearly extrapolated to $z=0$. Faint blue points show measurements from the N-body simulations, with error bars indicating Poisson uncertainties, while bold blue points indicate binned means of these results with error bars indicating the uncertainties on those means. Orange points show results from our model, while faint orange points show model results with added scatter to mimic the N-body data as described below.

While there is substantial scatter in the N-body results at fixed $\delta_\mathrm{e}$, there is a clear trend: the halo mass function amplitude is lower in more overdense regions, across all Symphony simulations. Such a trend is expected---denser regions of the universe evolve more rapidly, and so correspond to the mean halo mass function (i.e., the halo mass function averaged over all environments) at a later time. Since, as time progresses, mass accumulates into ever higher mass halos, the amplitude of the halo mass function at low masses must decrease to preserve the total amount of mass per unit Lagrangian volume. Our model, in which the environmental dependence is derived from the peak-background split approach, reproduces this trend very well, but shows substantially less scatter at fixed overdensity. The small scatter visible in our model arises because, at fixed $\delta_\mathrm{e}$, there is some variation in the environment mass, $M_\mathrm{e}$, which we compute as the mass contained within the convex hull within which we compute the halo mass function. Bold blue points indicate binned means of the N-body results (with error bars indicating the uncertainty on those means) and are in quite good agreement with the expectations from our model. The faint orange points show our model results when we sample them from an overdispersed Poisson distribution\footnote{Following \cite{2010MNRAS.406..896B}, this is modeled using a negative binomial distribution with intrinsic scatter of 5\%---less than the 18\% that \cite{2010MNRAS.406..896B} found for subhalo populations.}. This results in a scatter broadly consistent with that seen in the N-body data, suggesting that this scatter arises purely from random sampling of the halo mass function in the N-body realizations.

As discussed in \S\ref{sec:zoominAnalysis}, the use of a convex hull to estimate the environmental overdensity of our zoom-in simulations could lead to some bias and/or scatter in the estimated linear overdensities, in turn leading to bias and/or scatter in the predicted halo mass functions. The results shown in Figure~\ref{fig:environmentalDependence} suggest that any such effects are mild for the samples considered in this work. In particular, the mean trends with environmental overdensity (bold points in Figure~\ref{fig:environmentalDependence}) agree well with those predicted by our model, and the scatter in the amplitude of the halo mass function at a given environmental overdensity appears consistent with what would be expected from simple Poisson statistics. As noted in \S\ref{sec:zoominAnalysis}, alternatives to a convex hull-based approach exist, but are more complicated. A future study could explore such alternatives and ask which estimator of the environmental volume provides the best (i.e., that which minimizes both bias and scatter) predictor for the environmental dependence of the halo mass function.

\subsection{Artificial Halos}

In the previous subsections, we saw that our model for artificial halos is able to capture the overall trends seen for models with different transfer functions, but that it does not accurately capture all of the details, in particular, often underestimating the amplitude of the artificial halo contribution. Here, we explore further, considering how this model performs across resolutions and redshifts. 

\begin{figure*}
\includegraphics[width=7in]{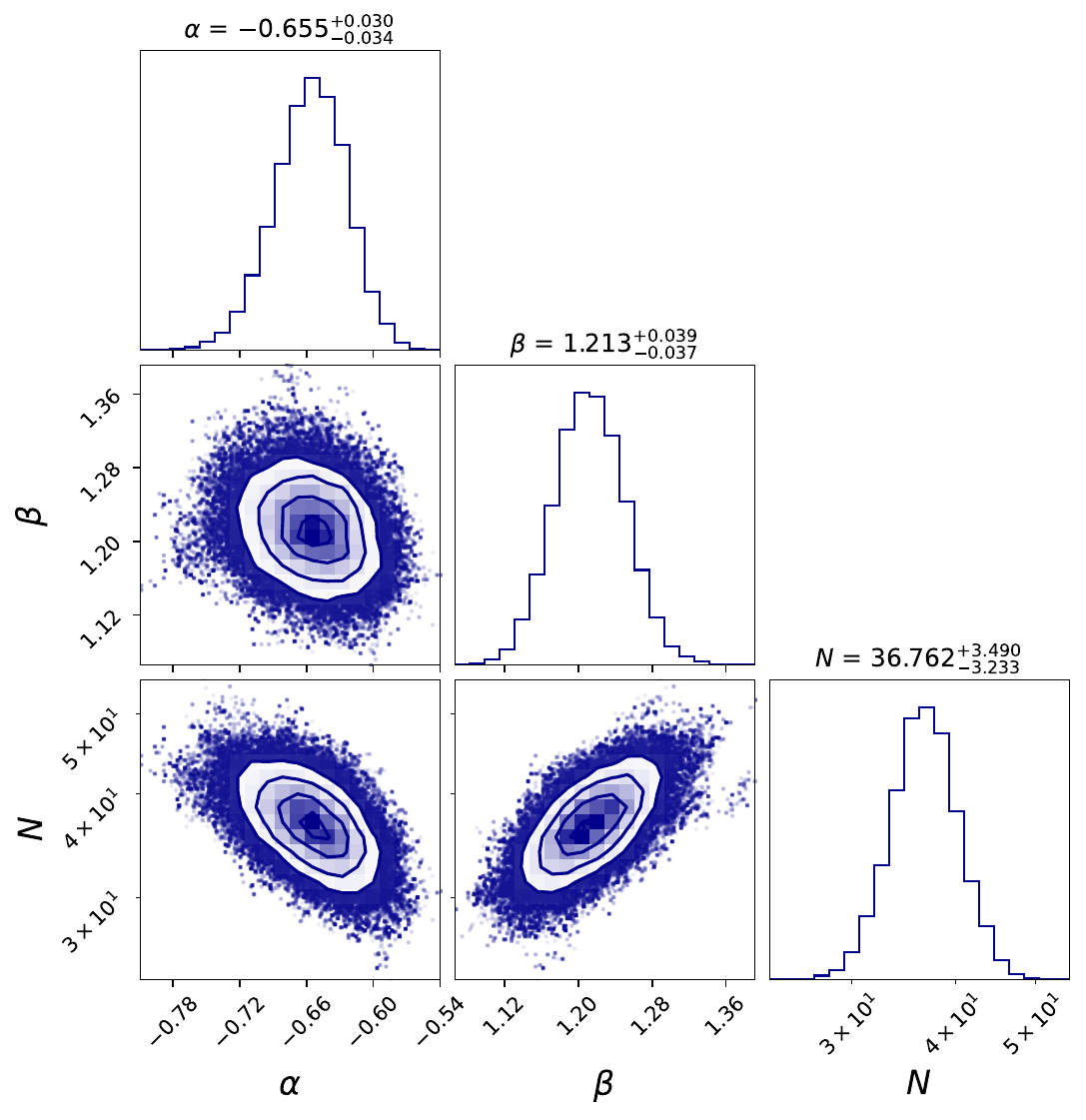} 
\caption{Posterior distributions for the parameters of the artificial halos model resulting from our MCMC simulation, following the format of Figure~\ref{fig:hmfCornerPlot}.}
\label{fig:pseudoCornerPlot}
\end{figure*}

Figure~\ref{fig:pseudoCornerPlot} shows the posterior distribution for the parameters of our artificial halo model.  These are all well-constrained, with only weak correlations between parameters. The normalization parameter, $N$, has a median of approximately 37---this is the number of particles at which the variance in the artificial halo model reaches 1 at $z=0$. The exponent of the variance function is tightly constrained to be $\alpha=-0.653^{+0.029}_{-0.031}$, and the exponent on the growth factor is constrained to by $\beta=1.216^{+0.039}_{-0.038}$, significantly different from 1, indicating the variance associated with artificial halos does not follow linear theory growth.

\begin{figure*}
\includegraphics[width=7in]{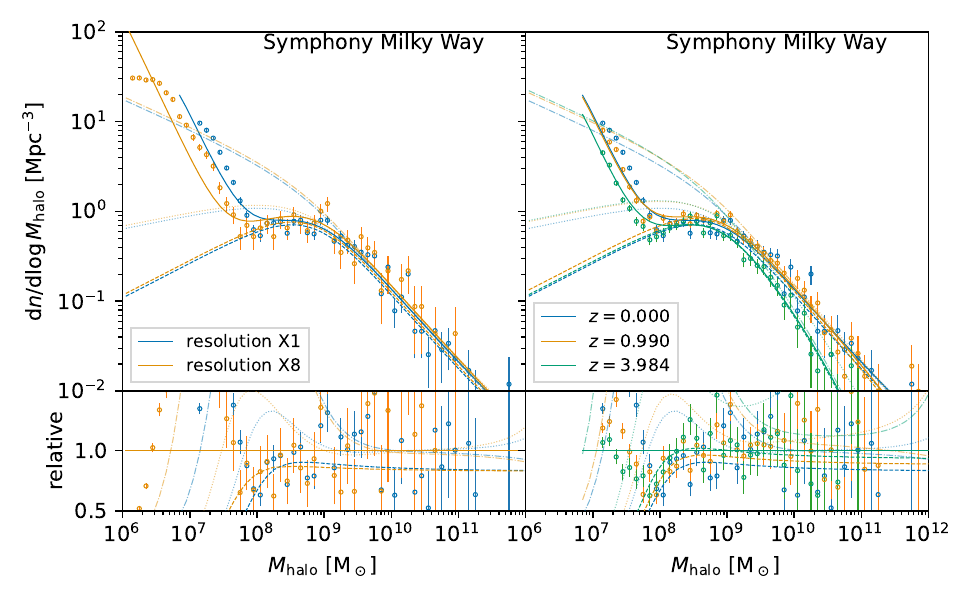} 
\caption{Halo mass functions at COZMIC FDM $25.9 \times 10^{-22}$~eV models. The left panel shows results at resolutions X1 and X8 at $z=0$, while the right panel shows results for resolution X1 at $z=0$, $z\approx 1$, and $z\approx 4$. Symbols and line types follow those in Figure~\ref{fig:hmfMDPLBoxes}. The lower panel shows the same halo mass functions but normalized to the model expectation for each Symphony sample, to emphasize the relative offsets between N-body data and our model.}
\label{fig:hmfArtificialHalos}
\end{figure*}

Figure~\ref{fig:hmfArtificialHalos} explores how the artificial halos model performs as a function of resolution (left panel) and redshift (right panel). We use the COZMIC FDM $25.9\times10^{-22}$~eV model in this figure as the artificial halo contribution is very clear in this model. Considering first the effect of resolution (left panel) it can be seen that, while our model for artificial halos somewhat underpredicts the artificial halo mass function, it scales with resolution in a way that closely matches that seen in the N-body simulation measurements. In the right panel, we explore the behavior as a function of redshift. In this case, our artificial halos model can be seen to be in quite good agreement with the N-body measurements at $z\approx 4$ and $z\approx 1$. However, it shows little evolution from $\approx 1$ to $z=0$ while the N-body data show a modest increase in the artificial halo mass function between these same redshifts, resulting in our model underpredicting the number of artificial halos at $z=0$. This suggests that further flexibility in the artificial halo mass function model may be needed to fully capture the behavior.

\cite{2025ApJ...986..127N} demonstrated that their results for \emph{sub}halo mass functions were converged for halo masses corresponding to at least 300 particles in their fiducial resolution simulations (what we refer to as the ``X1'' resolution here), and, in particular, that artificial halos did not significantly contaminate their results above this threshold. Considering the left panel of Figure~\ref{fig:hmfArtificialHalos}, and specifically the COZMIC FDM $25.9\times10^{-22}$~eV model at resolution X1, it can be seen that artificial halos are only just beginning to become important at this 300 particle threshold (which corresponds to a mass of $M_\mathrm{halo}=1.2\times 10^8\mathrm{M}_\odot$). Since subhalos have typically lost a large fraction of their mass to tidal forces, most of the 300 particle subhalos considered by \cite{2025ApJ...986..127N} would have begun as much more massive isolated halos, where the contribution of artificial halos is negligible. As such, our results are consistent with the findings of \cite{2025ApJ...986..127N} in this regard.

\subsection{Universal form}

\begin{figure}
\includegraphics[width=3.5in]{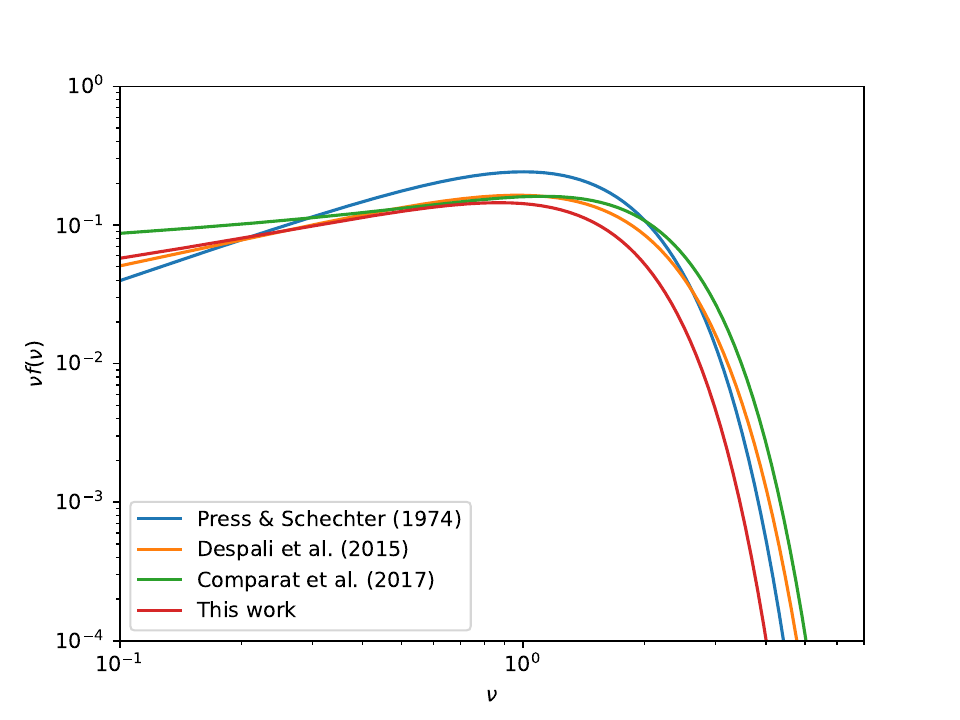} 
\caption{Solid lines show the global halo mass function, $\nu f(\nu)$, which, as shown by \citet{2016MNRAS.456.2486D} for example, is largely independent of cosmological parameters. In addition to the result from this work, we show, for comparison, the fits from \citet[][blue]{1974ApJ...187..425P}, \citet[][green]{2017MNRAS.469.4157C} and \citet[][orange]{2016MNRAS.456.2486D}.}
\label{fig:globalMassFunction}
\end{figure}

As discussed in \S\ref{sec:likelihood}, we perform out fits in the $M$--$n(M)$ space. Many previous studies \citep[e.g.][]{2016MNRAS.456.2486D} have performed fits in the $\nu$--$f(\nu)$ space where
\begin{equation}
\frac{\mathrm{d}n}{\mathrm{d}M} = \frac{\bar{\rho}}{M} \nu f(\nu) \frac{\mathrm{d}\log \nu}{\mathrm{d}M},
\end{equation}
finding that, when expressed in this form, the mass function $\nu f(\nu)$ is approximately universal (i.e. independent of cosmological parameters). Our intrinsic halo mass function is easily expressed in these coordinates, and Figure~\ref{fig:globalMassFunction} shows our best fit, together with a few representative fitting functions from the literature. An important point to keep in mind here is that these earlier studies made use of a top-hat window function, while in this work we adopt our ``extended ETHOS'' window function. This changes the mapping $M \rightarrow \sigma(M) \rightarrow \nu$. As such the discrepancies between our mass function and others in this plane is exaggerated compared to those in the $M$--$n(M)$ space (as seen in previous figure in this work).

\subsection{Posterior predictive checks}

To assess how well our model performs in explaining the N-body data, we perform posterior predictive checks. Specifically, for each individual N-body simulation, we use samples drawn from our converged MCMC chains to generate model expectations for the halo mass function. For each sampled mass function, we forward model a realization of a halo population by computing the mean number of halos expected in each bin of the mass function in the simulation volume, and then drawing an actual number of halos from a Poisson distribution with that mean. Using this set of forward modeled halo mass functions realizations, we can:
\begin{enumerate}
\item evaluate, in each bin, the median and other percentiles of interest;
\item compute a metric quantifying the deviation of each realization from the median, $\mathcal{H}$.
\end{enumerate}
We choose to evaluate a metric, $\mathcal{H}$, that is simply the squared difference between the realized halo mass function and the median, summed over each bin used in our likelihood analysis. We then evaluate $\mathcal{H}_\mathrm{t}$, the squared difference between the measured, N-body halo mass function and the median model expectation, summed over each bin used in our likelihood analysis. We can then evaluate the fraction of the halo mass functions' realizations which deviate from the median to a greater degree than does the N-body data, i.e., those for which $\mathcal{H} > \mathcal{H}_\mathrm{t}$. We interpret this fraction as the probability that a random realization of our model deviates from the median to a greater degree than the actual, N-body realization, $p(\mathcal{H} > \mathcal{H}_\mathrm{t})$. If $p(\mathcal{H} > \mathcal{H}_\mathrm{t})$ is very small, it is unlikely that the N-body halo mass function could have resulted from random draws from our model, suggesting that our model would \emph{not} be a good description of the N-body data. Conversely, if $p(\mathcal{H} > \mathcal{H}_\mathrm{t})$ is not very small, we can conclude that the N-body data is plausibly drawn from our model distribution, implying that our model \emph{is} a good description of the N-body data.

\begin{figure*}
\begin{tabular}{cc}
\includegraphics[width=3.5in]{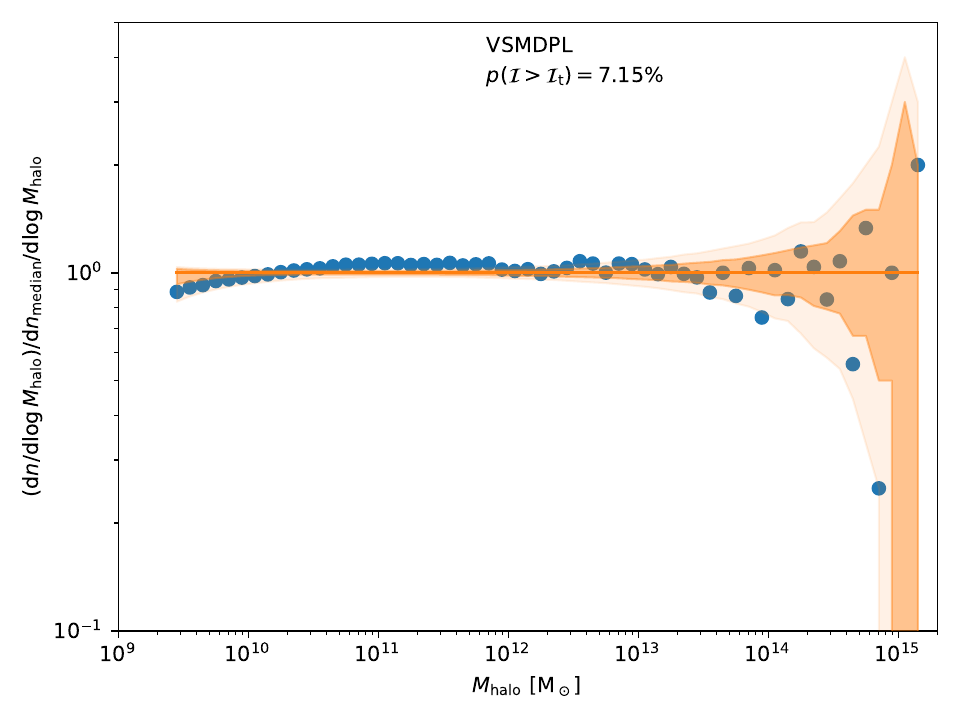} & \includegraphics[width=3.5in]{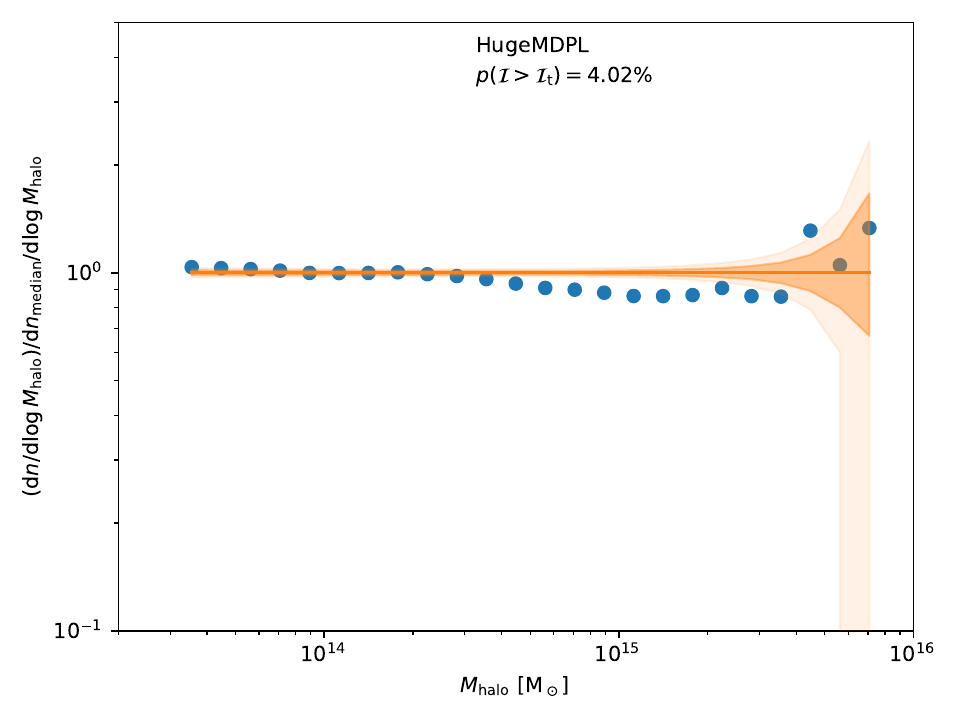}
\end{tabular}
\caption{Halo mass functions for the VSMDPL (left) and HugeMDPL (right) simulations, normalized to the median expectation from our model. Blue points show measured halo mass functions from the N-body simulations. The orange line shows the median ``measured'' halo mass function from our converged MCMC samples, with shaded bands showing the 15.9--84.1 (darker) and 2.3--97.7 (lighter) interquantile ranges from the same MCMC samples after forward modeling the number of halos in each sample assuming Poisson statistics. The model expectation is shown only over the range for which the model is fit to the N-body data. The probability that a randomly-generated halo mass function sample deviates from the median by more than does the measured N-body halo mass function, $p(\mathcal{H} > \mathcal{H}_\mathrm{t})$, is shown in the top right of each panel.}
\label{fig:ppdMDPL}
\end{figure*}

Figure~\ref{fig:ppdMDPL} shows halo mass functions for the VSMDPL (left) and HugeMDPL (right) simulations, normalized to the median expectation from our model. Blue points show measured halo mass functions from the N-body simulations. The orange line shows the median ``measured'' halo mass function from our converged MCMC samples, with shaded bands showing the 15.9--84.1 (darker) and 2.3--97.7 (lighter) interquantile ranges from the same MCMC samples after forward modeling the number of halos in each sample assuming Poisson statistics. The model expectation is shown only over the range for which the model is fit to the N-body data. The probability that a randomly-generated halo mass function sample deviates from the median by more than does the measured N-body halo mass function, $p(\mathcal{H} > \mathcal{H}_\mathrm{t})$, is shown in the top right of each panel.

In the case of these two MDPL simulations, we find $p(\mathcal{H} > \mathcal{H}_\mathrm{t})$ values of around 4--7\%. Thus, our model is marginally (but not strongly) consistent with the N-body data in these cases. Systematic deviations, at the level of 10--15\%, from the model realizations can be seen for both VSMDPL (for which the model overpredicts at the lowest masses, and underpredicts at intermediate masses) and HugeMDPL (which the model overpredicts at masses of order $10^{15}\mathrm{M}_\odot$) simulations.

\begin{figure}
\includegraphics[width=3.5in]{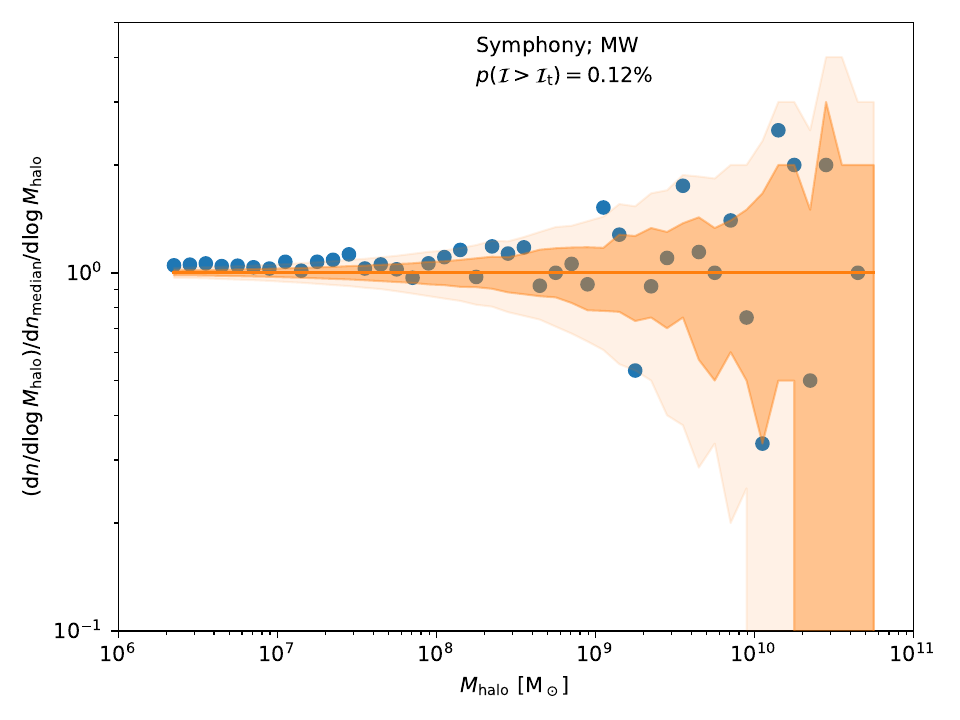}
\caption{Halo mass functions for the Symphony CDM X64 simulation. Blue points show measured halo mass functions from the N-body simulation, normalized to the median expectation from our model. The orange line shows the median ``measured'' halo mass function from our converged MCMC samples, with shaded bands showing the 15.9--84.1 (darker) and 2.3--97.7 (lighter) interquantile ranges from the same MCMC samples after forward modeling the number of halos in each sample assuming Poisson statistics. The model expectation is shown only over the range for which the model is fit to the N-body data. The probability that a randomly-generated halo mass function sample deviates from the median by more than does the measured N-body halo mass function, $p(\mathcal{H} > \mathcal{H}_\mathrm{t})$, is shown in the top right.}
\label{fig:ppdSymphony}
\end{figure}

\begin{figure}
\includegraphics[width=3.5in]{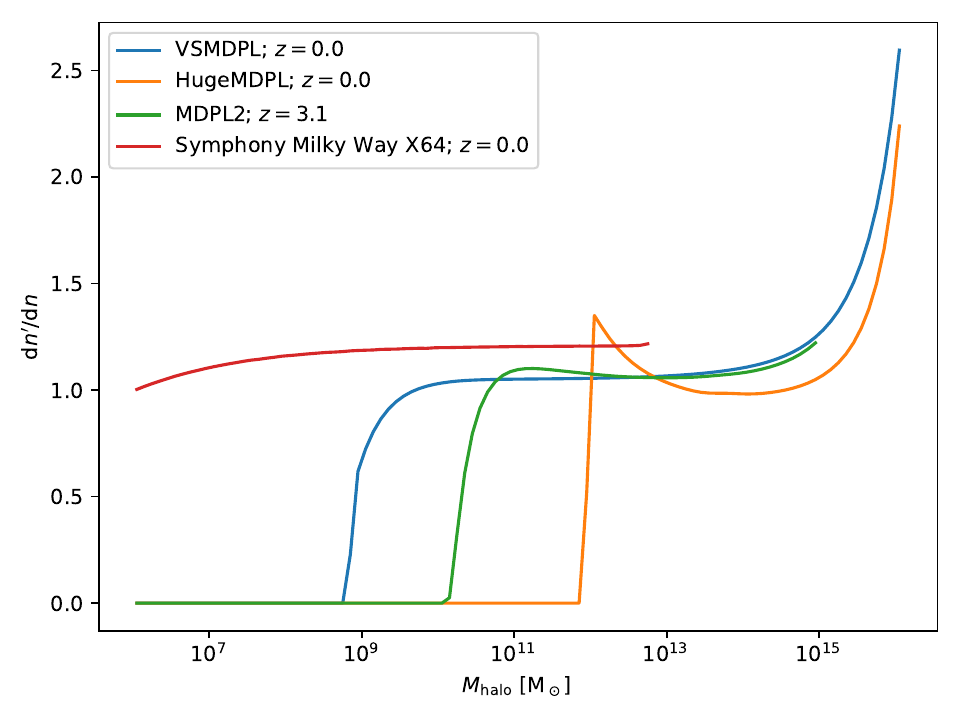}
\caption{The ratio of the fitted, $\mathrm{d}n^\prime/\mathrm{d}\log M_\mathrm{halo}$, to the intrinsic, $\mathrm{d}n/\mathrm{d}\log M_\mathrm{halo}$, halo mass function is shown for a representative subset of CDM simulations, spanning the range of masses and redshifts considered in this work.}
\label{fig:systematicsRatio}
\end{figure}

Figure~\ref{fig:ppdSymphony} shows the same analysis for the Symphony Milky Way X64 simulation. Overall, the model provides a reasonable fit to the data, with typical errors at the $\approx 10\%$ level at low halo masses. However, in detail, we find that $p(\mathcal{H} > \mathcal{H}_\mathrm{t})=0.12\%$, suggesting that our model is formally inconsistent with the N-body data. This is driven by the behavior at masses below around $10^7\mathrm{M}_\odot$, where the model underpredicts the number of halos found in the N-body simulation. Figure~\ref{fig:systematicsRatio} shows the ratio of the fitted, $\mathrm{d}n^\prime/\mathrm{d}\log M_\mathrm{halo}$, to the intrinsic, $\mathrm{d}n/\mathrm{d}\log M_\mathrm{halo}$, halo mass function for a representative subset of CDM simulations, spanning the range of masses and redshifts considered in this work. Notably, the effects of the detection efficiency (\S\ref{sec:detectionEfficiency}) can be clearly seen in all models, causing this ratio to decrease toward low halo masses (or, more precisely, toward low numbers of particles per halo). For the Symphony Milky Way CDM X64 simulation, this detection efficiency becomes important at the 10--20\% level below around $10^7\mathrm{M}_\odot$. It is therefore plausible that this effect might be driving the deviation seen in this simulation at these scales; it is of about the correct magnitude to do so, and this warrants future study, perhaps comparing different halo finders to quantify any differences in their detection efficiencies. Similarly, the possibility of bias or scatter in the estimate of the linear overdensity of the environment arising from our use of a convex hull (see \S\ref{sec:zoominAnalysis}) could also potentially affect the agreement here. Of course, we have only a single Milky Way X64 resolution simulation, so it is not warranted to draw strong conclusions from this. In the HugeMDPL simulation, the sharp cutoff at around $10^{12}\mathrm{M}_\odot$ arises from the very weak mass dependence of the detection efficiency model for this simulation ($\alpha=-0.008$) combined with the 10-particle minimum to which we evaluate the halo mass functions. This same weak mass dependence in the detection efficiency also results in the HugeMDPL model curve lying below 1 at masses of around $10^{13}$--$10^{15}\mathrm{M}_\odot$. This likely reflects a limitation in our model---either in the functional form of the detection efficiency, or some missed systematic effect in the N-body data that the MCMC then uses the freedoms of the detection efficiency model to attempt to fit.

\begin{figure*}
\begin{tabular}{cc}
\includegraphics[width=3.5in]{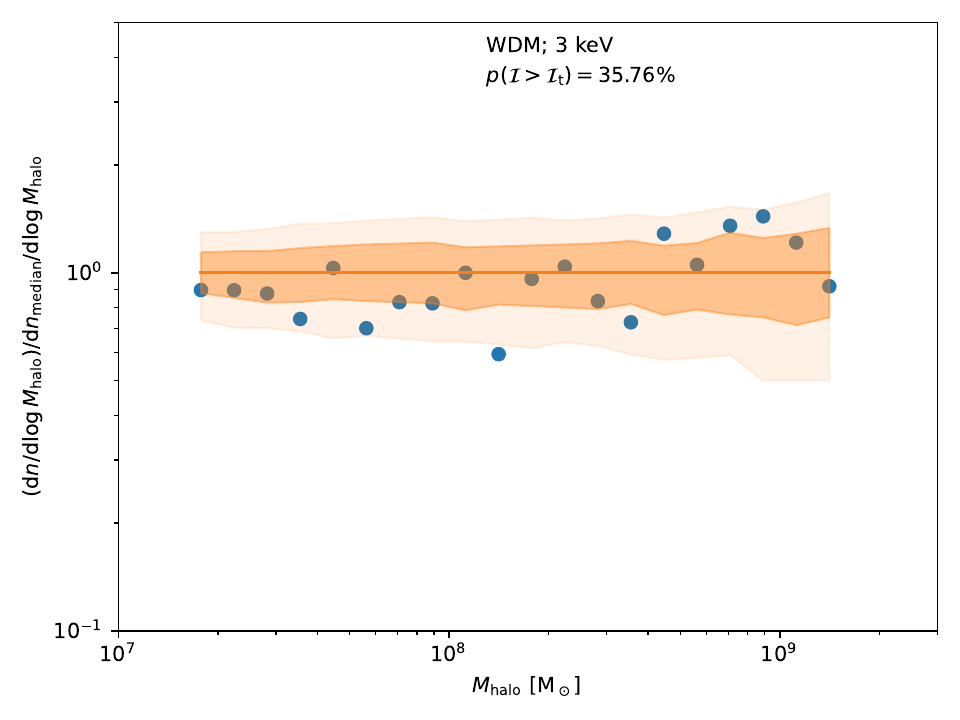} & \includegraphics[width=3.5in]{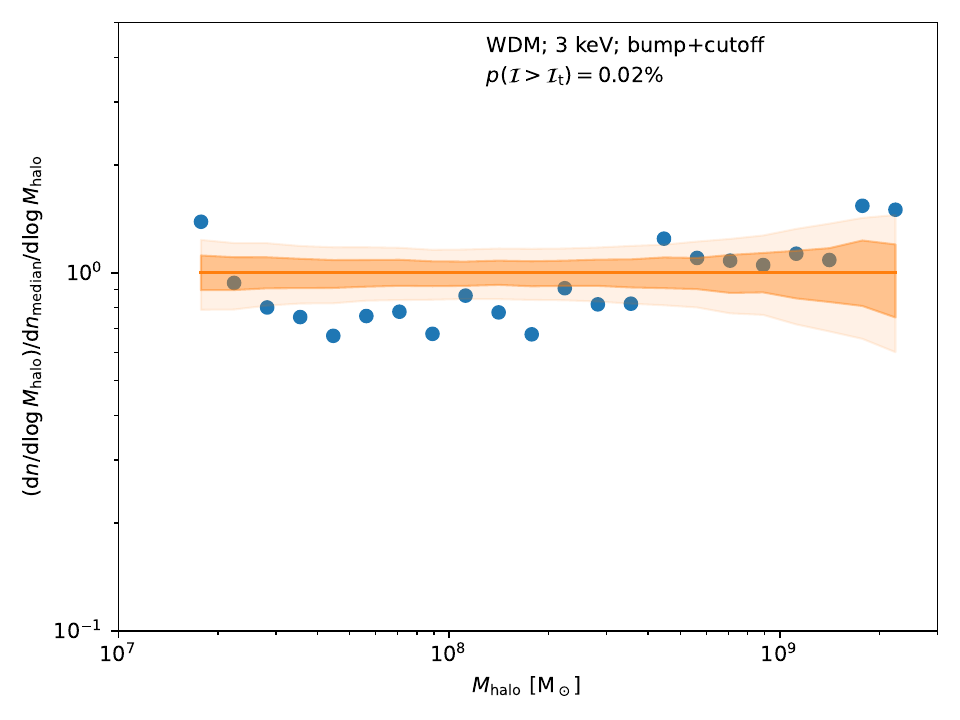} \\
\includegraphics[width=3.5in]{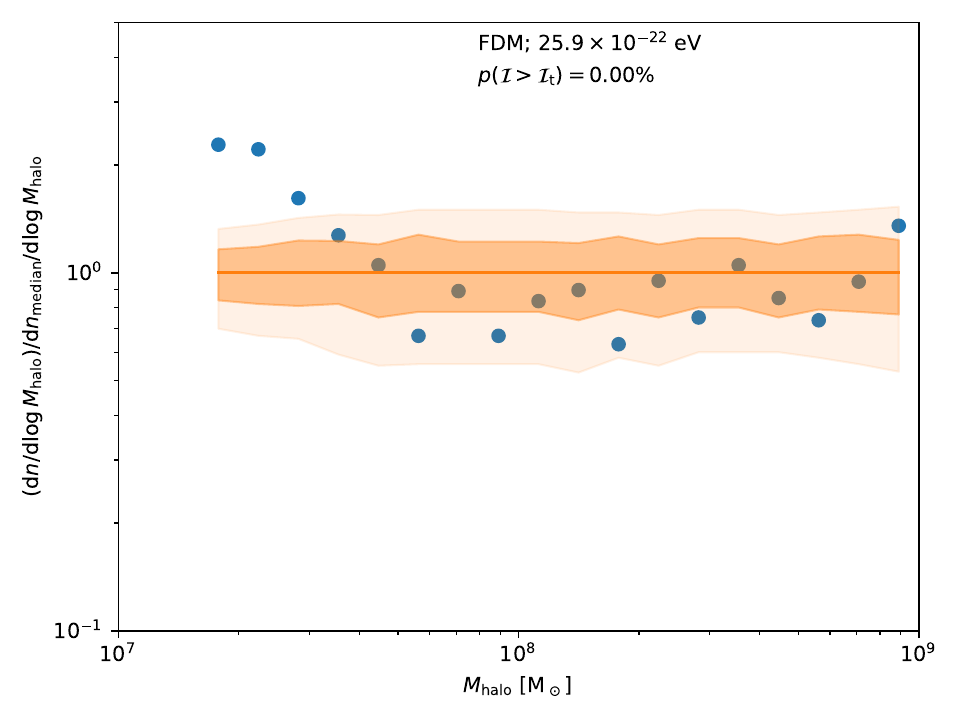} & \includegraphics[width=3.5in]{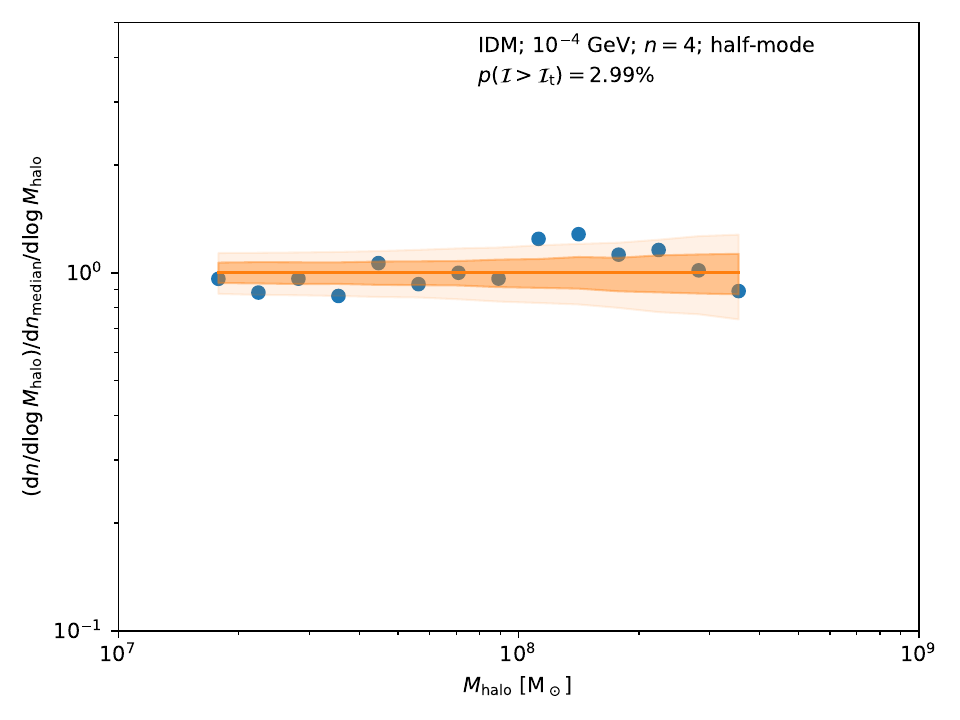}
\end{tabular}
\caption{Halo mass functions for representative COZMIC simulations, normalized to the median expectation from our model: 3~keV WDM (top left), 3~keV WDM plus a bump (top right), $25.9 \times 10^{-22}$~eV FDM (lower left), and $10^{-4}$~GeV, $n=2$, half-mode IDM (lower right). Blue points show measured halo mass functions from the N-body simulations. The orange line shows the median ``measured'' halo mass function from our converged MCMC samples, with shaded bands showing the 15.9--84.1 (darker) and 2.3--97.7 (lighter) interquantile ranges from the same MCMC samples after forward modeling the number of halos in each sample assuming Poisson statistics. The model expectation is shown only over the range for which the model is fit to the N-body data. The probability that a randomly-generated halo mass function sample deviates from the median by more than does the measured N-body halo mass function, $p(\mathcal{H} > \mathcal{H}_\mathrm{t})$, is shown in the top right of each panel.}
\label{fig:ppdCOZMIC}
\end{figure*}

Finally, Figure~\ref{fig:ppdCOZMIC} shows the same analysis for a few representative cases from the COZMIC suite: 3~keV WDM (top left), 3~keV WDM plus a bump (top right), $25.9 \times 10^{-22}$~eV FDM (lower left), and $10^{-4}$~GeV, $n=2$, half-mode IDM (lower right). In all cases, we again find that the model provides a reasonable fit to the data over most halo masses of interest, with a typical precision of $\approx 10\%$, consistent with our findings averaged over all beyond-CDM models presented previously. When applying the same statistical tests as in our comparisons to CDM simulations above, our model is formally consistent with the WDM simulation ($p(\mathcal{H} > \mathcal{H}_\mathrm{t})\approx 36\%$), and marginally consistent with the IDM simulation ($p(\mathcal{H} > \mathcal{H}_\mathrm{t})\approx 3\%$). In the WDM+bump and FDM cases shown, our model is not consistent with the N-body data ($p(\mathcal{H} > \mathcal{H}_\mathrm{t})=0.02\%$ and $<0.01\%$ respectively). In these cases, clear deviations of the N-body data from the expected model distribution can be seen---in the case of the WDM+bump simulation, our model overpredicts the number of halos around $10^8\mathrm{M}_\odot$ by around 25\%, while for the FDM simulation the model fails to capture the upturn in the mass function at the lowest masses (where artificial halos begin to contribute significantly).

\section{Discussion}\label{sec:discussion}

The halo mass function is a fundamental statistic quantifying the non-linear structure in the universe. It is used extensively in astrophysical and cosmological analyses. As such, having accurate models of the halo mass function is highly important. While high-resolution N-body simulations provide the ``gold standard'' for quantifying non-linear structure, they remain computationally expensive. This has motivated the development of simple fitting functions that can be calibrated to N-body simulation measures of the halo mass function to allow for rapid evaluation of the halo mass function. In this work, we have sought such a fitting function that works across halo mass, redshift, environment, and dark matter physics as encoded in the linear matter power spectrum. To do this, we make use of a suite of cosmological volume simulations (primarily to constrain the halo mass function at large masses) and a suite of cosmological zoom-in simulations, which constrain the mass function to very small halo masses, and across a variety of power spectrum shapes motivated by dark matter microphysics models. To our knowledge, this is the first time that individual zoom-in simulations have been used in this way to measure the halo mass function\footnote{The work of \cite{2020Natur.585...39W}, who measured halo mass functions from a set of nested zoom-in simulations of an initial 20~Mpc simulation volume, achieving extremely high resolution, is similar, but differs from the types of zoom-in simulations used here which specifically target specific classes halos of interest for small-scale structure analysis and their surroundings, including in non-CDM models.}. 

Our work demonstrates that it is possible to find a functional form for the halo mass function (and, equally importantly, the window function) that results in a reasonable match to measurements from N-body simulations over a mass range of $10^6\mathrm{M}_\odot$ to $10^{16}\mathrm{M}_\odot$, redshifts from $z=0$ to $z=4$, environmental overdensities from $\delta_\mathrm{e}=-1.5$ to $1.0$, and for linear matter power spectra spanning thermal warm dark matter, fuzzy dark matter, models with interactions in the dark sector, and mixed W/CDM models, thus representing a wide variety of dark matter microphysics. However, quantitative posterior predictive check analyses clearly show that our model is \emph{not} able to explain the shapes of halo mass functions across the entire range of masses and power spectra shapes that we consider---showing systematic deviations at the 10--20\% level.

Achieving this required the adoption of a window function that depends on the slope of the power spectrum. Other works have also explored more complex window functions. For example, \cite{2026arXiv260201320R} propose a variation on the smooth-$k$ window function of \cite{2018JCAP...04..010L}, with a variable slope, chosen to allow a match to both the small-scale suppression regime of the halo mass function and the intermediate regime where dark acoustic oscillations cause features in the halo mass function. While \cite{2026arXiv260201320R} demonstrate that this model performs well for models with dark acoustic oscillations, they do not explore its performance for other transfer functions, nor in the very high mass regime (i.e., above the exponential cut off in the halo mass function).

At low masses, models with truncations in their power spectra exhibit ``artificial halos'' forming from the particle noise in the initial conditions. We introduce a simple model for this population, which we simultaneously fit to the N-body data. We find that this model is able to match the dependence of the artificial halo mass function on N-body resolution, and its evolution with redshift at $z\approx 1$ and higher, although it fails to produce a sufficiently strong evolution at lower redshifts to match the N-body results at lower redshifts.

As elucidated by \cite{2016MNRAS.456.2486D}, the choice of how halo masses are defined is crucial to obtaining the most universal form of the halo mass function. \cite{2016MNRAS.456.2486D} recommend using halo masses defined using the spherical collapse, ``virial'' density contrast definition---a recommendation that we follow here. \cite{2022arXiv220711827G} propose defining a halo as ``\ldots the collection of particles orbiting in their own self-generated potential,'' and demonstrate that the mass function of halos defined in this way is close to the original Press-Schechter prediction, providing that the critical threshold for collapse, $\delta_\mathrm{c}$ is allowed to be a weak function of peak height. In future work, it may be interesting to explore if this definition allows for a simpler and/or more successful fitting function to be obtained across mass, redshift, environment, and dark matter physics models.

As discussed in \S\ref{sec:correlations}, we have ignored the effects of covariances in the halo mass function between redshifts within an individual simulation, due to the excessive computational cost of including these in our likelihood model. However, we \emph{can} estimate the effects of these correlations on our results, utilizing our estimated covariance matrix for the MDPL2 simulation. To assess what neglecting these covariances cost us, we have re-derived the parameter uncertainties under this estimated data covariance matrix without re-running the MCMC, by exploiting the fact that our MCMC sampler recorded the predicted mass function at every likelihood. This allows us to estimate the model's Jacobian, $\partial n / \partial \theta$, directly. Combining this with the Fisher information of the negative-binomial likelihood gives the parameter covariance under any assumed weighting of the data. We compute two quantities from this. The first is the covariance a correctly-weighted fit would have returned, which measures how much constraining power the covariances remove. The second is the ``sandwich'' covariance, the true sampling uncertainty of the mis-weighted estimator we in fact used, which measures how far our quoted error bars were optimistic. The two are different questions and, by the Gauss–Markov theorem, the second is never smaller than the first.

We find that introducing our bootstrap-estimated covariance for MDPL2 inflates the halo mass function parameter (i.e., $A$, $a$, $b$, $c$, $p$, and $q$) uncertainty volume by a factor of $1.15$. The effect is strongly concentrated: the $1\sigma$ uncertainty on $c$ increases by 62\%, on $A$ by 19\% and on $b$ by 12\%, while $p$ and $q$ are essentially unaffected. This factor of $1.15$ is the ratio of the \emph{true} statistical error of the fit we performed to the posterior width that the fit reported. We further find that, had we repeated the fit with the covariance properly included, we would have found a posterior width close to the one we currently find. In other words, the correlations do not reduce the information the simulations contain about the halo mass function; it is the mis-weighting of that information which costs roughly 15\% in precision, and performing the fit correctly would recover it. We stress that these figures should be considered as lower bounds, for two reasons. Only one of the five MDPL constraints has a measured covariance, and the remaining four carry a further 18\% of the halo mass function parameter information; and our bootstrap estimator,
which resamples merger-tree root halos, captures the correlations induced by shared progenitors but is by construction blind to those induced by large-scale clustering.

Our likelihood function effectively weights each simulation by its expected variance, assuming Poisson statistics (plus our $\mathcal{C}_\mathrm{disc}$ model discrepancy term, which then results in the negative binomial likelihood that we actually use). We explored other weighting choices (e.g. down-weighting the MDPL simulations---which carry high statistical significance---to allow the Symphony/COZMIC simulations to have a stronger effect on the posterior), finding that these typically did not significantly affect the results, beyond changing the posterior volume\footnote{Of course, excluding \emph{all} COZMIC models with a small-scale suppression of power results in almost no constraints on the artificial halo model parameters, and greatly weakened constraints on the window function parameters. These remain formally consistent with the maximum posterior likelihood values that we report, however.}. Because Fisher information is additive over independent constraints, we can use the analysis described above to rank our 232 constraints by how much each contributes. Considering just the parameters of the intrinsic halo mass function ($A$, $a$, $b$, $c$, $p$, and $q$) we find that the MDPL boxes contribute around 29\% of the constraining power (MDPL2 is the largest individual contributor at 11\%), with Symphony/COZMIC zoom ins providing the remaining 71\% (each individual zoom in provides less than 2.5\%, but there are many independent realizations).

Breaking this down further, we find that for the $a$ and $b$ parameters, the constraints are strongest from the large-volume MDPL boxes (since these best constrain the exponential cutoff where these parameters matter most). Interestingly, $c$ is best constrained by the smaller-volume MDPL boxes (despite appearing in the exponential)---as these help to break the degeneracy between $c$ and $b$. The $p$ and $q$ parameters are dominated by the Symphony constraints; they affect the halo mass function primarily at lower masses, where Symphony provides the greatest lever arm. The normalization, $A$, is constrained much more uniformly by all simulations, as might be expected for an overall normalization parameter. The constraining power is therefore quite broadly distributed across our entire set of simulations.

\section{Conclusions}\label{sec:conclusion}

In this work, we have constructed and calibrated a fitting function for the dark matter halo mass function that is effective across a broad range of halo masses, redshifts, and environments, and which can be applied to power spectra arising from a wide range of dark matter microphysics models. This is, to our knowledge, the first time that halo mass functions measured from N-body zoom-in simulations of individual halos have been used to constrain such a fitting function. Furthermore, we have shown that our fitting function outperforms previous models across the broad range of masses and power spectra considered here, although prior fitting functions can perform marginally better for the specific scenarios for which they were constructed. Nevertheless, quantitative analysis shows that our model can not fully explain the shapes of halo mass functions across all masses and power spectra shapes that we consider, with systematic deviations at the 10--20\% level. While more complicated functional forms, or machine learning approaches, could, of course, result in better matches, the real test of any model of the halo mass function will be its ability to correctly predict behavior in regions of mass, redshift, environment, and power spectrum shape beyond those for which it was calibrated.

Accurate knowledge of the halo mass function is crucial for a wide variety of astrophysical and cosmological analyses. We expect that the fitting function presented in this work will be particularly valuable for studies which aim to probe dark matter microphysics on very small mass scales \citep[e.g.,][]{2021PhRvL.126i1101N,2021ApJ...917....7N,2021MNRAS.507.2432G,2023MNRAS.524.6159K}.

\section*{Acknowledgments}

This material is based upon work supported by the National Science Foundation (NSF) under Grant No.\ 2509561 (E.O.N. and A.B.) and No.\ 2407380 (V.G.). X.D. acknowledges support from the National Science Foundation through Grants No. NSF-AST-1836016 and No. NSF-AST-2205100, and by the Gordon and Betty Moore Foundation through Grant No. 8548. V.G. also acknowledges the support from the NSF CAREER Grant No. PHY-2239205, from the Research Corporation for Science Advancement under the Cottrell Scholar Program, the IBM Einstein Fellowship at the Institute for Advanced Study, and from Grant 63667 from the John Templeton Foundation. The opinions expressed in this publication are those of the author(s) and do not necessarily reflect the views of the John Templeton Foundation nor the NSF.

This work used data from the \emph{Symphony} and \emph{Milky Way-est} suites of simulations, hosted at \url{https://web.stanford.edu/group/gfc/gfcsims/}, which were supported by the Kavli Institute for Particle Astrophysics and Cosmology at Stanford, SLAC National Accelerator Laboratory, and the U.S.\ Department of Energy under contract number DE-AC02-76SF00515 to SLAC.

This work used data from COZMIC I~\citep{2025ApJ...986..127N}, II~\citep{2025ApJ...986..128A}, III~\citep{2025ApJ...986..129N}, and enhanced $P(k)$ COZMIC resimulations~\citep{2025ApJ...993...17N}. Associated data are publicly available on Zenodo at the following DOIs: COZMIC I at 10.5281/zenodo.14649137~\citep{nadler_2025_14649137}; COZMIC II at 10.5281/zenodo.14663119~\citep{nadler_2025_14663119}; COZMIC III at 10.5281/zenodo.14666735~\citep{nadler_2025_14666735}; and enhanced $P(k)$ resimulations at 10.5281/zenodo.16915369~\citep{nadler_2025_16915369}.

The CosmoSim database used in this paper is a service by the Leibniz-Institute for Astrophysics Potsdam (AIP). The MultiDark database was developed in cooperation with the Spanish MultiDark Consolider Project CSD2009-00064. The authors gratefully acknowledge the Gauss Centre for Supercomputing e.V. (\href{www.gauss-centre.eu}{\tt www.gauss-centre.eu}) and the Partnership for Advanced Supercomputing in Europe (PRACE, \href{www.prace-ri.eu}{\tt www.prace-ri.eu}) for funding the MultiDark simulation project by providing computing time on the GCS Supercomputer SuperMUC at Leibniz Supercomputing Centre (LRZ, \href{www.lrz.de}{\tt www.lrz.de}). The Bolshoi simulations have been performed within the Bolshoi project of the University of California High-Performance AstroComputing Center (UC-HiPACC) and were run at the NASA Ames Research Center.

MDPL merger tree data is available through \href{https://www.cosmosim.org/}{\tt cosmosim.org}.

%%%%%%%%%%%%%%%%%%%%%%%%%%%%%%%%%%%%%%%%%%%%%%%%%%
\section*{Data Availability}

All halo mass functions measured from N-body simulations used in this work are distributed as part of the Galacticus \href{https://github.com/galacticusorg/datasets}{datasets} repo. All code used in the analysis is part of Galacticus and is available from the Galacticus \href{https://github.com/galacticusorg/galacticus}{repo}. Relevant parameter files and details of how to run the N-body analysis and MCMC can be found at \href{https://github.com/galacticusorg/galacticusPublications/blob/master/2026/Unified\%20Halo\%20Mass\%20Function/ReadMe.md}{here}. MCMC chains from this work are available via Zenodo at the following DOI: \href{https://doi.org/10.5281/zenodo.22098931}{10.5281/zenodo.22098931}.

%%%%%%%%%%%%%%%%%%%% REFERENCES %%%%%%%%%%%%%%%%%%

% The best way to enter references is to use BibTeX:

\bibliographystyle{mnras}
\bibliography{example} % if your bibtex file is called example.bib

%%%%%%%%%%%%%%%%%%%%%%%%%%%%%%%%%%%%%%%%%%%%%%%%%%

\appendix

\section{Correlations in the Halo Mass Function}\label{app:correlations}

To account for correlations between redshifts in the measured N-body halo mass functions, we would require a joint likelihood function for the number of halos in each bin of the halo mass functions across all redshifts, one allowing us to account for correlations. As the likelihood function assumes a negative binomial distribution for each bin, we considered two possible joint likelihood functions:
\begin{enumerate}
\item Retaining the negative binomial distributions as the marginal distribution for each bin, and using a copula\footnote{A copula is a multivariate distribution function on the unit (hyper)cube for which every marginal distribution is uniform. This allows any general multivariate distribution function to be described by its marginal distributions plus a copula to describe correlations \citep{sklar1959}. This allows the correlation structure to be separated from the marginal distributions.} to include the correlations. We considered a Gaussian copula, which has the advantage of straightforwardly allowing any desired correlation structure to be applied. However, evaluation of the likelihood then requires integration over hyper-rectangles in the resulting multivariate normal distribution. Given the dimensionality and number of halos (in the MDPL simulations in particular), we were not able to find a numerically stable method to compute these integrals.
\item Using the multivariate negative binomial distribution proposed by \cite{famoye2014}. While this approach allows for a reasonably simple form for the likelihood function, incorporating correlations while retaining negative binomial distribution marginals, there is an upper limit on the strength of correlations that can be included. For the halo counts and stopping parameters relevant in this work, that upper limit is typically very small, and so this approach does not allow us to adequately capture the correlations.
\end{enumerate}

\section{MCMC convergence}\label{app:convergence}

\begin{figure*}
\begin{tabular}{cc}
\includegraphics[width=85mm]{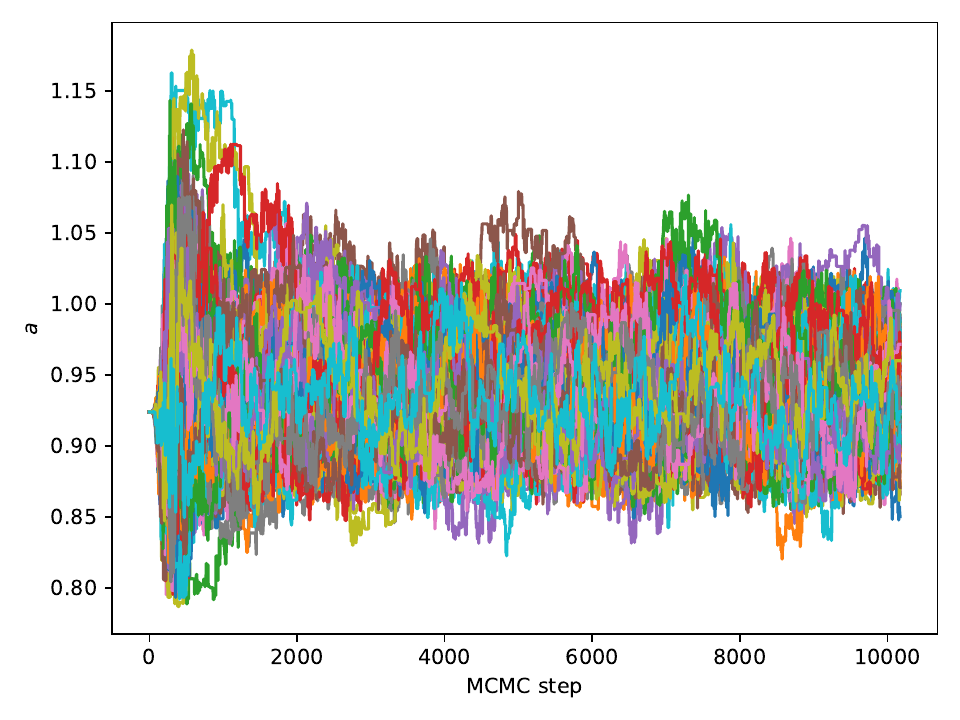} & \includegraphics[width=85mm]{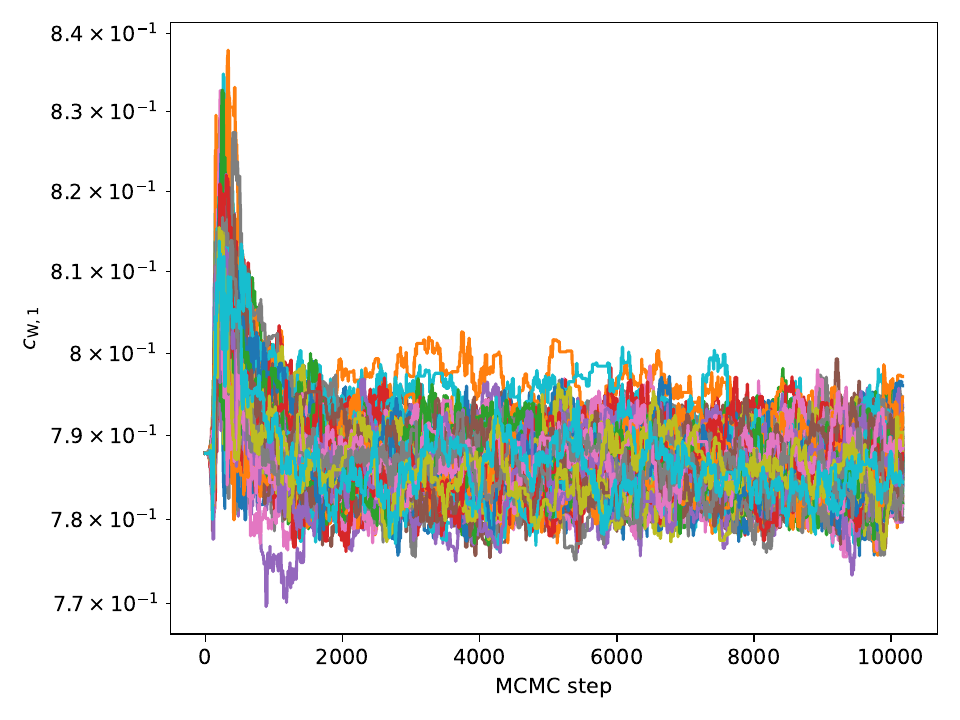} \\\includegraphics[width=85mm]{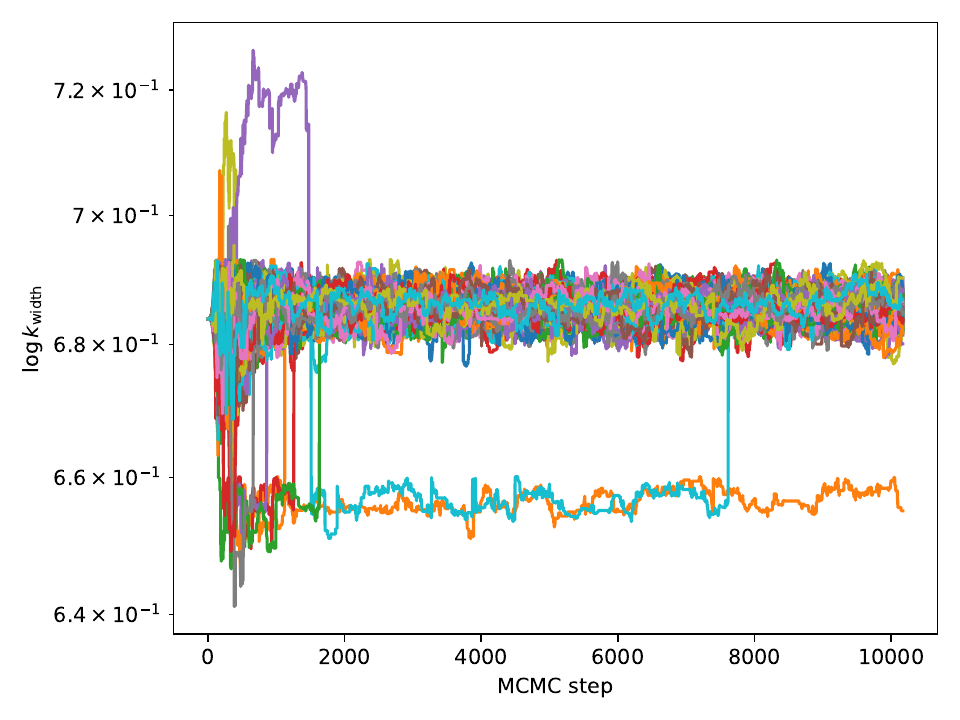} & \includegraphics[width=85mm]{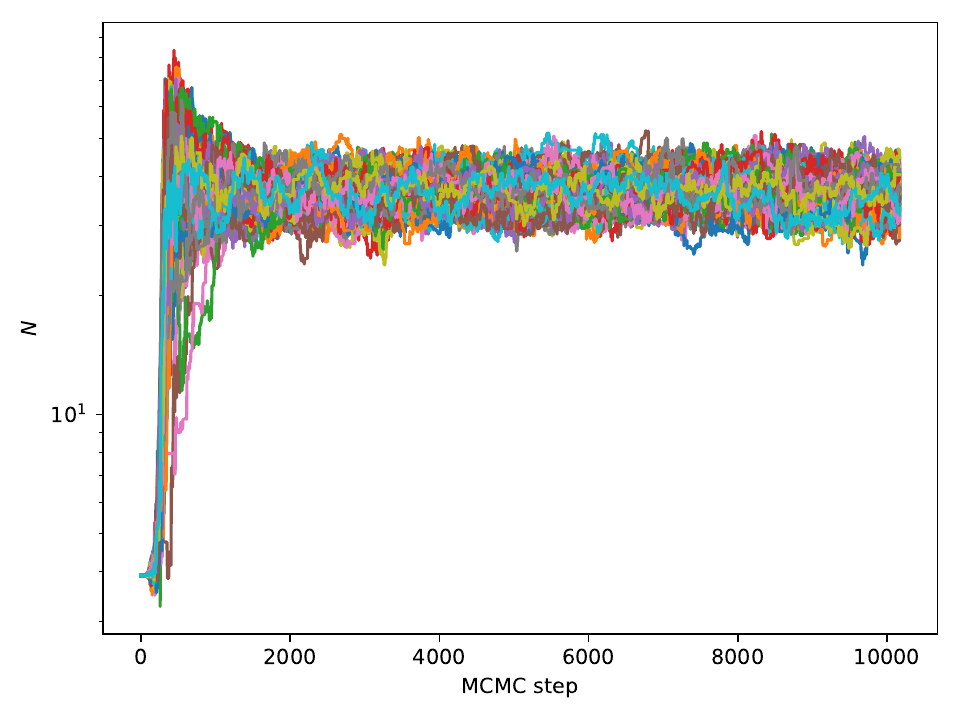}
\end{tabular}
\caption{Values of parameters for all 160 chains of our MCMC as a function of MCMC simulation step. Four representative parameters are displayed. In each panel, colored lines indicate the value of the parameter in each MCMC chain as a function of MCMC simulation step.}
\label{fig:mcmcConvergence}
\end{figure*}

We start our MCMC chains in a highly under-dispersed state (a compact, Gaussian sphere), allowing them to diffuse out to fill the high probability regions of the posterior distribution. This prevents the use of convergence statistics such as the Gelman-Rubin $\hat{R}$ parameter \citep{gelman_a._inference_1992}, which requires the chains to be initialized in an overdispersed state. Therefore, we instead rely on visual inspection of the chains to assess when the MCMC simulation has converged. Figure~\ref{fig:mcmcConvergence} shows the states of our MCMC chains for four different parameters, chosen to illustrate typical behaviors across all parameters in our model. Each parameter shows an initial period during which the chains rapidly diffuse out from their initial compact distribution---this lasts for around between a few 100, up to 1,000 steps (for the parameter $A$, lower right panel in Figure~\ref{fig:mcmcConvergence}, the chains are clearly still diffusing out from their starting point until step 1,000). In most cases, the chains appear to be well-converged after this. There are some exceptions, however. For example, in the parameter $\log k_\mathrm{width}$ (lower right panel of Figure~\ref{fig:mcmcConvergence}), two chains remain outliers for many steps. One chain eventually converges after around 7,600 steps, but the other remains an outlier. As such, we burn 7,600 steps and then simply remove the remaining non-converged chain from our analysis.

\section{Numerical/selection bias models}

In \S\ref{sec:hmfModel} we describe models for various numerical and selection effects that can bias the halo mass function recovered from N-body simulations. In this appendix, we explore the constraints on the parameters of these models.

\begin{figure*}
\includegraphics[width=7in]{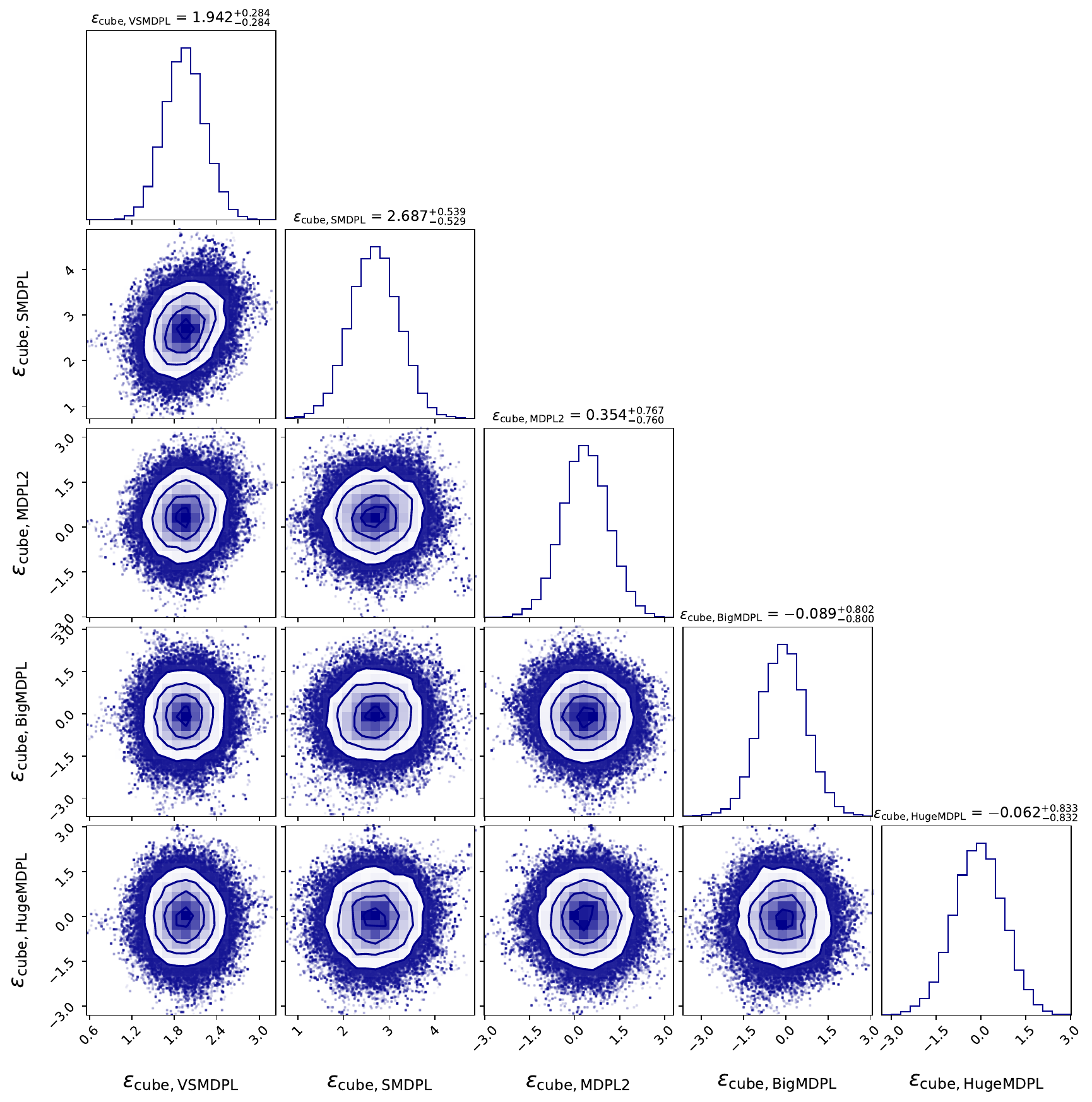} 
\caption{Posterior distributions for the parameters controlling perturbations to the halo mass functions in MDPL simulations due to missing modes from our MCMC simulation. Panels on the diagonal show the 1D marginalized distributions for each parameter (with the median and $16^\mathrm{th}$ and $84^\mathrm{th}$ percentiles shown above each panel), while off-diagonal panels show joint distributions for each pair of parameters. All posterior distributions are compact and unimodal.}
\label{fig:perturbationsCornerPlot}
\end{figure*}

We begin by considering the model for the effects of the finite box size in the MDPL cosmological simulations. In this model, we compute the expected variance, $\sigma^2$, in the box due to the finite number of modes contained within the simulation cube. We then allow the halo mass function to be perturbed by an amount $\epsilon \sigma$, where $\epsilon$ is a parameter to be constrained by our MCMC simulation. Through this construction, for an ensemble of simulations cubes we expect $\epsilon$ to be a random variable following standard normal distribution and so adopt that as a prior for these parameters. The actual values recovered then describe the specific deviation found in each MDPL simulation. Figure~\ref{fig:perturbationsCornerPlot} shows the posterior distributions for these parameters. For the larger volumes (MDPL2, BigMDPL, and HugeMDPL), the posteriors are essentially consistent with the priors---there is no constraint on the $\epsilon$ parameter. In these large volumes, the missing modes have little influence on the halo mass function. In the two smaller volumes (VSMDPL and SMDPL), we find posterior distributions that are peaked away from $\epsilon=0$, with median values of $\epsilon_\mathrm{cube,VSMDPL}=1.88$ and $\epsilon_\mathrm{cube,SMDPL}=2.61$ and with posterior distributions that are inconsistent with zero, indicating that missing modes have a measurable effect on the halo mass functions. When averaged over the posterior, the probably to find such a large deviation in VSMDPL is 3\%, and in SMDPL is 1\%. Together this is very unlikely, suggesting that these parameters may be absorbing other limitations of our modeling (e.g. the lack of any covariance structure in our likelihood functions). It should be noted that, while this effect is measurable, it is still small in an absolute sense---in Figure~\ref{fig:hmfMDPLBoxes} the difference between the fitted mass function (solid lines) and the intrinsic mass function (which removes this perturbation due to missing modes; dashed line) is tiny.

\begin{figure*}
\includegraphics[width=7in]{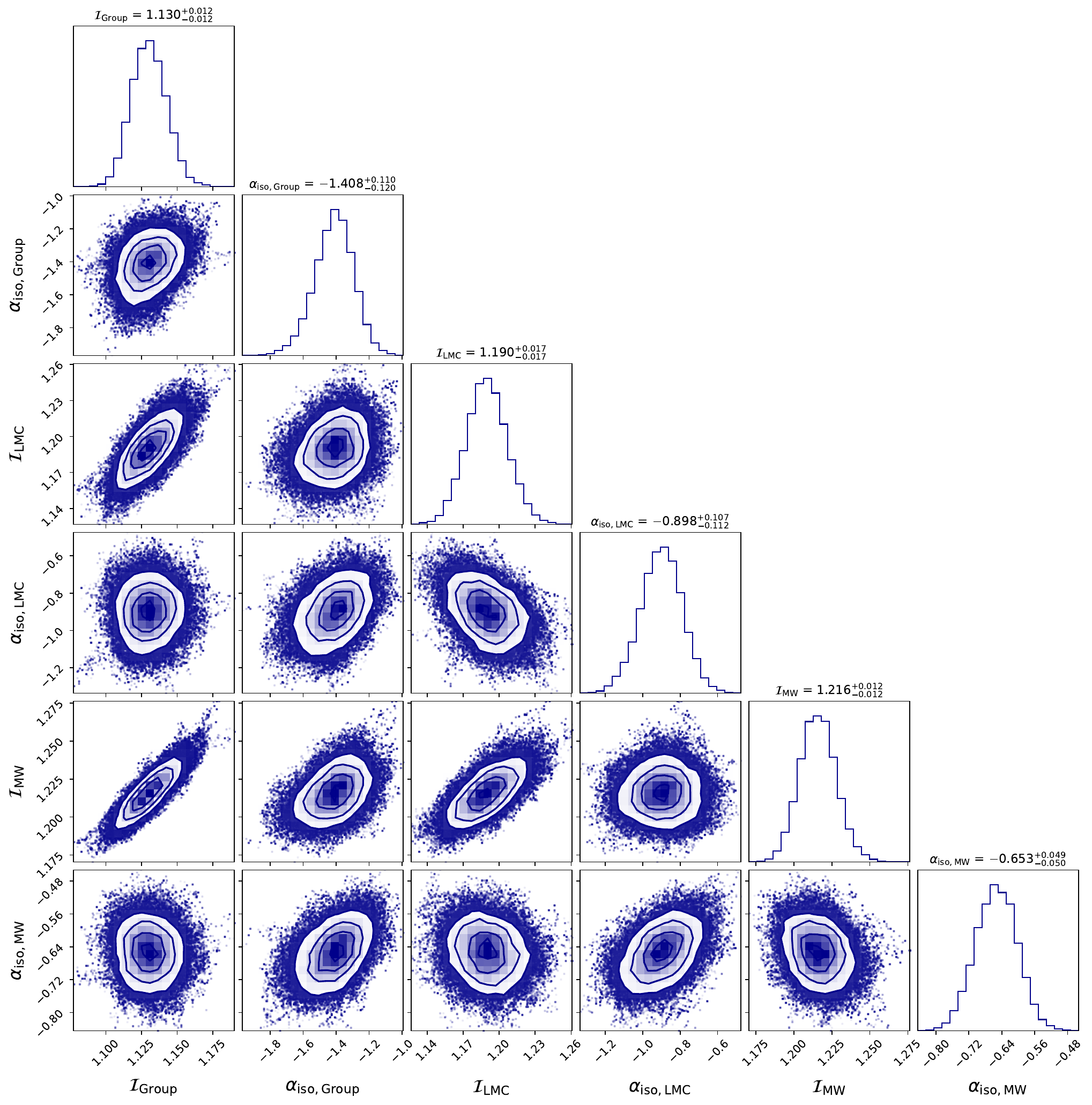} 
\caption{Posterior distributions for the parameters controlling the isolation bias model from our MCMC simulation. Panels on the diagonal show the 1D marginalized distributions for each parameter (with the median and $16^\mathrm{th}$ and $84^\mathrm{th}$ percentiles shown above each panel), while off-diagonal panels show joint distributions for each pair of parameters. All posterior distributions are compact and unimodal.}
\label{fig:isolationCornerPlot}
\end{figure*}

Figure~\ref{fig:isolationCornerPlot} shows posterior distributions for parameters of our isolation bias model, which aims to account for any bias introduced into the halo mass functions of Symphony/Milky Way-est/COZMIC simulations due to the isolation criteria imposed when selecting halos for resimulation. Recall that for this model we adopted a normal prior for $\log \mathcal{I}$ (the isolation bias at $z=0$), centered at $\mathcal{I}=1$, and a uniform prior between $-3$ and $0$ for $\alpha_\mathrm{iso}$, the redshift dependence of this bias. We find that, in all cases, values of $\mathcal{I}\approx 1.1$--1.2 are preferred, suggesting that isolation results in a small positive bias in the halo mass functions. The $\alpha_\mathrm{iso}$ parameters, controlling the redshift dependence, show more diversity (approximately $-0.9$ for LMC, $-0.7$ for Milky Way, and $-1.4$ for Group simulations.

\begin{figure*}
\includegraphics[width=7in]{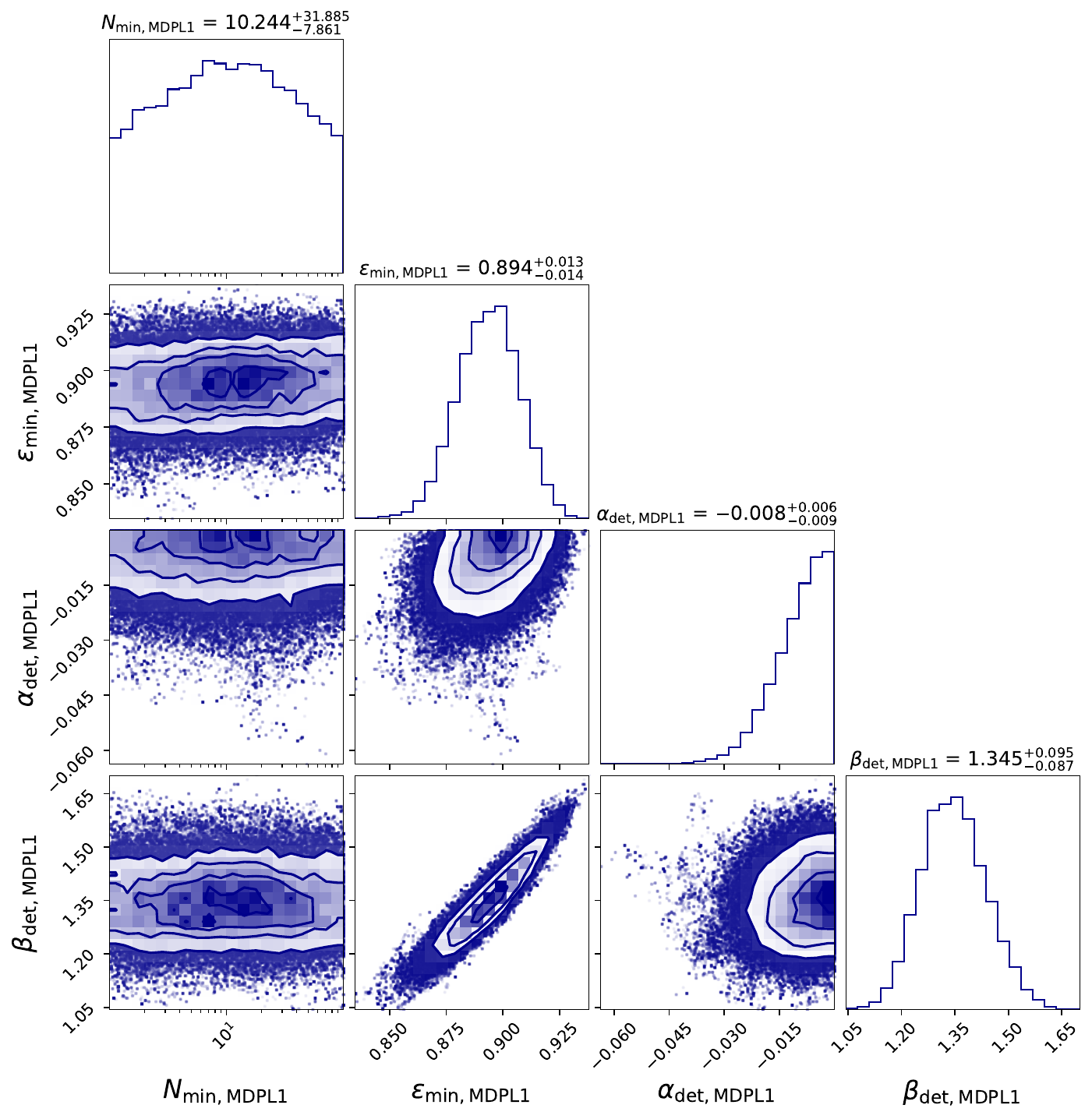} 
\caption{Posterior distributions for the parameters controlling the halo detection efficiency model for Rockstar MDPL type-1 from our MCMC simulation. Panels on the diagonal show the 1D marginalized distributions for each parameter (with the median and $16^\mathrm{th}$ and $84^\mathrm{th}$ percentiles shown above each panel), while off-diagonal panels show joint distributions for each pair of parameters.}
\label{fig:detection1CornerPlot}
\end{figure*}

\begin{figure*}
\includegraphics[width=7in]{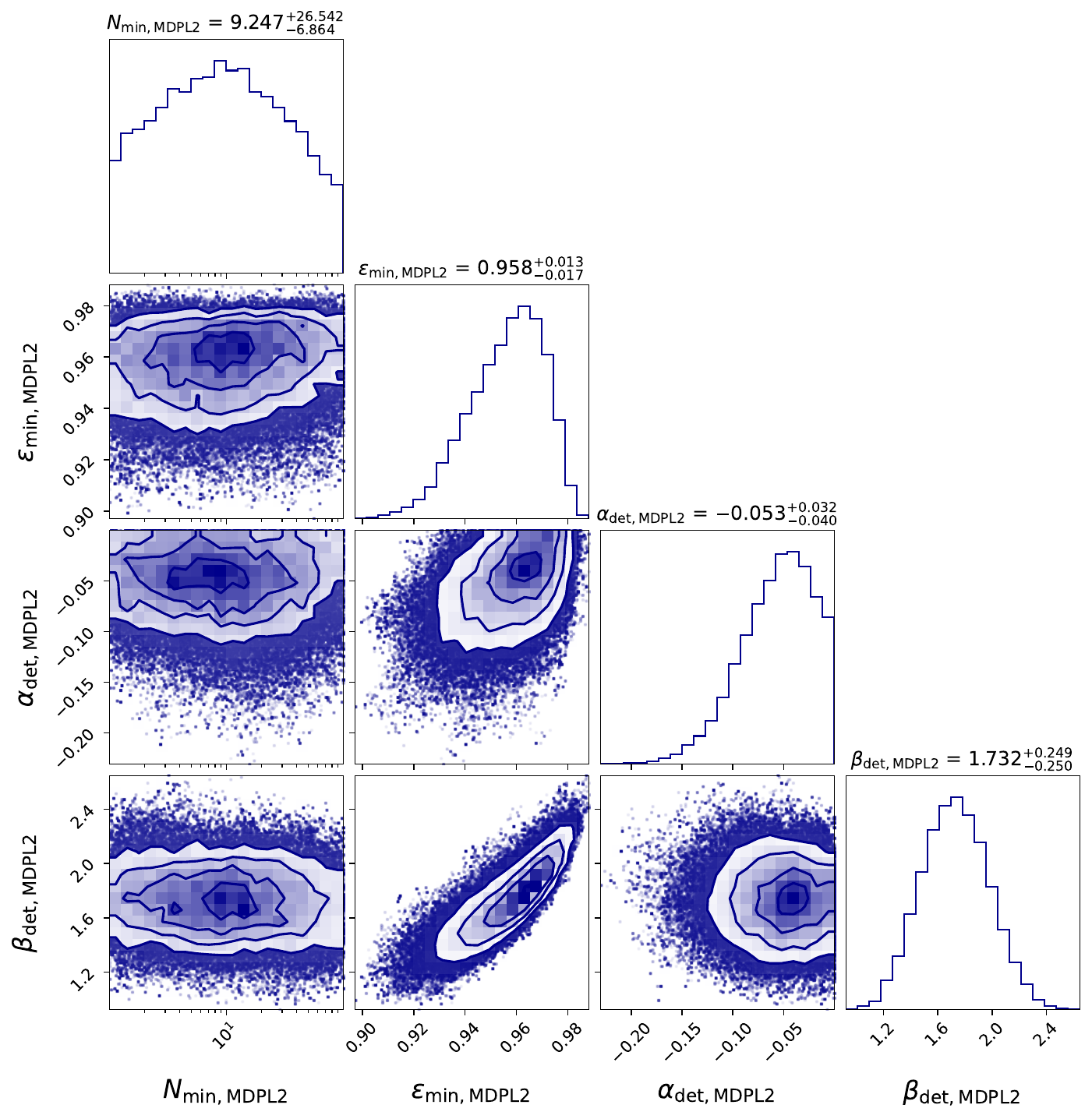} 
\caption{Posterior distributions for the parameters controlling the halo detection efficiency model for Rockstar MDPL type-2 from our MCMC simulation. Panels on the diagonal show the 1D marginalized distributions for each parameter (with the median and $16^\mathrm{th}$ and $84^\mathrm{th}$ percentiles shown above each panel), while off-diagonal panels show joint distributions for each pair of parameters.}
\label{fig:detection2CornerPlot}
\end{figure*}

\begin{figure*}
\includegraphics[width=7in]{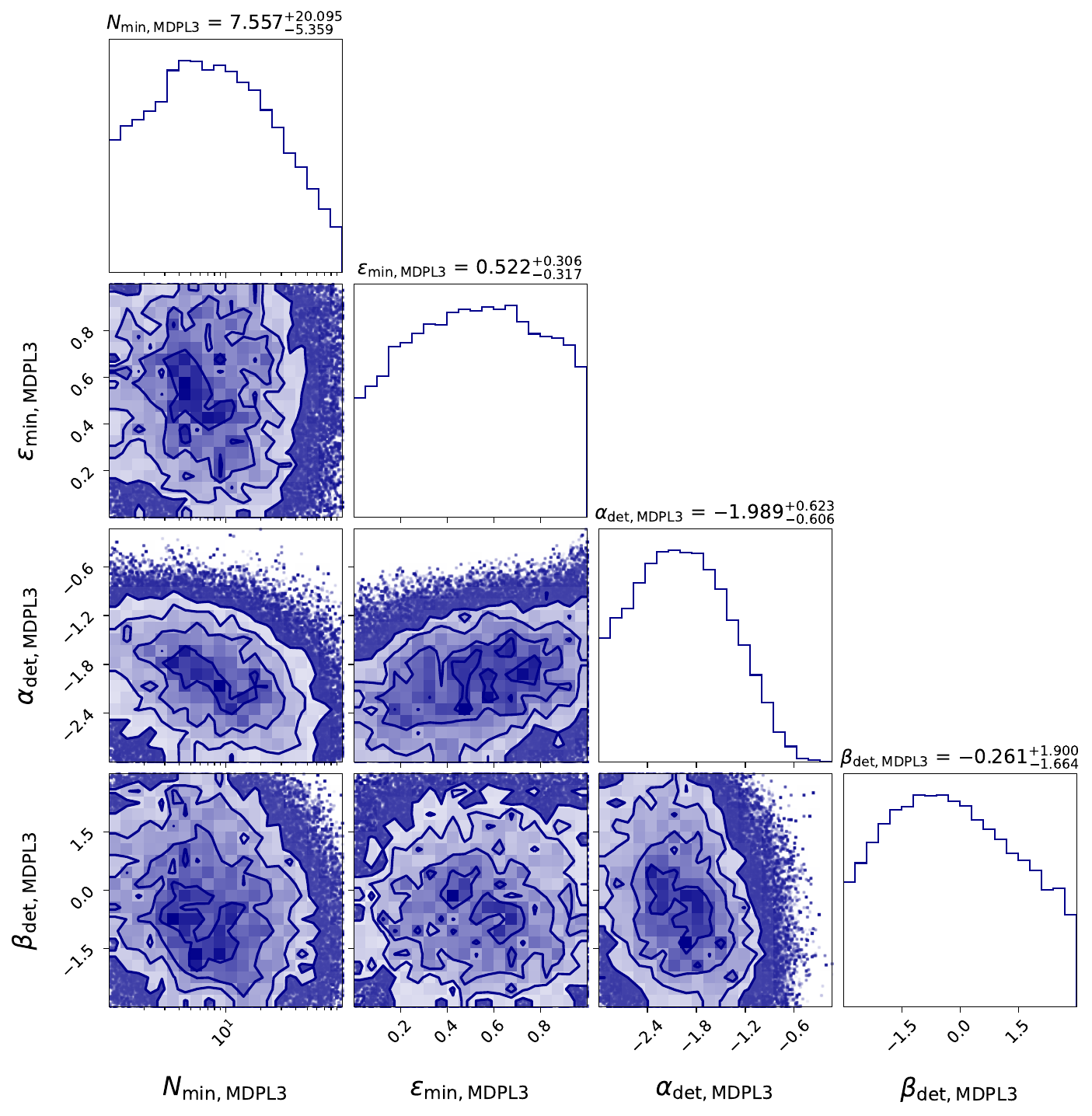} 
\caption{Posterior distributions for the parameters controlling the halo detection efficiency model for Rockstar MDPL type-3 from our MCMC simulation. Panels on the diagonal show the 1D marginalized distributions for each parameter (with the median and $16^\mathrm{th}$ and $84^\mathrm{th}$ percentiles shown above each panel), while off-diagonal panels show joint distributions for each pair of parameters.}
\label{fig:detection3CornerPlot}
\end{figure*}

\begin{figure*}
\includegraphics[width=7in]{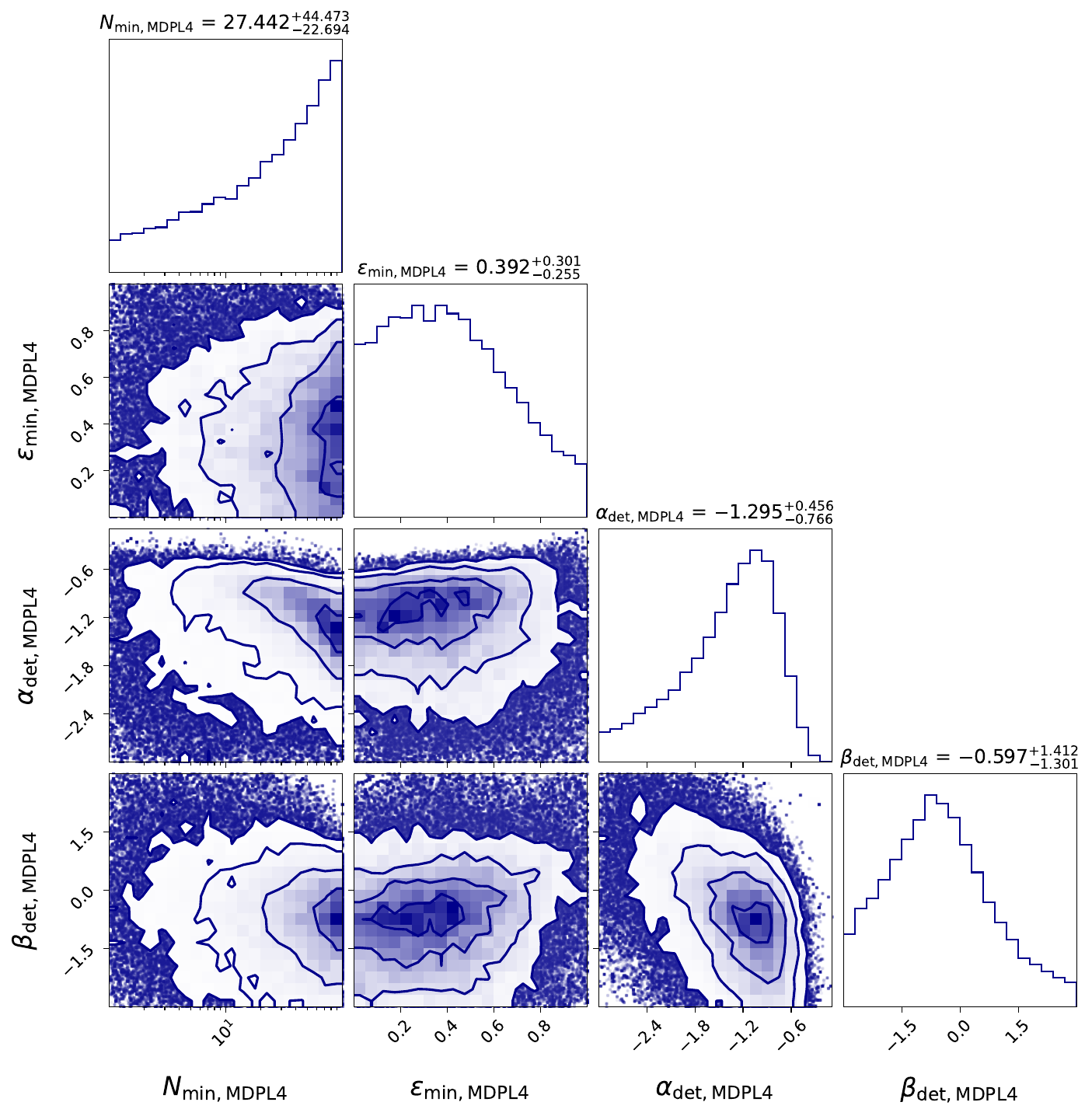} 
\caption{Posterior distributions for the parameters controlling the halo detection efficiency model for Rockstar MDPL type-4 from our MCMC simulation. Panels on the diagonal show the 1D marginalized distributions for each parameter (with the median and $16^\mathrm{th}$ and $84^\mathrm{th}$ percentiles shown above each panel), while off-diagonal panels show joint distributions for each pair of parameters.}
\label{fig:detection4CornerPlot}
\end{figure*}

\begin{figure*}
\includegraphics[width=7in]{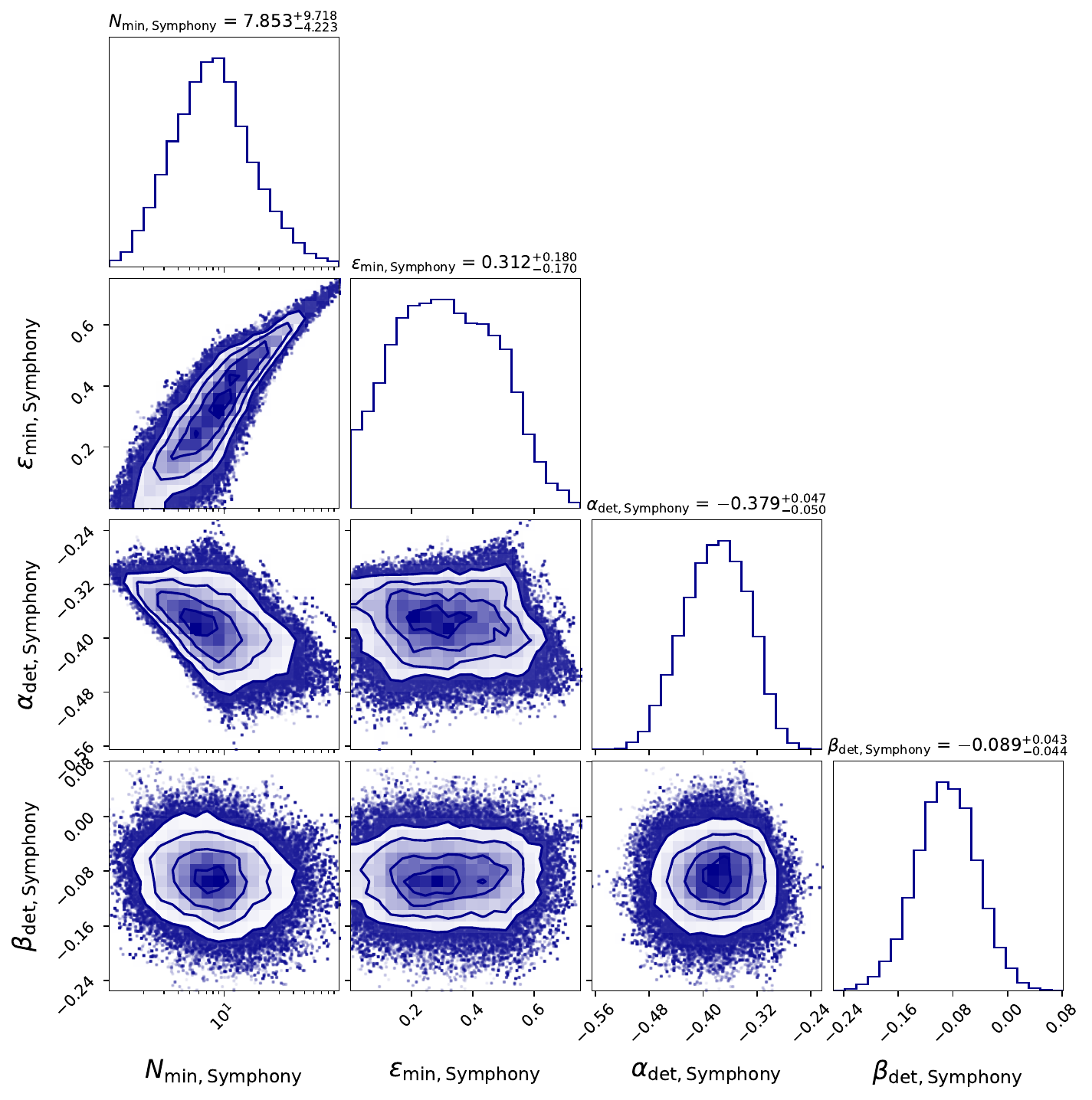} 
\caption{Posterior distributions for the parameters controlling the halo detection efficiency model for Rockstar Symphony from our MCMC simulation. Panels on the diagonal show the 1D marginalized distributions for each parameter (with the median and $16^\mathrm{th}$ and $84^\mathrm{th}$ percentiles shown above each panel), while off-diagonal panels show joint distributions for each pair of parameters.}
\label{fig:detectionSymphonyCornerPlot}
\end{figure*}

Finally, we examine the posterior distributions of the parameters of our halo detection efficiency model. Recall that we adopt independent parameters for four separate Rockstar versions used in MDPL, and another for Symphony/Milky Way-est/COZMIC. Posterior distributions for these are shown in Figures~\ref{fig:detection1CornerPlot} through \ref{fig:detectionSymphonyCornerPlot}. We find that $N_\mathrm{min}$ is typically constrained to be around 10, with a broad distribution, although for the ``MDPL4'' type, larger values are preferred. The detection efficiency parameter, $\epsilon_\mathrm{min}$, is typically found to be around $0.9$, although it is smaller for the ``MDPL3'' and ``MDPL4'' types, and is strongly correlated with $\beta_\mathrm{det}$ in many cases. 

\end{document}